\documentclass[letterpaper,10pt]{article}
\usepackage[utf8x]{inputenc}
\usepackage{multirow}
\usepackage[table]{xcolor}
\usepackage{graphicx}
\usepackage{palatino,longtable}
\usepackage{array}
\usepackage{rotating}
\usepackage{calc}
\usepackage{subfigure}
\usepackage[normalem]{ulem}
\newcommand{\MyCaption}[3]{\caption[#1]{#1 #2\label{#3}}}

\newcommand{\MyEquation}[1]{\begin{equation}
\ensuremath{\input{#1}}
\end{equation}
}

\newcommand{\MyEquationInline}[1]{\ensuremath{\input{#1}}}

\newcommand{\MyEquationLabeled}[2]{\begin{equation}
\ensuremath{\input{#1}}
\label{#2}
\end{equation}
}

\newcommand{\MyConstraintReference}[1]{ (see Constraint \ref{#1}) }

\newcommand{\MyEquationReference}[1]{ (see Equation \ref{#1}) }

\newcommand{\MyFigureReference}[1]{ (see Figure \ref{#1}) }

\newcommand{\MyHline}[0]{\noalign{\hrule height 2pt}}

\newcommand{\MyIgnore}[1]{}

\newcommand{\MySectionReference}[1]{ (see Section \ref{#1}) }

\newcommand{\MyTableReference}[1]{ (see Table \ref{#1}) }

\newcommand{\MyPageSetup}[0]{
\newlength{\myrightmargin}
\newlength{\myleftmargin}
\newlength{\mytopmargin}
\newlength{\mybottommargin}

\setlength{\myrightmargin}{1.0in}
\setlength{\myleftmargin}{1.0in}
\setlength{\mytopmargin}{1.0in}    
\setlength{\mybottommargin}{1.0in}
\setlength{\oddsidemargin}{0.0in}   % extra room on inside side

\setlength{\evensidemargin}{0 in}
\setlength{\marginparsep}{0 in}
\setlength{\marginparwidth}{0 in}
\setlength{\hoffset}{\myleftmargin - 1.0in}
\setlength{\textwidth}{8.5in -\myleftmargin -\myrightmargin -\oddsidemargin}

\setlength{\voffset}{\mytopmargin -1.0in}
\setlength{\topmargin}{0 in}
\setlength{\headheight}{12pt}
\setlength{\headsep}{20pt}
\setlength{\footskip}{36pt}
\setlength{\textheight}  {11.0in-\mytopmargin-\mybottommargin-\headheight-\headsep-\footskip} 
}

\MyIgnore{
\newcommand{\MyShortNameA}[0]{1000}
\newcommand{\MyShortNameB}[0]{990 --- 1}
\newcommand{\MyShortNameC}[0]{900 --- 100}
\newcommand{\MyShortNameD}[0]{900 ---  10}
\newcommand{\MyShortNameE}[0]{500 --- 500}
\newcommand{\MyShortNameF}[0]{500 --- 499}
\newcommand{\MyShortNameG}[0]{500 --- 400}
\newcommand{\MyShortNameH}[0]{500 --- 300}
\newcommand{\MyShortNameI}[0]{500 ---  50}
\newcommand{\MyShortNameJ}[0]{200 --- 200}
\newcommand{\MyShortNameK}[0]{165 ---  17}
\newcommand{\MyShortNameL}[0]{100 ---  100}
\newcommand{\MyShortNameM}[0]{91 ---  90}
\newcommand{\MyShortNameN}[0]{1 ---  1}
}
\newcommand{\MyShortNameFormat}[1]{\textit{#1}~}
\newcommand{\MyShortNameA}[0]{\MyShortNameFormat{100}}
\newcommand{\MyShortNameB}[0]{\MyShortNameFormat{90,10}}
\newcommand{\MyShortNameC}[0]{\MyShortNameFormat{90\ldots1}}
\newcommand{\MyShortNameD}[0]{\MyShortNameFormat{80\ldots2}}
\newcommand{\MyShortNameE}[0]{\MyShortNameFormat{50,50}}
\newcommand{\MyShortNameF}[0]{\MyShortNameFormat{50,49,1}}
\newcommand{\MyShortNameG}[0]{\MyShortNameFormat{50,40,10}}
\newcommand{\MyShortNameH}[0]{\MyShortNameFormat{50,30,10,10}}
\newcommand{\MyShortNameI}[0]{\MyShortNameFormat{50\ldots5}}
\newcommand{\MyShortNameJ}[0]{\MyShortNameFormat{20\ldots20}}
\newcommand{\MyShortNameK}[0]{\MyShortNameFormat{16\ldots1}}
\newcommand{\MyShortNameL}[0]{\MyShortNameFormat{10\ldots10}}
\newcommand{\MyShortNameM}[0]{\MyShortNameFormat{10\ldots9}}
\newcommand{\MyShortNameN}[0]{\MyShortNameFormat{1\ldots 1}}

\newcommand{\MyShortNameAAJB}[0]{Not defined}
\newcommand{\MyShortNameBAJB}[0]{Not defined}
\newcommand{\MyShortNameCAJB}[0]{Not defined}
\newcommand{\MyShortNameDAJB}[0]{Not defined}
\newcommand{\MyShortNameEAJB}[0]{Not defined}
\newcommand{\MyShortNameFAJB}[0]{Not defined}
\newcommand{\MyShortNameGAJB}[0]{Not defined}
\newcommand{\MyShortNameHAJB}[0]{Not defined}
\newcommand{\MyShortNameIAJB}[0]{Not defined}
\newcommand{\MyShortNameJAJB}[0]{Not defined}
\newcommand{\MyShortNameKAJB}[0]{Not defined}
\newcommand{\MyShortNameLAJB}[0]{Not defined}
\newcommand{\MyShortNameMAJB}[0]{Not defined}
\newcommand{\MyShortNameNAJB}[0]{Not defined}

\newcommand{\MyShortNameACLC}[0]{Not defined}
\newcommand{\MyShortNameBCLC}[0]{Not defined}
\newcommand{\MyShortNameCCLC}[0]{Not defined}
\newcommand{\MyShortNameDCLC}[0]{Not defined}
\newcommand{\MyShortNameECLC}[0]{Not defined}
\newcommand{\MyShortNameFCLC}[0]{Not defined}
\newcommand{\MyShortNameGCLC}[0]{Not defined}
\newcommand{\MyShortNameHCLC}[0]{Not defined}
\newcommand{\MyShortNameICLC}[0]{Not defined}
\newcommand{\MyShortNameJCLC}[0]{Not defined}
\newcommand{\MyShortNameKCLC}[0]{Not defined}
\newcommand{\MyShortNameLCLC}[0]{Not defined}
\newcommand{\MyShortNameMCLC}[0]{Not defined}
\newcommand{\MyShortNameNCLC}[0]{Not defined}

\newcommand{\MyLittleS}[0]{\ensuremath{s}~}
\newcommand{\MyLargeS}[0]{\ensuremath{S}~}
\newcommand{\AverageNonLCC}{\ensuremath{ \MyLittleS \stackrel{}{=} \frac{n- \MyCloseLCC}{m-1}}~}
\newcommand{\MyMath}{cyan}
\newcommand{\MyLCC}{orange}
\newcommand{\MyM}{red}
\newcommand{\MyERef}[1]{\mbox { (E\ref{#1})}}
\newcommand{\MyCRef}[1]{\mbox { (C\ref{#1})}\cellcolor{\MyLCC}}
\newcommand{\MyAttackNotation}[2]{\ensuremath{A_{#1,#2}}~}
\renewcommand{\MyCaption}[3]{\caption[#1]{\textbf{#1} #2\label{#3}}}

\newcommand{\mc}[3]{\multicolumn{#1}{#2}{#3}}
\newcommand{\mr}[3]{\multirow{#1}{#2}{#3}}

\newcommand{\MyExplanation}[3] {Equation \ref{#3} shows the estimation of \AverageNonLCC ~
  for  test case  \ensuremath{m = #1}~and \ensuremath{\MyCloseLCC = #2}~
  for large values of $n$:}

\newcommand{\MyCloseLCC}{\ensuremath {\mathclose \mid LCC  \mathclose \mid}~}
\newcommand{\MyLocalSubfigure}[3]{\subfigure[#1] {\label{#2}  \includegraphics*[viewport = 180mm 220mm 20mm 55mm, clip, width = 3.0in, angle = 90]{#3}}}

\title{
Connectivity Damage to a Graph \\
by the Removal of an Edge or a Vertex
}

\author{{Charles L. Cartledge and Michael L. Nelson} \\
{\{ccartled,mln\}@cs.odu.edu} \\
Old Dominion University, Department of Computer Science \\
Norfolk, VA 23529 USA
}

\MyPageSetup
\begin{document}

\maketitle

\begin{abstract}
The approach of quantifying the damage inflicted on a graph in Albert, Jeong and Barab{\'a}si's (AJB) 
report ``Error and Attack Tolerance of Complex Networks''  using
the size of the largest connected component and the average size of the remaining components does not
capture our intuitive idea of the damage to a graph caused by disconnections. 
We evaluate an alternative metric based on average inverse path lengths
(AIPLs) that better fits our intuition that a graph can still be
reasonably functional even when it is disconnected.
We compare our metric with AJB's using a test set of graphs and report the differences.  
AJB's report should not be confused with a report by  Crucitti et al. with
the same name.

Based on our analysis of graphs of different sizes and types, and using various
numerical and statistical tools; the ratio of the average inverse path lengths
of a connected graph of the same size as the sum of the size of the
fragments of the  disconnected graph
can be used as a metric about the damage of a graph by the removal of an edge
or a node.  This damage is reported in the range (0,1) where 0 means that the
removal had no effect on the graph's capability to perform its functions.  A
1 means that the graph is totally dysfunctional.  We exercise our metric on a
collection of sample graphs that have been subjected to various attack profiles
that focus on edge, node or degree betweenness values.

We believe that this metric can be used to quantify the damage done to the graph by an attacker,
and that it can be used in evaluating the positive effect of adding additional
edges to an existing graph.

\end{abstract}

%  \tableofcontents
%  \listoffigures
%  \listoftables

\section{Introduction}\label{sec:introduction}
The likelihood of a graph, or network to remain functional in the face of random failures and
directed attacks has been the interest to many different authors.  In attempting
to understand the problem and their root causes; we reviewed
 \cite{Albert2000, crucitti2004error, goh:classification_of_scale-free_networks,
dekker2004simulating, holme2002attack, wang2003cns,criado2005effective,netotea2006evolution,beygelzimer2005improving,zio-modeling,lee2006rnt,newth2004evolving,lee-attack,cohen2001breakdown}.
 Our  desire is to have a single value that could be used across graphs as an indicator of the graph's
``damage,'' ``robustness,'' or  ``general health.''  This value would be applicable whether or not
the graph was \emph{connected} or \emph{disconnected}.

The paper documents the approach used to arrive at a single metric that can be used to
report the damage caused to the graph by the removal of either an 
edge or a vertex.  The complement of this damage estimate would be the ``health'' of 
a graph by the addition of an edge.
Included are
the supporting equations, the data
used to test main stream and ``corner''  test cases   and an analysis of the results.

Our sense is that a graph may still perform most of its duties (i.e., communicate
between nodes, maintain data in a node, respond to queries, etc.) even when it
may not be able to perform those functions between any arbitrary nodes $u$ and $v$.
In this sense, a  graph may be \emph{connected} or \emph{disconnected}.

There are a number of metrics that can be used to quantify different aspects of a graph
(see sections \ref{sec:connected} and \ref{sec:disconnected}).
Most of these metrics do not have a meaning when the graph
is disconnected.  But still; our intuition says that a disconnected graph  may still be able
to perform most of its functions.  Our intuition is captured in Figure \ref{fig:notionalGraphs}
where representative connected and disconnected graphs are presented.  If an attacker
intent on disrupting the functions of a graph, then it is probably reasonable to assume that
the attacker would not be content with simply the disconnection of the graph. The attacker
would probably want to cause greater damage.
\begin{figure}
\centering
\subfigure[Representative connected graph]{\label{fig:notationalConnected}   \includegraphics[width=2.0in, angle=-90]{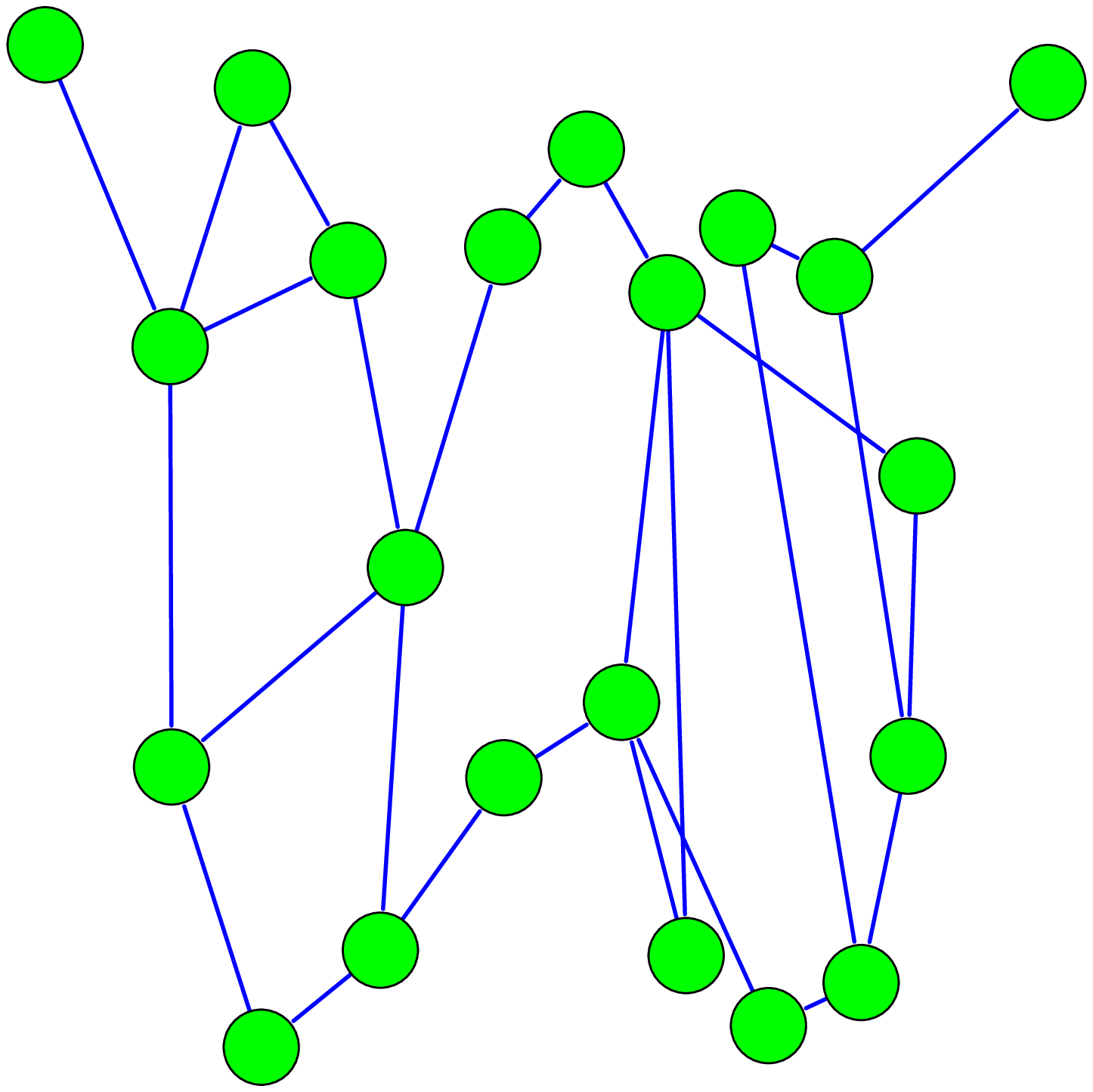}}
\subfigure[Representative disconnected graph]{\label{fig:notationalDeleted}   \includegraphics[width=2.0in, angle=-90]{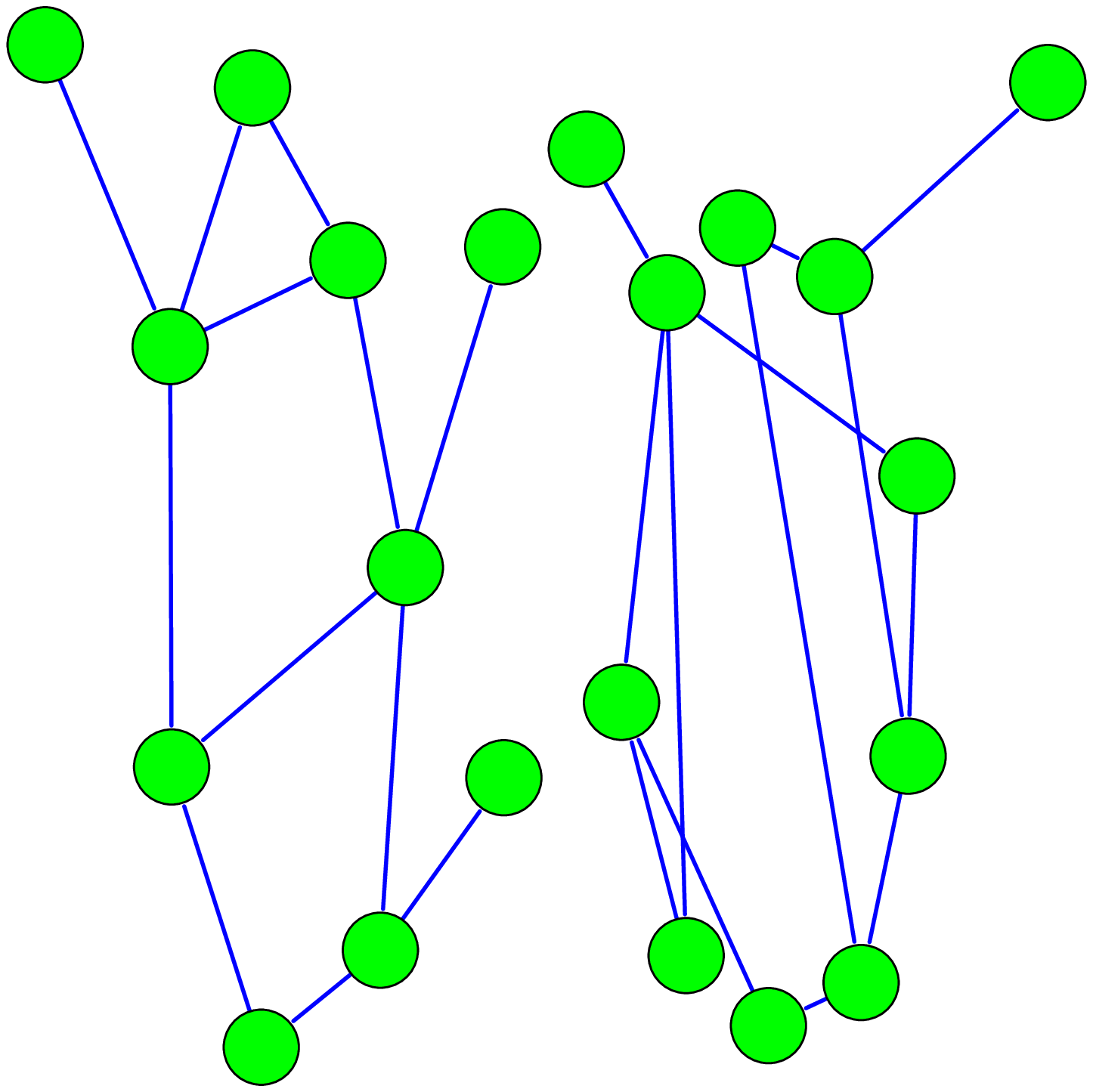}}
\MyCaption{Representative graphs.}
{The graph in Figure \ref{fig:notationalConnected} is connected, while the one in Figure \ref{fig:notationalDeleted} is not.
Yet, our intuition is that both graphs can perform much of their functions and meet 
most of their responsibilities even though they may not be able to meet all of them. }
{fig:notionalGraphs}
\end{figure}

Our quest is to derive a metric  for a graph (that is either connected or disconnected)
that reports the inability of a graph to perform its functions.  The inverse of this metric
would  report its ability to perform its functions,  thusly how healthy the graph is.  We intend for
this metric to form the basis for a ``game'' where an attacker selects a graph component (either an
edge or a vertex) for removal based on the amount of damage that the removal will cause to
the graph.  Additionally, the graph will be able to ``repair'' itself through the addition 
of new edges in between selected nodes that would result in a ``less damaged'' graph.  The metric
could be used as part of a ``game'' where an attacker and the graph alternated turns.  The attacker
could be given the equivalent of a some number of ``bullets'' to damage a graph and then
the graph would be given the same number of repair opportunities to ``repair'' itself.

Section \ref{sec:relwork} presents related work by Albert, Jeong and  Barab{\'a}si, and Criado, Flores, et al., and
Holme and  Kim among others.  A brief synopsis of relevant papers from these authors is given and how
the metric that we intuit exists is different than that put forth by the authors.  
Section \ref{sec:alt} presents the criteria that our metric must exhibit.  
Section \ref{sec:comparison} investigates how a collection of different metrics performs against
a series of sample graphs.  These proposed metrics are use against a small sample graph to 
triage candidate metrics.  Once our metric is identified, we introduce a series of larger graphs
and continue our investigation. Section \ref{sec:analysis} provides a summary analysis of
Albert, Jeong and  Barab{\'a}si's paper. Section \ref{sec:conclusion} contains our conclusion.  
Appendix \ref{sec:metrics} contains
a comparison of various graph related metrics applicable to both connected and disconnected
graphs. 
Appendix \ref{sec:derivation} contains a more detailed analysis  of
Albert, Jeong and  Barab{\'a}si's paper.
Appendix \ref{sec:attackProfiles} has a series of profiles that an attacker could 
use when seeking to damage a graph.  These profiles contain techniques that can focus on either
edges or vertices and then summarizes which profile is most effective.

\section{Related work}\label{sec:relwork}

Albert, Jeong and  Barab{\'a}si's (AJB) paper \cite{Albert2000} looks at the effect on the average (or expected)
path length for a graph (specifically  snapshots of the Internet and the WWW) when the highest degreed
node (be it an Internet router, or a well connected HTML page) is removed from the graph.
Within their context, the Internet is a graph where routers equate to nodes and communications links
equate to edges.  Also  the WWW is a graph where pages equate to nodes and HTML links equate to
edges.
They
proposed a tuple metric $(LCC, S, s)$ based on the proportion of the graph represented by the ratio of
largest connected component
$LCC$ to the entire graph \MyLargeS
and the mean size of all remaining fragments \MyLittleS.

Klau and Weiskircher \cite{klau2005robustness} formalized AJB's idea
into a two argument tuple $(S, s)$.
Holme and Kim et al. \cite{holme2002attack} took AJB's paper and
expanded it by introducing the idea of using the \emph{average inverse path length} (AIPL) as an
approach to measure the vulnerability of a graph to different types of attacks.
Crucitti,  Latora, et al. \cite{crucitti2004error} published a  paper
with the same title as AJB's, dealing with the same general topic, but
proposing a  metric they called \emph{global efficiency}.  Their global efficiency is
AIPL, but with a different name.
Notetea and Pongor
\cite{netotea2006evolution} proposed measuring the ``robustness'' of a network by computing the
AIPL before and after a change is made to a graph under consideration. If the robustness
of the graph is improved, then the change becomes permanent.  If the robustness decreases then
the change is reverted.
Criado, Flores et al. in \cite{criado2005effective}  propose to quantify the vulnerability of a graph
based on the number of nodes, number of edges and the standard deviation of the degrees of the
nodes.  Ideas from these and other authors are expanded upon in the following sections.

\subsection{Ideas from Albert, Jeong and  Barab{\'{a}}si}
Equations \ref{equ:n} through \ref{equ:s} were derived from Albert, Jeong and  Barab{\'a}si \cite{Albert2000},
and are the basic  definitions for the number of nodes $n$ in the graph at any point in time.
At that point in time, there is a set of clusters \MyLittleS in the graph.  If the graph is connected then
there is one cluster.
In \cite{Albert2000}, the node with the highest degree is removed (along with
its adjacent edges) and all values are computed again.
  $n$ starts at an initial value and is decremented at each time step until all nodes are disconnected.

Equation \ref{equ:m} is the number of clusters (components) in the set of clusters $c$.
Equation \ref{equ:LCC} identifies the size of the largest connected component $LCC$  in $c$.
Equation \ref{equ:S} is the ratio (percentage) of the size of $LCC$ to the current $n$.
Equation \ref{equ:s} is the mean size of all the remaining clusters (i.e., less the $LCC$) in the graph.
The minimal values of \MyLittleS under differing
conditions (\MyLittleS$=f(n, LCC, m)$) are shown in Table \ref{tbl:s2}.
\begin{equation}
n  \stackrel{\mbox{\tiny{def}}}{=} \mbox{ number of nodes in } G
\label{equ:n}
\end{equation}

\begin{equation}
c  \stackrel{\mbox{\tiny{def}}}{=} \mbox{ set of clusters in } G
\label{equ:c}
\end{equation}

\begin{equation}
m \stackrel{}{=}  \mathclose \mid c  \mathclose \mid
\label{equ:m}
\end{equation}

\begin{equation}
LCC \stackrel{}{=} max(\mid \mathclose{<}c \mathclose{>} \mathclose \mid)
\label{equ:LCC}
\end{equation}

\begin{equation}
S \stackrel{}{=} \frac{\MyCloseLCC}{n}
\label{equ:S}
\end{equation}

\begin{equation}
\AverageNonLCC
\label{equ:s}
\end{equation}

The various characteristics in   equations \ref{equ:n} through \ref{equ:s}
are subject to some mathematical constraints.  These constraints are:
\begin{equation}
1 \leq  \MyCloseLCC \leq n
\label{equ:lcc}
\end{equation}

\begin{equation}
m_{min}= \left \{ \begin{array}{ll}
  1 & \mbox{when }  \MyCloseLCC == n \\
  2 & \mbox{otherwise}
\end{array}
\right.
\end{equation}

\begin{equation}
m_{max}= \left \{ \begin{array}{ll}
  1 & \mbox{when }  \MyCloseLCC == n \\
  n-LCC & \mbox{otherwise}
\end{array}
\right.
\end{equation}

\begin{equation}
m_{min} \leq m \leq m_{max}
\label{equ:M}
\end{equation}

\begin{equation}
1 \leq j \leq m
\label{equ:j}
\end{equation}

In addition to the mathematical constraints, there are a series of logical constraints.  These constraints are:
\begin{enumerate}
\item \MyLittleS$ <  \MyCloseLCC$\label{cont:5} \MyEquationReference{equ:s}
\item  \MyLargeS will always be in the range $\frac {1}{n} \leq S \leq 1$ \label{cont:6} \MyEquationReference{equ:S}
\item If \MyCloseLCC == 1 then $\forall c: \mid c_{i} = 1 \Longrightarrow m = n$ meaning that anytime where $m == n$
and \MyCloseLCC $\neq 1$~is a contradiction and can not happen.\label{cont:1}
\item If $\MyCloseLCC == \frac{n}{2} \Longrightarrow m_{max}  = \frac {n}{2} $ where $ \forall c_{i}:\mathclose \mid c_{i} \mathclose \mid == 1$.\label{cont:2}
\item If $\MyCloseLCC == \frac{n}{j} \Longrightarrow  m_{max} = \frac {n}{j} $ where $ \forall c_{i}:\mathclose \mid c_{i} \mathclose \mid == 1$.\label{cont:4}
\item If $\MyCloseLCC == (n-1) \Longrightarrow m = 2$.\label{cont:3}
\end{enumerate}
Constraint \ref{cont:6} limits \MyCloseLCC between $n$ and 1.  The \MyCloseLCC~ will equal $n$
 when the graph is  connected (i.e., the graph has not been fragmented).
$LCC$ will equal 1 when the graph is totally disconnected (i.e., the graph is composed of only nodes and no edges).  Equation \ref{equ:M} limits  
the number of fragments  $m$ to
between 1 and $n$.
Equation \ref{equ:j} limits the  number of fragments to the greater of 1 (when the graph is totally connected;
 i.e. one cluster) or $n$ (when the graph is totally disconnected).
AJB were interested in the fraction $f$ of their graphs that had to be removed to cross a 
percolation threshold that would cause the graph to become severely fragmented.  We are interested
in the continuum of the graph's performance while it is connected and after it 
is disconnected.  The percolation threshold is of passing interest, while the ideas that they espouse
serve as starting point for our investigation.
\subsection{Ideas from Criado, Flores, et al.}
Criado, Flores et al. in \cite{criado2005effective}  propose to quantify the vulnerability of a graph
based on the number of nodes, number of edges and the standard deviation of the degrees of the
nodes.  Perhaps most importantly, they define  the attributes of a vulnerability function
in terms of the graph.

 Their definition is:
\begin{itemize}
\item[]Let $\mathcal{G}$ be the set of all possible graphs with a finite number of vertices.  A
\emph{vulnerability function} $v$ is a function $v:\mathcal{G}\rightarrow [0,1]$~ verifying the following
properties:
\begin{enumerate}
\item  $v$ is invariant under isomorphisms.
\item $v(G') \leq v(G)$ if $G'$ is obtained from $G$ by adding edges.
\item  $v(G)$ is computable in polynomial time with respect to the number
of vertices of $G$.
\end{enumerate}
\end{itemize}

The equation they present to meet their definitions is:
\MyEquationLabeled{vulnerability}{vulnerability}

Supported by:
\MyEquation{vulnerabilityDegreeStdDev}

Equation \ref{vulnerability} evaluates to the interval [0,1].  A value of
0 means that the graph is very robust (low vulnerability), while a value of 1
means that the graph is very vulnerable (not robust).  Using equation \ref{vulnerability}
before and after a modification to a graph can be used as a way to measure
what effect the change has had on the graph's vulnerability.  If the
vulnerability increases, then probably the change should not be finalized.
While their system of equations meets their requirements, the equations do not report
the type of damage that we are interested in measuring.  Their definition of the attributes
of a metric are in harmony with our intuition.
\subsection{Ideas from Holme and  Kim}
Holme and Kim in \cite{holme2002attack} looked at how an attacker could maximize the damage
to a graph by following one of two approaches:
\begin{enumerate}
\item To remove the vertex with the highest initial degree (ID)
 \MyEquation{degreeUndirectedDegree}
\item Or, the vertex with the highest normalized in-betweenness
centrality (IB) \MyEquation{betweennessVertex}
\end{enumerate}

For these approaches, they allowed the attacker two different options.  The options are:
\begin{enumerate}
\item To attack the graph (remove a vertex) based on the ordering of the vertices when a series of attacks started, or
\item To recompute ID and IB after a vertex has been removed.
\end{enumerate}
This second option took into account that the characteristics of the graph change when a vertex
is removed and therefore the ID and IB ordering would change.  Recomputed ID and IB were called
RD and RB respectively.

They used their ID, IB, RD and RB attack profiles on the hep-lat e-print archive,
a snapshot of the Internet autonomous system connections over a 24 hour period and
Erd{\"{o}s-R{\'{e}nyi random, Watts-Strogatz small-world, and
Barab{\'{a}si-Albert scale-free graphs.  They concluded that each of the different types of
graphs respond (as in how the AIPL responds) differently and that the attacker
should use the RB approach to maximize the impact as measured by AIPL.

Holme and Kim used AIPL as their metric to assess the functionality of the current
graph.  They did not use AIPL  to assess how the most recent
attack affected the graph's ability to perform.
\subsection{Ideas from Crucitti, Latora, Marchiori and  Rapisarda}
Crucitti et al. in \cite{crucitti2004error} look at the behavior of a network (i.e., a graph
that has a measurable flow along an edge) when a node or an edge is removed.  Their
premise is that the flow between nodes will always take the lowest cost path.  In their
models, each edge has a capacity and a tolerance factor.  As edges/nodes are removed, the
flow that was going through the removed component is spread out to other edges.  The
removal of a critical edge (high flow) and the redistribution of the flow through
adjacent edges can result in a cascade of failures as the increased flow causes additional
edges to reach saturation.

They investigated these phenomena for Erd{\"{o}s-R{\'{e}nyi random graphs and
Barab{\'{a}si-Albert scale-free graphs using the same ideas of ID, IB, RD and RB as introduced
in by Holme in \cite{holme2002attack}.  Crucitti introduces the idea of
\emph{global efficiency} that has the same form and character as AIPL.
\MyEquation{globalEfficiency2}
 Crucitti computes global
efficiency after a node or an edge is removed, but they do not compare the current
efficiency versus a connected graph's efficiency.
\subsection{Ideas from Netotea and  Pongor}
Netotea and Pongor in \cite{netotea2006evolution} focus on the evolution of a graph towards
a new organization that is more robust or efficient.  Their definition of efficiency $E$  is
AIPL and their definition of robustness $R$ is the ratio of the current  efficiency $E_{t}$
divided by the
the previous efficiency $R = \frac{E_{t}}{E}$.

Netotea and Pongor use a genetic algorithm that starts with a random graph (100 nodes and 120 edges)
and mutates and crossovers the graph until it reaches a ``steady state'' condition.  A steady
state was achieved when the goals of $E$, $R$ and the maximum percentage of periphery nodes (those
nodes with a degreeness of 1) was reached.  $E_{t}$ was computed after either 1 or 5 of the
highest betweenness nodes were removed.

Netotea and Pongor's idea of \emph{robustness} $R$ comes close to capturing our idea of a
single number that measures the \emph{health} of a graph.  Health is the inverse
of our idea of damage.
\subsection{Ideas from others}
Lee and Kim in \cite{lee-attack} look at the effects of node and path failure on the Internet and
report on the percentage of nodes that are required for disconnection. While they model failure of
the graph, they do not report on how damaged the graph is when attempting to perform its functions.

Cohen et al. in \cite{cohen2001breakdown} focus on the modeling the failure of the Internet
when the most connected routers (highest degreed nodes) are removed.  While they look towards
quantifying the percolation value $p$ where the Internet and scale-free graphs become disconnected,
they do not report on the graph's ability to perform.

 Newth and Ash in \cite{newth2004evolving} look at
cascading failures in a complex network.  They extend the work of Crucitti et al. in \cite{crucitti2004error}
by manipulating their graph by: (1) adding a new edge, or (2) deleting an existing edge, or (3)
changing one end of an existing edge.  If the graph becomes disconnected during any of these
operations, the change in rejected.

Beygelzimer et al. in \cite{beygelzimer2005improving} use AIPL
as their metric for the robustness of a graph.  They take an existing graph,
 rewire it using a number of different schemes and look at the robustness after each modification.
They disallow any rewiring that would disconnect the graph.

Zio and  Sansavini in \cite{zio-modeling} look at how the failure of a node or an edge may
cause a failure in adjacent components as the load of the failed component cascades to its neighbors.
These failures may be the result of random acts or targeted attacks.  They do not use
transfer of load as a metric of the damage done to the graph.

Lee et al. in \cite{lee2006rnt} look at how the topology of the graph affects which type of attack
profile would be most effective.  They propose a new metric, called \emph{attack power} to
quantify the effect of any of their attack profiles.  They measure damage to their graph
using degree distribution, average path length and vertex cover.  They enumerate some interesting
attack profiles, but their approach does not address a disconnected graph.
Klau and Ren{\'e} Weiskircher
in \cite{klau2005robustness} (a chapter in \cite{brandes2005network})
provide a very nice survey of robustness and resilience metrics
and ideas that have been advocated by various authors.  None of the approaches provide
a single unit-less value that describes the damage inflicted on a graph by the removal
of an edge or node and the possible disconnection of the graph.
Dekker in \cite{dekker2004simulating} introduces the idea of \emph{intelligence} of a graph related
to the quality of a sensor and the time delay associated with the data from the sensor.  The intelligence
of the graph starts to loose its meaning when the graph becomes disconnected.  While the idea of
intelligence in the graph is appealing to our sense that a graph can still perform when it is
fragmented, Dekker's metric does not speak to the total graph.

{\'A}goston et al. in \cite{agoston2005multiple} enumerate a series of attack profiles including:
\begin{enumerate}
  \item \textit{Complete knockout} --- meaning the removal of a node and its adjacent edges,
\item \textit{Partial knockout} --- meaning the removal of a set of edges (but not all) adjacent to a node,
\item \textit{Attenuation} --- meaning that the amount of traffic that an edge can support is decreased, 
subsequently, the total cost of a path that uses that edge from a source node \MyLittleS to a terminus node $t$ is increased,
\item \textit{Distributed knockout} --- meaning that a set of edges, not sharing a common node, are removed,
\item \textit{Distributed attenuation} --- meaning that the amount of traffic that the set of edges can support 
is decreased.
\end{enumerate}
These attack profiles are used in simulated attacks on \textit{Escherichia coli} and 
\textit{Saccharomyces cerevisiae} transcriptional regulatory networks.  Their conclusion is that 
multiple partial attacks causes more damage.  Our interests are slightly different because edges
in our network of DOs are really communications links vice edges that have a measurable 
capacity.  DOs in our network can either send messages via these communications links or they can not.
This difference in edge utilization and modeling eliminates the attenuation and distributed attenuation
profiles.  In our network, a DO exists or it does not and therefore all of its adjacent edges (communications
links) are valid, or not.  This approach matches {\'A}goston's complete knockout profile.  
We view partial  and distributed knockouts as being repeated application of removing
single edges in our network.

Yin et al. in \cite{yan2008multiple} take the ideas from {\'A}goston in \cite{agoston2005multiple} and apply
them to scale-free and random graphs.  Yin et al. apply weights to the edges in their graphs and use
AIPL as a metric to quantify the effect of each attack profile.  Their results confirm that scale-free
networks are relatively immune to random attacks, but very sensitive to targeted attacks.  While both
random and targeted attacks on random graphs have relatively the same effect.

Lee et al. in \cite{lee2006rnt} use the autonomic system (AS) connectivity graphs from National Laboratory for Applied
Network Research as their test graph.  Based on this graph, they apply weights to each of the
edges in the graph based on the amount of traffic along that edge.  They then focus on three different
types of failures. \textit{Node failure} where an AS is lost due to some sort of hardware failure (i.e., 
power supply failure, accidental or deliberate misconfiguration, etc.).  \textit{Link failure} where 
adjacent ASes are not able to communicate because of hardware failure (such as the cutting of a cable), or
electronic failure (such as DNS hacking, routing table poisoning, etc.).  \textit{Path failure} including
DoS and routing table loops, resulting in a flooding of the path with packets to the extent that the 
communications links are unusable.  Lee et al. then create different attack profiles based on 
these types of failures.  Their attack profiles are:
\begin{enumerate}
\item \textit{Random AS attack} --- randomly choose an AS and  and remove it,
\item \textit{Min-degree AS attack} --- order the ASes by their degree connectivity and then start removing 
them from low degree to high degree order,
\item \textit{Max-degree AS attack} --- order the ASes by the degree connectivity and then start removing
them from high degree to low degree order,
\item \textit{Random edge attack} --- randomly choose an edge and  remove it,
\item \textit{Min-weight edge attack} --- order the edges by their weight and then start removing 
them from low weight to high weight order,
\item \textit{Max-weight edge attack} --- order the edges by their weight and then start removing
them from high weight to low weight order,
\item \textit{Random path attack} --- randomly choose a path and remove it,
\item \textit{Max-weight edge attack} --- order all paths by weight and then remove paths in order
from heaviest to lightest, and
\item \textit{Max-length path attack} --- order all paths by length and then remove paths in order
from longest to shortest.
\end{enumerate}
After each attack, the effect on the  graph is quantified by a metric they labeled as ``attack power'' 
Attack power reports  the effect of each attack on the number of components that fail in the system.  We treat
Lee's path failure as a limited case of our edge failure \MySectionReference{sec:edgeSelection}.  Path failure is based on the path at the start of the
attack where the path meets some sort of criteria and then a series of edges are removed based on these criteria.
The limitation is that the set of criteria used to identify the path in the first place, may not be valid after
the removal of the first edge in the path.  We select an edge based on some criteria, remove the edge and then
reevaluate the entire graph to select the next edge.  We do not base future actions on information that
may be stale or obsolete.

Latora et al. in \cite{latora2005vulnerability} look at the vulnerability of complex networks to three different attack
profiles and then provide a method to reduce the vulnerability of the network by the addition
of edges between selected nodes.  Their attack profiles are: \textit{loss of a single cable connection} (loss of an
edge), \textit{loss of a single Internet router} (loss of a single node) and \textit{loss of two Internet routers} 
(loss of two nodes).  They assume that for the system $S$ there exists a \emph{performance} metric 
$\MyEquationInline{latoraPerformanceLHS}>0$ that characterizes
the performance of the graph and that this metric increases in value when the graph is damaged $D$.  Therefore

\MyEquationLabeled{latoraVulnerability}{equ:latora:vuln}

Where \MyEquationInline{latoraDamageWorst} is the worst possible damage that can happen to the graph
based on a specific attack profile.   They use  \MyEquationInline{latoraVulnerabilityLHS} as a  metric 
to quantify the efficacy of an attack.
The same metric is used to evaluate the effect of adding a communications link (an edge) between any two nodes
in order to improve (i.e., reduce the \emph{vulnerability}) of the system. 
Our approach is different in that we we are explicit about the metric that we will use
to measure the ``performance'' of the graph and we are currently focusing on attacking
the graph vice repairing it.  Our approach could be used to evaluate graph repair alternatives.
\section{An alternative approach}\label{sec:alt}
After looking at the different approaches in Section \ref{sec:relwork} and thinking about what it is that the \emph{damage} metric
is trying to capture, we do not feel that individually any of them fit the bill.

The attributes of the \emph{damage} metric should be:
\begin{enumerate}
\item Different fragmentation  cases should  result in different  numerical value,
\item Test cases where the size of the fragments have been scaled,
and the entire graph (for instance, increased by a factor of 10 or 0.1) should result in  the same value
\item The value should be useful without additional information about the graph (i.e.,  the
metric is  graph independent and does not require knowledge of the graph in a
different state),
\item The metric should be \emph{unitless}.  The approach and equations from
AJB's paper \cite{Albert2000} have some function of
 $node$. The units of \MyLargeS or \MyCloseLCC and \MyLittleS is $nodes$.  \MyEquationInline{fScoreLHS} \MyEquationReference{equ:fscore} and the generalized \MyEquationInline{fScoreBetaLHS}
 \MyEquationReference{equ:fscorebeta} metrics have units of $nodes$.  Geometric \MyEquationReference{equ:geomean} and quadratic
mean  \MyEquationReference{equ:quadmean} and ratio (\MyLittleS$/$\MyLargeS) are unit-less and therefore attractive.
\end{enumerate}

The desire/need to have a unit-less and scale-free description of the fragmentation and damage of a graph points to using a different
type of metric.  One that appears popular is based on the average inverse average path length (AIPL)
\MyEquationReference{equ:02c}.  There are a couple of variations on Equation \ref{equ:02c},
such as Equation \ref{equ:avi2} from \cite{netotea2006evolution} and Equation \ref{equ:avi3} from
\cite{crucitti2004error}.  Equation \ref{equ:avi2} is applicable to a graph that has directed edges
 and permits self loops.  Equation \ref{equ:avi3} is applicable to a graph that has directed edges
and does not permit self loops.
\MyEquationLabeled{globalEfficiency}{equ:avi2}
\MyEquationLabeled{globalEfficiency2}{equ:avi3}
AIPL equations are used to compute the AIPL between any
pair of nodes in a graph, even if the graph is disconnected.  Use of the AIPL can be counter intuitive, in that a larger AIPL is
better than a smaller AIPL because a smaller AIPL means that the average path length is increasing.

At the core of the \MyEquationInline{damageLHS} metric is the ratio of two AIPLs.  One
of the damaged/fragmented graph and the other an unfragmented artificial graph.
\MyEquationLabeled{damage-02}{equ:damageLocal}
The unfragmented artificial graph is constructed by sorting the original graph fragments
by their size and repeatedly connecting the nodes of two largest fragments with the highest centrality
value \MyEquationReference{equ:04c} until the graph is connected.
Conceptually, the artificial graph could have been existed in the fragmented graph's past and the
current fragmented graph is the result from losing edges.  The edges could have been lost due
to error or attack.

\section{Comparison and evaluation of various metrics}\label{sec:comparison}
\subsection{Small test case}
We create a small graph with 21 nodes and 27 edges \MyFigureReference{fig:original},
 and use it to show the effects
on ``classical'' graph metrics by using different attack profiles.
 Damage will be inflected on the
graph  by targeting either
the edge (\MyAttackNotation{E}{*})  or the vertex (\MyAttackNotation{V}{*}) based on its betweenness centrality measurement.
The betweenness centrality measurement is a count (or normalized value)
 of the number
of geodesic paths that use either an edge or a vertex, hence
edge or vertex centrality to the graph.

For \MyAttackNotation{E}{*} or  \MyAttackNotation{V}{*}, the appropriate centrality measurement
is computed and the component (edge or vertex as applicable) is removed from the graph.
Various graph metrics are computed and reported after each removal.
This targeted attack is repeated until the graph becomes disconnected.
After
targeting the edges, the graph will be restored to its initial condition
prior to targeting the vertices.

Removing graph components (either an edge or a vertex) may result in the
graph becoming disconnected, or fragmented.  Sometimes this fragmentation 
will result in a graph that is divided in half and whose $LCC$ is approximately
the same size as the non $LCC$.  A different choice in which component to remove
(a different attack criteria), might result in a graph whose $LCC$ contains 
all the remaining edges and all but one node.
\begin{figure}
\centering
\includegraphics*[viewport = 180mm 220mm 20mm 55mm, clip, width = 5.0in, angle = 90]{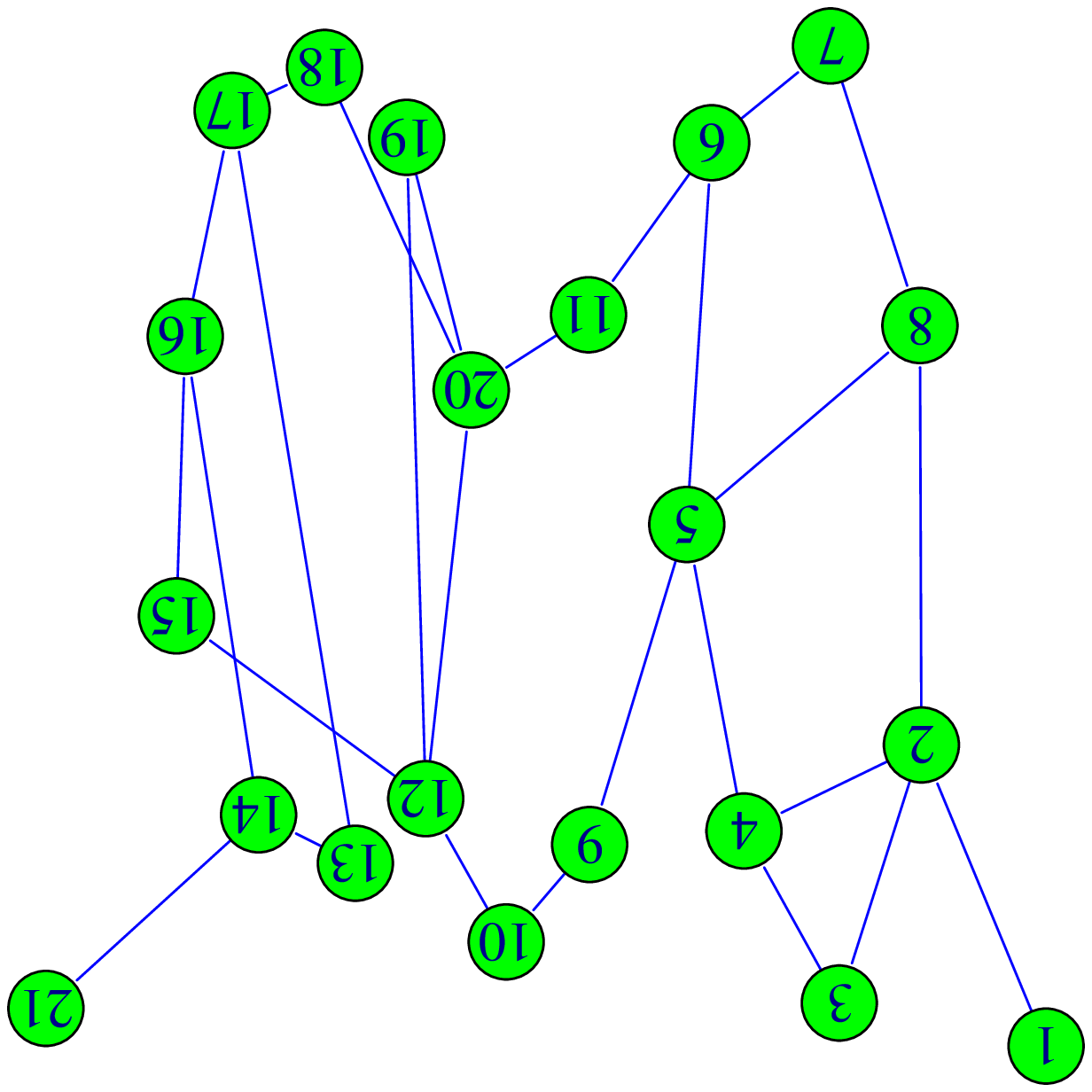}
\MyCaption{Small test graph used to show the effects of different attack profiles.}
{The graph has 21 nodes and 27 edges.  It clearly shows 2 groupings of nodes that are
connected by 2 separate sets of edges.}
{fig:original}
\end{figure}
\newcommand{\MyToyTabular}[0]{\begin{tabular}{|*{8}{p{0.50in}|}}}
\newcommand{\MyToyHeader}[1] {&APL&AIPL&{#1}& Clustering coefficient & Diameter & Eccentricity & Radius\\}

\subsubsection{Removal of vertices}
The effect of the \MyAttackNotation{V}{H} profile is 
tabulated in Table \ref{tbl:vertex:removal} and shown diagrammatically in 
Figure \ref{fig:vertexRemoval}.  Table \ref{tbl:vertex:removal} lists 
selected ``classic'' graph metric values and some are invalid after 
removing the second vertex.  Prior to the first removal, the centrality measurement
for all vertices is computed.  The vertex with the highest value is then removed and
all centrality values are recomputed so that the new highest valued
vertex can be identified.   In Figure \ref{fig:vertexRemoval}, each vertex is
labeled with its centrality value and the one with the highest value is 
drawn in red.

The graph is disconnected after removing two vertices.  Removing the vertices
or edges with the highest centrality measurement results in a disconnected graph
after two removals, but the choice of with type of component to remove results
in two different graphs (compare Figure \ref{fig:vertexRemoval} and Figure \ref{fig:edgeRemoval}).

\begin{table}
\newcommand{\MyMc}[1]{\multicolumn{1}{p{1.0in}|}{\textbf{#1}}}
\setlength{\extrarowheight}{3pt}
\centering
\begin{tabular}{|p{1.0in}|*{3}{r|}}
\MyHline
\textbf{Metric name} &\MyMc{Original values} & \MyMc{After removal of the  vertex with the 69 centrality measurement} & \MyMc{After removal of the  vertex with the 98 centrality measurement} \\
\hline
Highest vertex centrality & 69 & 98 & 28 \\
APL & 3.86 & 4.54 & --- \\
AIPL & 0.38 &0.35 & 0.24 \\
Clustering coefficient & 0.12 & 0.16 & 0.12 \\
Diameter & 10.00 & 11.00 & --- \\
Eccentricity & 10.00 & 11.00 & --- \\
Radius & 1.00 & 1.00 & --- \\
\MyHline
\end{tabular}
\MyCaption{Effects of an \MyAttackNotation{V}{H} attack profile on the sample graph.}
{Various ``classic'' graph values are computed using the original graph, including the 
vertex centrality of all vertices.  The vertex with the highest centrality is then removed \MyFigureReference{fig:vertex:1}
and the values are recomputed.  Again, the vertex with the highest centrality is removed \MyFigureReference{fig:vertex:3}
and values are computed.  The marker --- is used to indicate that the  graph metric is not computable  because the
graph is disconnected.}
{tbl:vertex:removal}
\end{table}
\begin{figure}
\centering
\MyLocalSubfigure{Original graph labeled with vertex betweenness values}{fig:vertex:orig-labled}{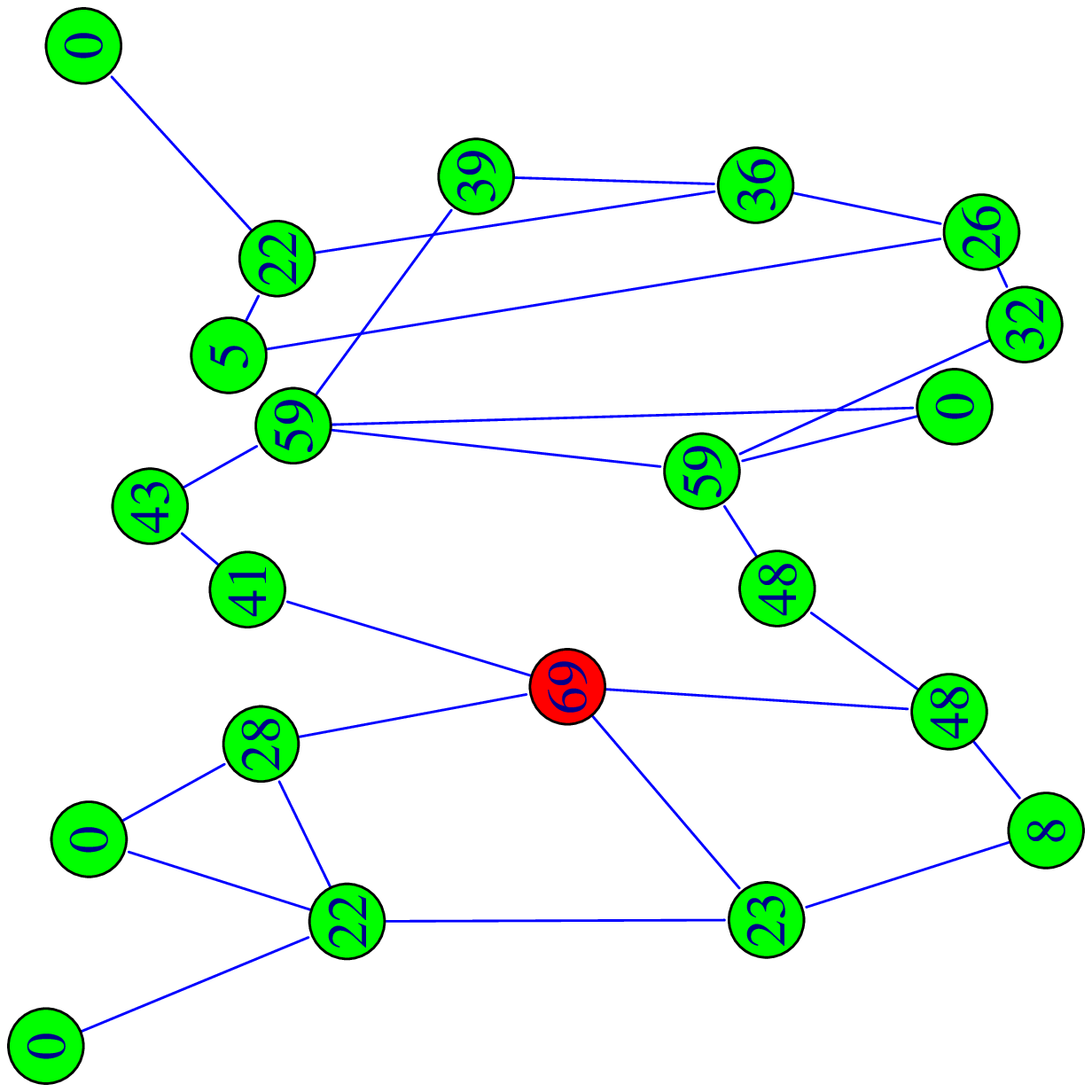}
\MyLocalSubfigure{Identifying and labeling first highest valued  vertex}{fig:vertex:1}{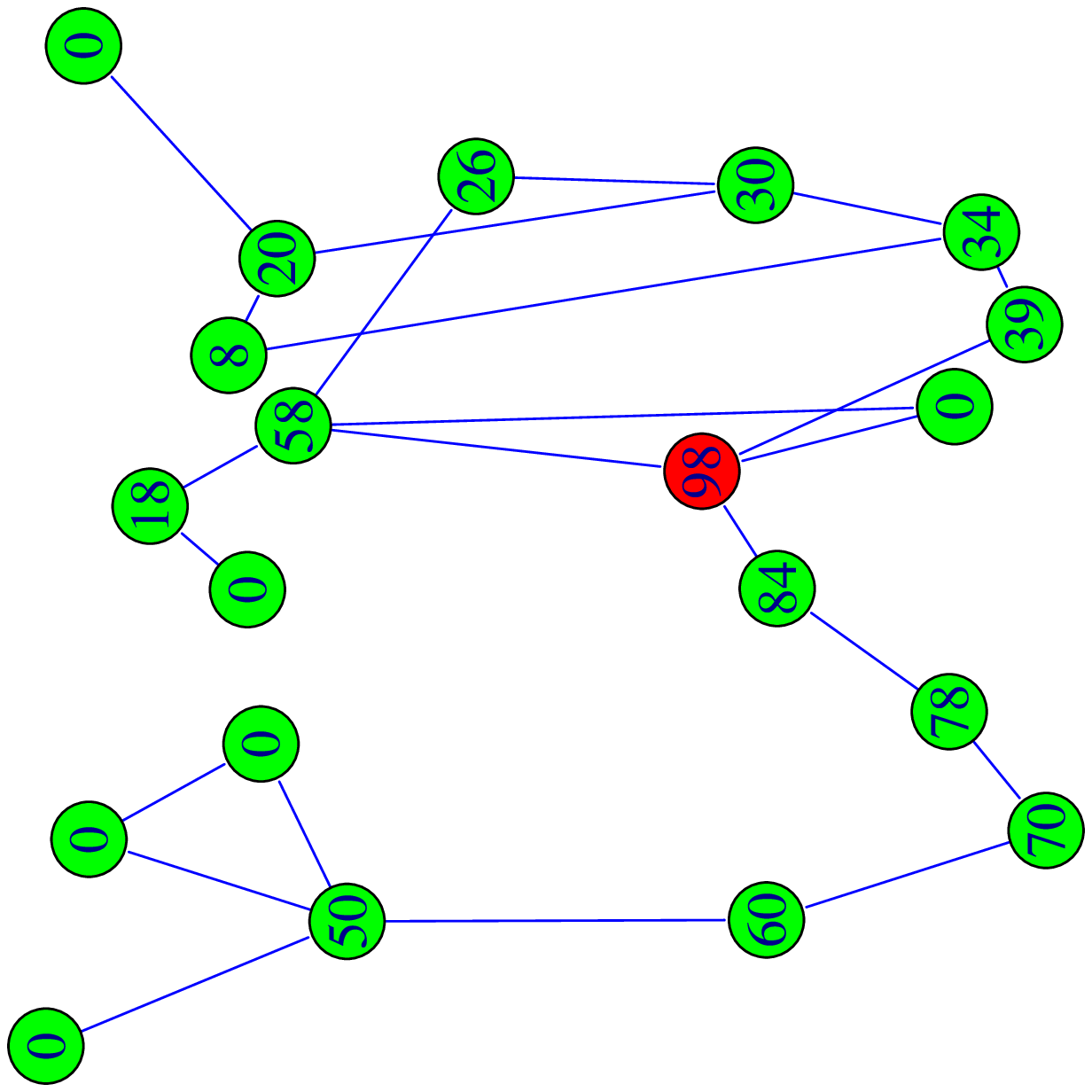}
\MyLocalSubfigure{Identifying and labeling new highest valued vertex after removing initial highest valued vertex}{fig:vertex:3}{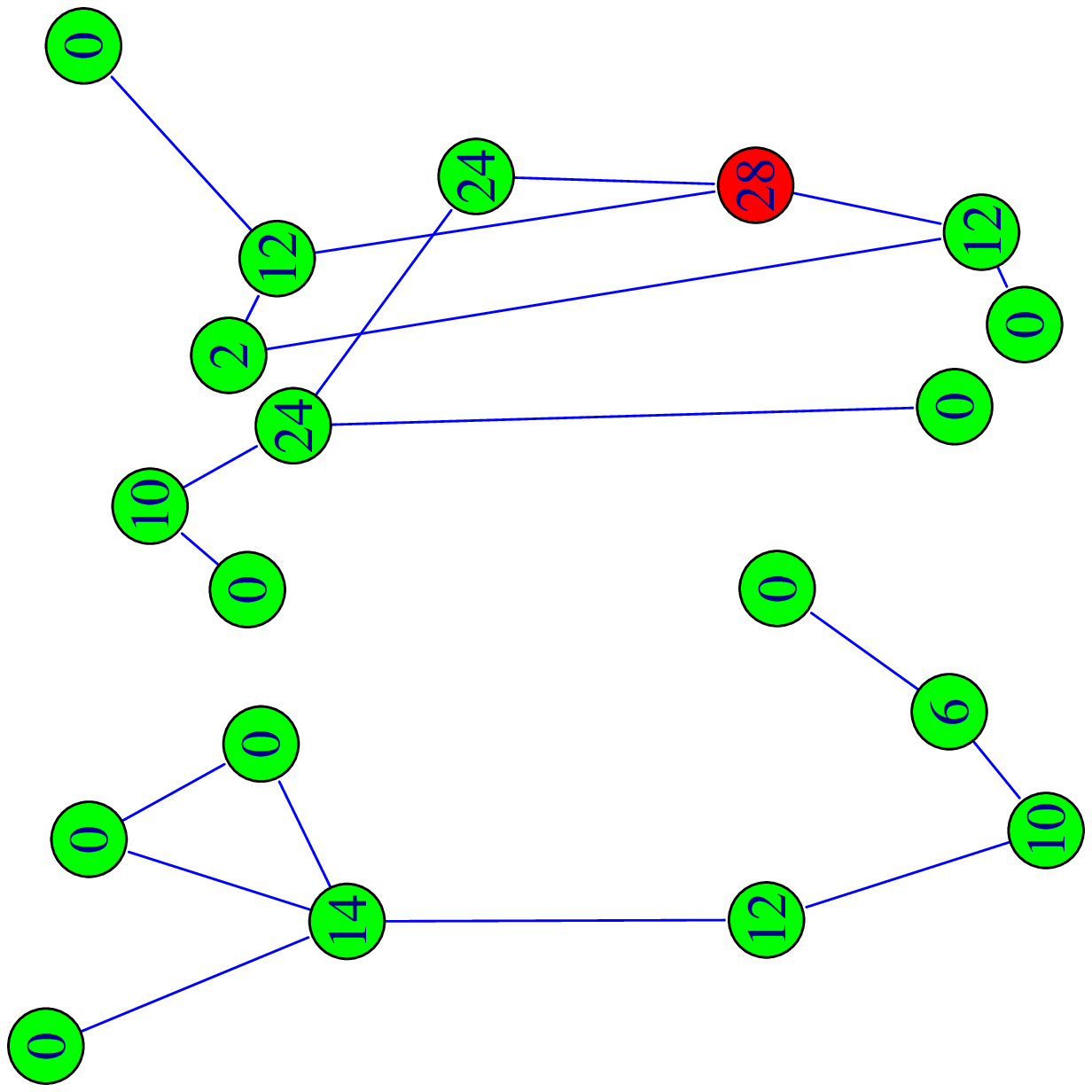}
\MyLocalSubfigure{The graph after removing two vertices}{fig:vertex:4}{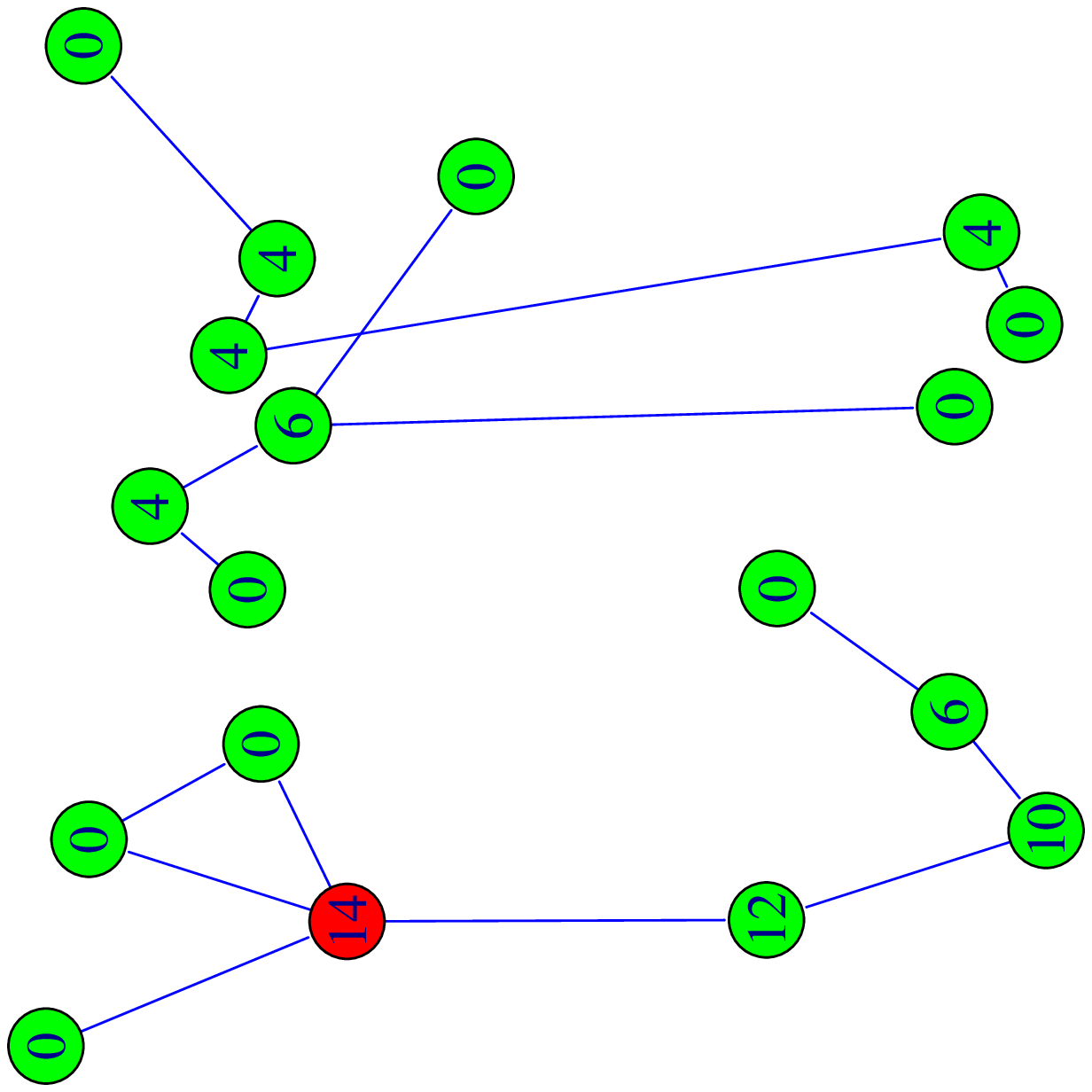}
\MyCaption{Damage to a graph by the \MyAttackNotation{V}{H} profile.}
{Each vertex in the original graph is labeled with its  centrality value \MyFigureReference{fig:edge:orig-labeled}.  The
vertex with the highest centrality measurement is selected and highlighted prior to its removal \MyFigureReference{fig:vertex:1}.
After the removal of the first vertex, all vertex centrality values are recomputed and again
the vertex with the highest value is selected for removal \MyFigureReference{fig:vertex:3}.  The graph is disconnected
after the removal of the second vertex \MyFigureReference{fig:vertex:4}.}
{fig:vertexRemoval}
\end{figure}

\subsubsection{Removal of edges}
The effect of the \MyAttackNotation{E}{H} profile is
tabulated in Table \ref{tbl:edge:removal} and shown diagrammatically in 
Figure \ref{fig:edgeRemoval}.  Table \ref{tbl:edge:removal} lists 
selected ``classic'' graph metric values and some are non valid after 
removing the second edge.  Prior to the first removal, the centrality measurement
for all edges is computed.  The edge with the highest value is then removed and
all centrality values are recomputed so that the new highest valued
edge can be identified.  Note that this is different than an attack
on a path because a path based attack does not recompute a new set
of paths after each removal.  In Figure \ref{fig:edgeRemoval}, each edge is
labeled with its centrality value and the one with the highest value is 
drawn with a wide red stroke.

The graph is disconnected after removing two edges.  Removing the vertices
or edges with the highest centrality measurement results in a disconnected graph
after two removals, but the choice of with type of component to remove results
in two different graphs (compare Figure \ref{fig:vertexRemoval} and Figure \ref{fig:edgeRemoval}).

\begin{table}
\setlength{\extrarowheight}{3pt}
\newcommand{\MyMc}[1]{\multicolumn{1}{p{1.0in}|}{\textbf{#1}}}
\centering
\begin{tabular}{|p{1.0in}|*{3}{r|}}
\MyHline
\textbf{Metric name} &\MyMc{Original values} & \MyMc{After removal of the  edge with the 59 centrality measurement} & \MyMc{After removal of the  edge with the 110 centrality measurement} \\
\hline
Highest edge centrality & 59 &110 &18\\
APL & 3.86 & 4.34 & --- \\
AIPL & 0.38 &0.36 & 0.26 \\
Clustering coefficient & 0.12 & 0.13 & 0.13 \\
Diameter & 10.00 & 11.00 & --- \\
Eccentricity & 10.00 & 11.00 & --- \\
Radius & 1.00 & 1.00 & --- \\
\MyHline
\end{tabular}
\MyCaption{Effects of an \MyAttackNotation{E}{H} attack profile on the sample graph.}
{Various ``classic'' graph values are computed using the original graph, including the 
edge centrality of all edges.  The  edge with the highest centrality is then removed \MyFigureReference{fig:edge:1}
and the values are recomputed.  Again, the edge with the highest centrality is removed \MyFigureReference{fig:edge:2}
and values are computed.  The marker --- is used to indicate that the  graph metric is not computable  because the
graph is disconnected.}
{tbl:edge:removal}
\end{table}
\begin{figure}
\centering
\MyLocalSubfigure{Original graph labeled with edge betweenness values} {fig:edge:orig-labeled}  {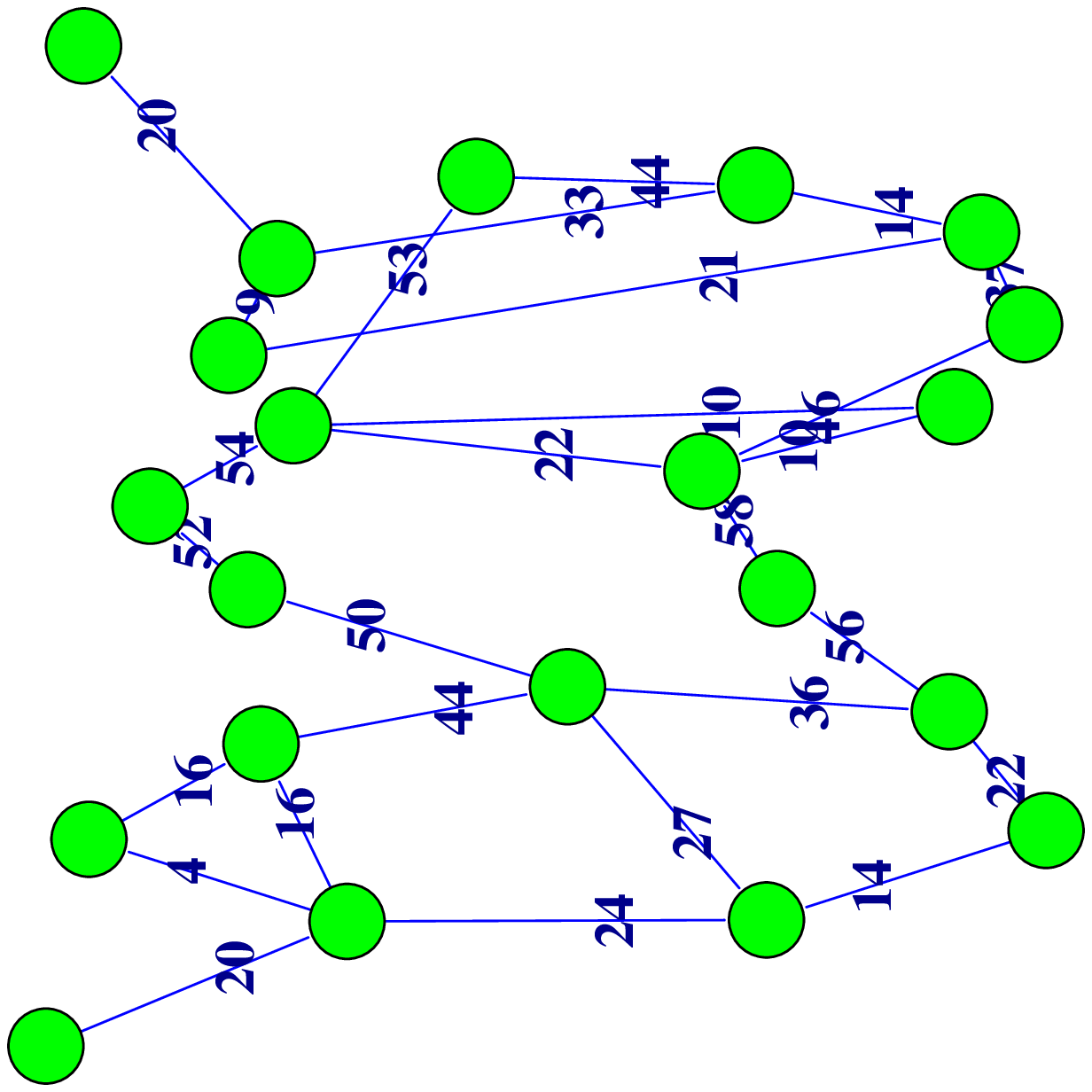}
\MyLocalSubfigure{Identifying and labeling first highest centrality valued edge} {fig:edge:1} {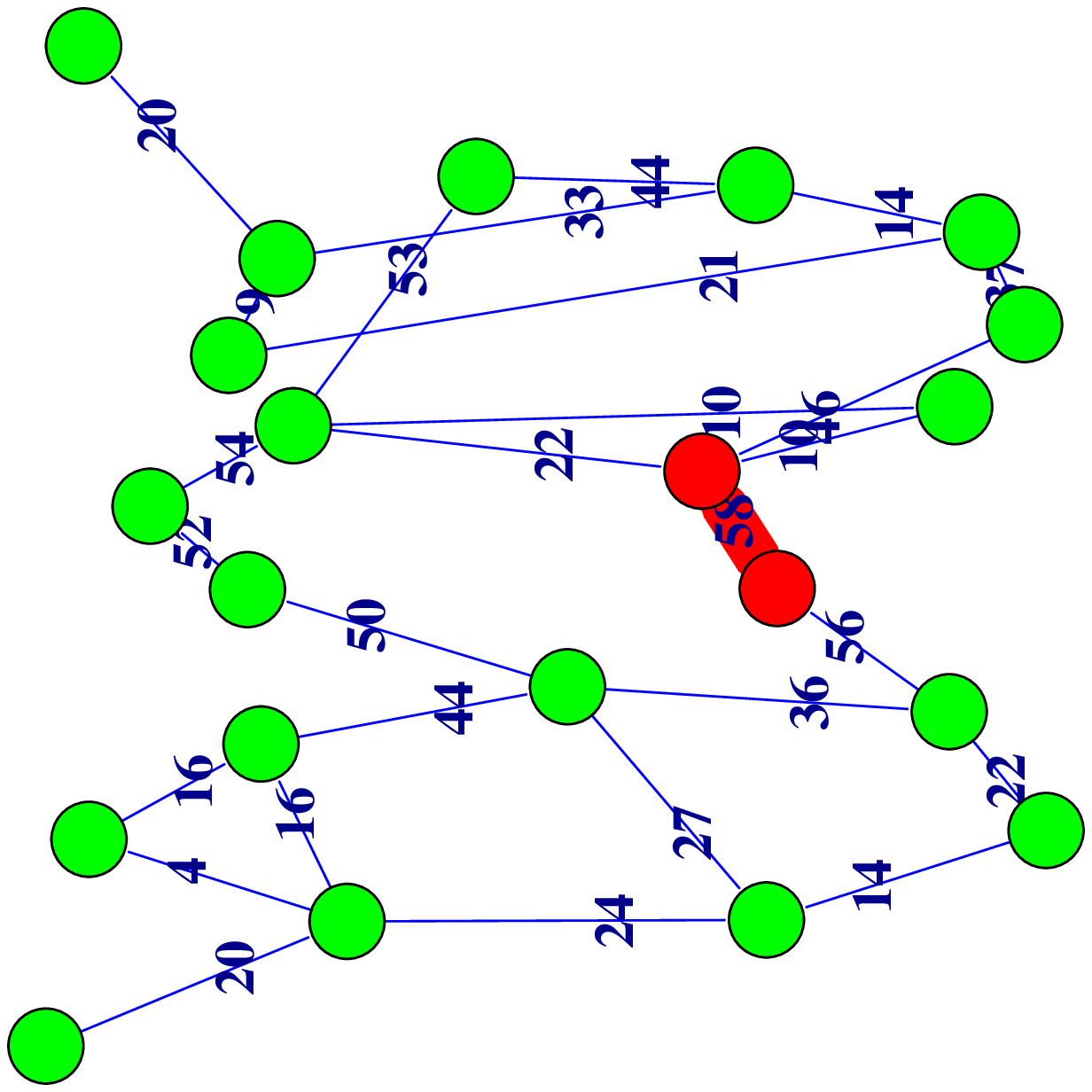}
\MyLocalSubfigure{Identifying and labeling new  highest centrality valued edge after removing initial highest valued edge} {fig:edge:2} {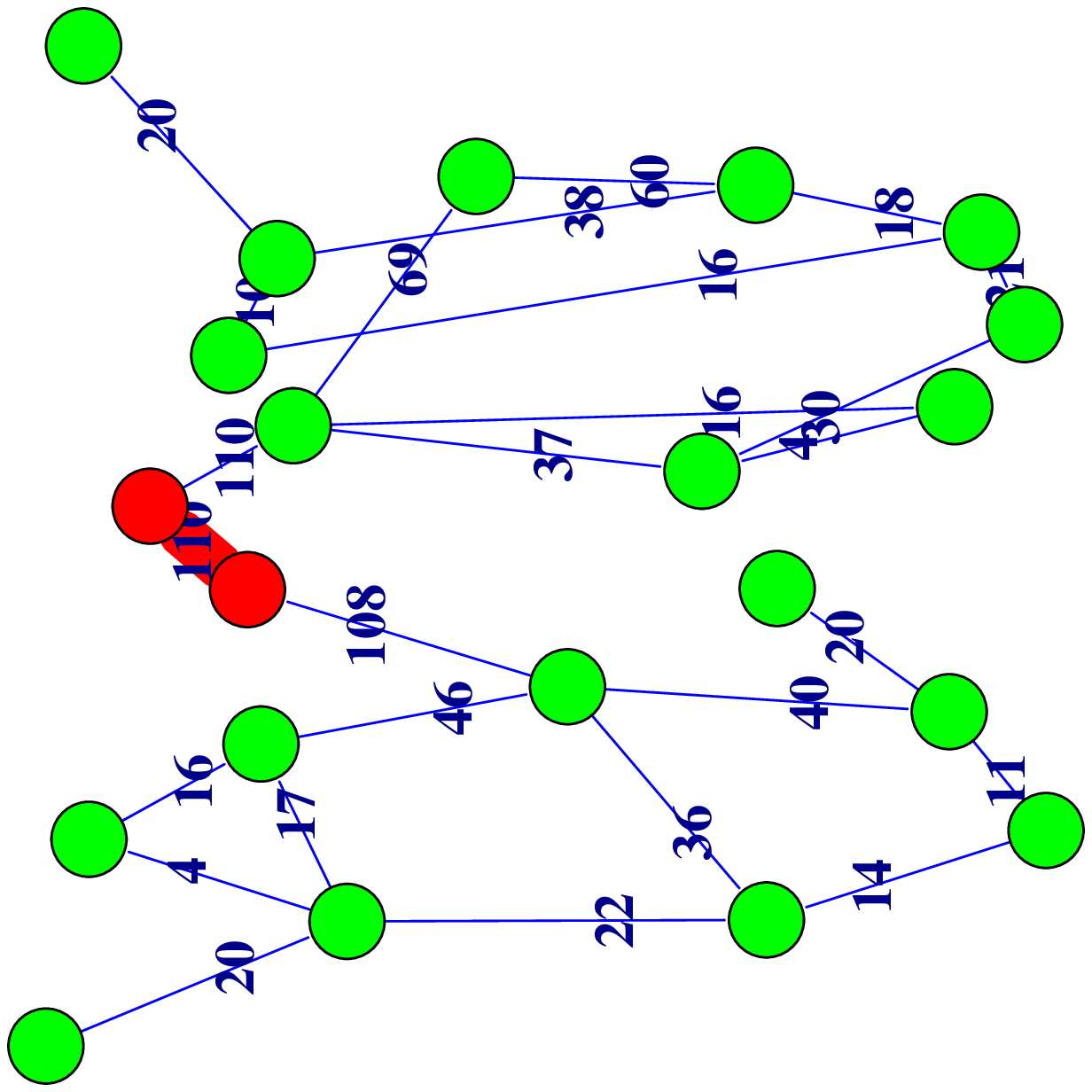}
\MyLocalSubfigure{The graph after removing two edges} {fig:edge:3} {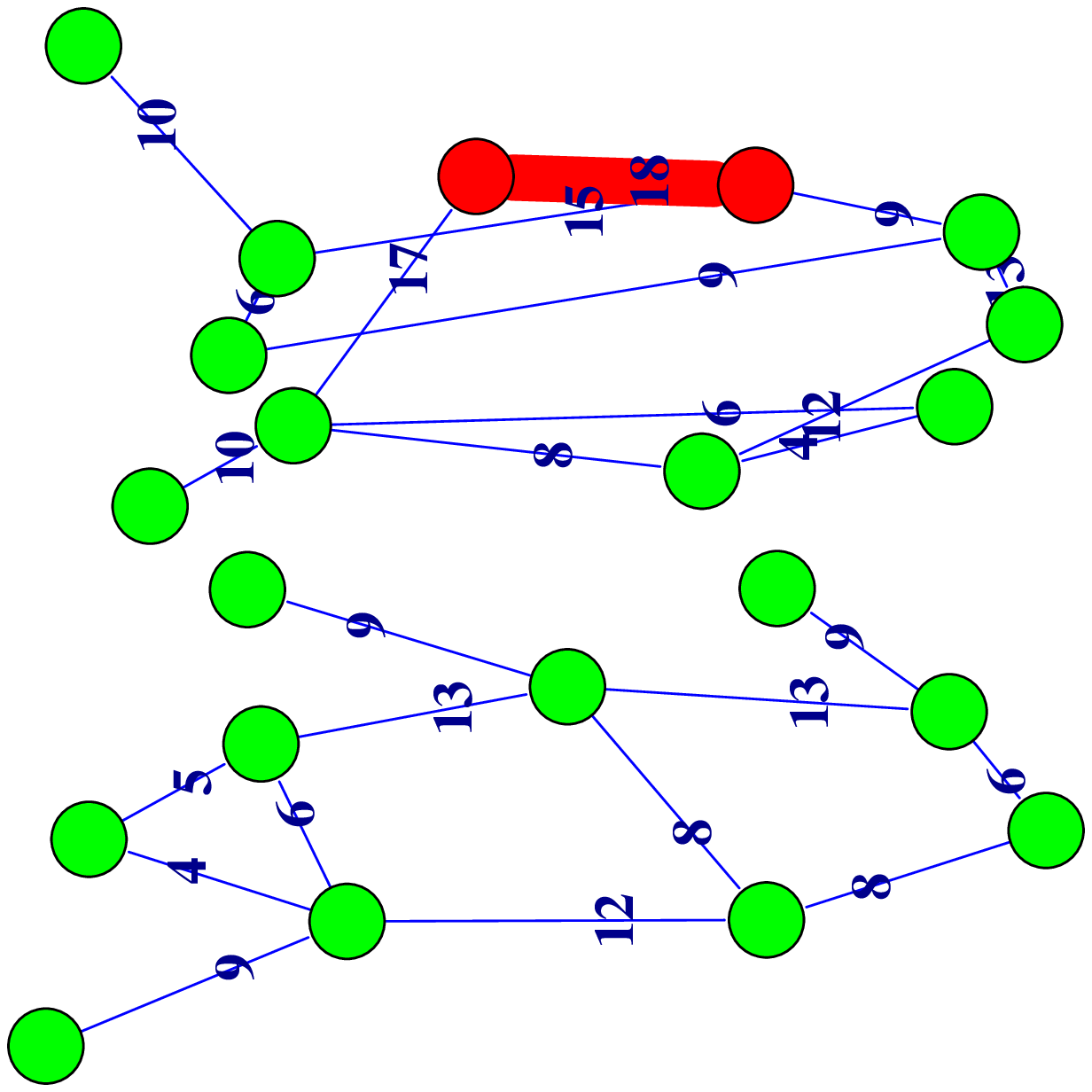}
\MyCaption{Damage to a graph by the \MyAttackNotation{E}{H} profile.}
{Each edge in the original graph is labeled with its edge centrality value \MyFigureReference{fig:edge:orig-labeled}.  The
edge with the highest centrality measurement is selected and highlighted prior to its removal \MyFigureReference{fig:edge:1}.
After the removal of the first edge, all edge centrality values are recomputed and again
the edge with the highest value is selected for removal \MyFigureReference{fig:edge:2}.  The graph is disconnected
after the removal of the second edge \MyFigureReference{fig:edge:3}.}
{fig:edgeRemoval}
\end{figure}

\subsubsection{Comparing \MyAttackNotation{E}{H} and \MyAttackNotation{V}{H} profiles}
Both the \MyAttackNotation{E}{H} and the \MyAttackNotation{V}{H} profiles result
in a disconnected graph after to removals.  But the two profiles result in different 
graphs at time point of disconnection (compare Figure \ref{fig:vertex:4} and Figure \ref{fig:edge:3}).
\subsection{A change in notation}
Figure \ref{fig:original} is small and sparse enough that it is practical to draw 
and label the complete graph and still be able to understand its structure.  As graphs
get larger, and more interesting it is not practical to draw and label every component.
Therefore, we introduce a different notation style that is more in keeping with the
aspects of the graph that are if interest to our research.

We are interested in how the graph functions, its connectivity as it becomes more
and more fragmented.  The internal connectivity (how many edges are in a fragment) 
is of less interest than the fact that the graph is fragmented, and that the numbers
and relative sizes of these fragments can be used as a metric to describe how well
the fragmented graph ``operates'' when compared to the unfragmented graph.

Specific graph instances will have names such as \MyShortNameB whose \MyCloseLCC and number and 
size of any fragments are shown in Table \ref{tbl:cases}. Tables \ref{tbl:circles.1} through \ref{tbl:circles.4} 
provide notational diagrams of the graph instances.

\subsection{Larger test cases}
A series of  test cases were constructed to exercise the different approaches
proposed by Albert, Jeong and  Barab{\'{a}}si and ourselves.
Each test case consists of some
number of fragments (a.k.a., components) between 1 and 11.  The test cases are intuitively ordered from least  to most
damaged.  The test cases are described numerically in Table \ref{tbl:cases}, and shown 
 diagrammatically in  Tables \ref{tbl:circles.1} through \ref{tbl:circles.4}.
\begin{table}
\centering
\scriptsize
\setlength{\extrarowheight}{3pt}
\begin{tabular} {|c|*{11}{r|}}
\MyHline
\textbf{Name} & \textbf {\MyCloseLCC} &
Frag. 2 &
Frag. 3 &
Frag. 4 &
Frag. 5 &
Frag. 6 &
Frag. 7 &
Frag. 8 &
Frag. 9 &
Frag. 10 &
Frag. 11 \\
\hline
\MyShortNameA & 100 & 
--- &
--- &
--- &
--- &
--- &
--- &
--- &
--- &
--- &
--- \\
\MyShortNameB & 90 & 
10 &
--- &
--- &
--- &
--- &
--- &
--- &
--- &
--- &
--- \\
\MyShortNameC & 90 & 
1 &
1 &
1 &
1 &
1 &
1 &
1 &
1 &
1 &
1 \\
\MyShortNameD & 80 & 
2 &
2 &
2 &
2 &
2 &
2 &
2 &
2 &
2 &
2 \\
\MyShortNameE & 50 & 
50 &
--- &
--- &
--- &
--- &
--- &
--- &
--- &
--- &
--- \\
\MyShortNameF & 50 & 
49 &
1 &
--- &
--- &
--- &
--- &
--- &
--- &
--- &
--- \\
\MyShortNameG & 50 & 
40 &
10 &
--- &
--- &
--- &
--- &
--- &
--- &
--- &
--- \\
\MyShortNameH & 50 & 
30 &
10 &
10 &
--- &
--- &
--- &
--- &
--- &
--- &
--- \\
\MyShortNameI & 50 & 
5 &
5 &
5 &
5 &
5 &
5 &
5 &
5 &
5 &
5 \\
\MyShortNameJ & 20 & 
20 &
20 &
20 &
20 &
--- &
--- &
--- &
--- &
--- &
--- \\
\MyShortNameK & 16 & 
15 &
14 &
13 &
10 &
9 &
8 &
7 &
4 &
3 &
1 \\
\MyShortNameL & 10 & 
10 &
10 &
10 &
10 &
10 &
10 &
10 &
10 &
10 &
--- \\
\MyShortNameM & 10 & 
9 &
9 &
9 &
9 &
9 &
9 &
9 &
9 &
9 &
9 \\
\MyShortNameN & 1 & 
1 &
1 &
1 &
1 &
1 &
1 &
1 &
1 &
1 &
1 \\
\MyHline
\end{tabular}
\MyCaption{A collection of connected and disconnected graphs used as test cases.}
{This is a set of graphs (some of which are connected and others that are not)  used to test various metrics and report how well the metric matches our  intuition of damage to the graph.  Each graph has 100 nodes. The test cases  are ordered by \MyCloseLCC.}
{tbl:cases}

\end{table}

\renewcommand{\MyShortNameAAJB}[0]{NaN}   \renewcommand{\MyShortNameACLC}[0]{0.00}
\renewcommand{\MyShortNameBAJB}[0]{10.00}   \renewcommand{\MyShortNameBCLC}[0]{0.14}
\renewcommand{\MyShortNameCAJB}[0]{1.00}   \renewcommand{\MyShortNameCCLC}[0]{0.16}
\renewcommand{\MyShortNameDAJB}[0]{2.00}   \renewcommand{\MyShortNameDCLC}[0]{0.31}
\renewcommand{\MyShortNameEAJB}[0]{50.00}   \renewcommand{\MyShortNameECLC}[0]{0.39}
\renewcommand{\MyShortNameFAJB}[0]{25.00}   \renewcommand{\MyShortNameFCLC}[0]{0.40}
\renewcommand{\MyShortNameGAJB}[0]{25.00}   \renewcommand{\MyShortNameGCLC}[0]{0.46}
\renewcommand{\MyShortNameHAJB}[0]{16.67}   \renewcommand{\MyShortNameHCLC}[0]{0.52}
\renewcommand{\MyShortNameIAJB}[0]{5.00}   \renewcommand{\MyShortNameICLC}[0]{0.64}
\renewcommand{\MyShortNameJAJB}[0]{20.00}   \renewcommand{\MyShortNameJCLC}[0]{0.66}
\renewcommand{\MyShortNameKAJB}[0]{8.40}   \renewcommand{\MyShortNameKCLC}[0]{0.78}
\renewcommand{\MyShortNameLAJB}[0]{10.00}   \renewcommand{\MyShortNameLCLC}[0]{0.81}
\renewcommand{\MyShortNameMAJB}[0]{9.00}   \renewcommand{\MyShortNameMCLC}[0]{0.82}
\renewcommand{\MyShortNameNAJB}[0]{1.00}   \renewcommand{\MyShortNameNCLC}[0]{1.00}

\begin{table}
\centering
\setlength{\extrarowheight}{3pt}
\begin{tabular}{|c|r|r|}
\MyHline
\multicolumn{1} {|c} {\textbf{Name}} & 
\multicolumn{1} {|c} {\MyLittleS} & 
\multicolumn{1} {|c|}{\MyEquationInline{damageLHS}} \\
\hline
\MyShortNameA&\MyShortNameAAJB&\MyShortNameACLC\\
\MyShortNameB&\MyShortNameBAJB&\MyShortNameBCLC\\
\MyShortNameC&\MyShortNameCAJB&\MyShortNameCCLC\\
\MyShortNameD&\MyShortNameDAJB&\MyShortNameDCLC\\
\MyShortNameE&\MyShortNameEAJB&\MyShortNameECLC\\
\MyShortNameF&\MyShortNameFAJB&\MyShortNameFCLC\\
\MyShortNameG&\MyShortNameGAJB&\MyShortNameGCLC\\
\MyShortNameH&\MyShortNameHAJB&\MyShortNameHCLC\\
\MyShortNameI&\MyShortNameIAJB&\MyShortNameICLC\\
\MyShortNameJ&\MyShortNameJAJB&\MyShortNameJCLC\\
\MyShortNameK&\MyShortNameKAJB&\MyShortNameKCLC\\
\MyShortNameL&\MyShortNameLAJB&\MyShortNameLCLC\\
\MyShortNameM&\MyShortNameMAJB&\MyShortNameMCLC\\
\MyShortNameN&\MyShortNameNAJB&\MyShortNameNCLC\\
\MyHline
\end{tabular}
\MyCaption{Comparing AJB's raw \MyLittleS to our proposed metric for the test graphs.}
{Raw \MyLittleS and  \MyEquationInline{damageLHS} are being evaluated as surrogates for the
 ``health'' of the graph. A healthy graph would have a value close to 0, 
while a totally disconnected graph would have a value of 1.
 Normalizing \MyLittleS to either the size of the graph, or to \MyCloseLCC does not meet these desired criteria.}
%{}
{tbl:comparison}
\end{table}

\newcommand{\MyCircleExplanation}[0]{The entire graph is contained with in the square.  
The LCC is represented by the large inner circle.  
While the smaller fragments are represented by the outer circles.  Within each square, the
circles represent the relative sizes of the different fragments.}
\newcommand{\MySubfigureFragments}[3]{#1 diagram, $\MyLittleS = #2$ , $\MyEquationInline{damageLHS} = #3$}
\begin{table}
\centering
\begin{tabular}{c c}
\setlength{\unitlength}{1.524000mm}
\begin{picture}(50.000000, 50.000000)
\put(0, 0){\line(0,1){50.000000}}
\put(0, 50.000000){\line(1,0){50.000000}}
\put(50.000000, 50.000000){\line(0,-1){50.000000}}
\put(50.000000, 0){\line(-1,0){50.000000}}
\put(25.000000, 25.000000){\circle{15.915494}}
\end{picture} & \setlength{\unitlength}{1.524000mm}
\begin{picture}(50.000000, 50.000000)
\put(0, 0){\line(0,1){50.000000}}
\put(0, 50.000000){\line(1,0){50.000000}}
\put(50.000000, 50.000000){\line(0,-1){50.000000}}
\put(50.000000, 0){\line(-1,0){50.000000}}
\put(25.000000, 25.000000){\circle{14.323945}}
\put(41.156345, 26.583955){\circle{1.591549}}
\end{picture} \\
\MySubfigureFragments{\MyShortNameA}{\MyShortNameAAJB}{\MyShortNameACLC} & 
\MySubfigureFragments{\MyShortNameB}{\MyShortNameBAJB}{\MyShortNameBCLC}\\
%\MyShortNameA notional diagram & \MyShortNameB notional diagram \\
\setlength{\unitlength}{1.524000mm}
\begin{picture}(50.000000, 50.000000)
\put(0, 0){\line(0,1){50.000000}}
\put(0, 50.000000){\line(1,0){50.000000}}
\put(50.000000, 50.000000){\line(0,-1){50.000000}}
\put(50.000000, 0){\line(-1,0){50.000000}}
\put(25.000000, 25.000000){\circle{14.323945}}
\put(39.514058, 25.159145){\circle{0.159155}}
\put(39.507079, 25.477360){\circle{0.159155}}
\put(39.493124, 25.795344){\circle{0.159155}}
\put(39.472200, 26.112947){\circle{0.159155}}
\put(39.444316, 26.430014){\circle{0.159155}}
\put(39.409487, 26.746393){\circle{0.159155}}
\put(39.367729, 27.061933){\circle{0.159155}}
\put(39.319063, 27.376481){\circle{0.159155}}
\put(39.263510, 27.689886){\circle{0.159155}}
\put(39.201099, 28.001998){\circle{0.159155}}
\end{picture} & \setlength{\unitlength}{1.524000mm}
\begin{picture}(50.000000, 50.000000)
\put(0, 0){\line(0,1){50.000000}}
\put(0, 50.000000){\line(1,0){50.000000}}
\put(50.000000, 50.000000){\line(0,-1){50.000000}}
\put(50.000000, 0){\line(-1,0){50.000000}}
\put(25.000000, 25.000000){\circle{12.732395}}
\put(38.110506, 25.318216){\circle{0.318310}}
\put(38.079629, 25.953899){\circle{0.318310}}
\put(38.017949, 26.587335){\circle{0.318310}}
\put(37.925610, 27.217033){\circle{0.318310}}
\put(37.802830, 27.841510){\circle{0.318310}}
\put(37.649898, 28.459295){\circle{0.318310}}
\put(37.467174, 29.068932){\circle{0.318310}}
\put(37.255088, 29.668987){\circle{0.318310}}
\put(37.014141, 30.258046){\circle{0.318310}}
\put(36.744899, 30.834722){\circle{0.318310}}
\end{picture} \\
\MySubfigureFragments{\MyShortNameC}{\MyShortNameCAJB}{\MyShortNameCCLC} & 
\MySubfigureFragments{\MyShortNameD}{\MyShortNameDAJB}{\MyShortNameDCLC}\\
%\MyShortNameC notional diagram &  \MyShortNameD notional diagram\\
\end{tabular}
\MyCaption{Notional diagrams for test cases \MyShortNameA, \MyShortNameB, \MyShortNameC  and \MyShortNameD.}{\MyCircleExplanation}{tbl:circles.1}
\end{table}
\begin{table}
\centering
\begin{tabular}{c c}
\setlength{\unitlength}{1.524000mm}
\begin{picture}(50.000000, 50.000000)
\put(0, 0){\line(0,1){50.000000}}
\put(0, 50.000000){\line(1,0){50.000000}}
\put(50.000000, 50.000000){\line(0,-1){50.000000}}
\put(50.000000, 0){\line(-1,0){50.000000}}
\put(25.000000, 25.000000){\circle{7.957747}}
\put(40.937826, 32.244466){\circle{7.957747}}
\end{picture} & \setlength{\unitlength}{1.524000mm}
\begin{picture}(50.000000, 50.000000)
\put(0, 0){\line(0,1){50.000000}}
\put(0, 50.000000){\line(1,0){50.000000}}
\put(50.000000, 50.000000){\line(0,-1){50.000000}}
\put(50.000000, 0){\line(-1,0){50.000000}}
\put(25.000000, 25.000000){\circle{7.957747}}
\put(40.788713, 32.110725){\circle{7.798592}}
\put(36.356450, 38.071989){\circle{0.159155}}
\end{picture} \\
\MySubfigureFragments{\MyShortNameE}{\MyShortNameEAJB}{\MyShortNameECLC} & 
\MySubfigureFragments{\MyShortNameF}{\MyShortNameFAJB}{\MyShortNameFCLC}\\
%\MyShortNameE notional diagram & \MyShortNameF notional diagram \\
\setlength{\unitlength}{1.524000mm}
\begin{picture}(50.000000, 50.000000)
\put(0, 0){\line(0,1){50.000000}}
\put(0, 50.000000){\line(1,0){50.000000}}
\put(50.000000, 50.000000){\line(0,-1){50.000000}}
\put(50.000000, 0){\line(-1,0){50.000000}}
\put(25.000000, 25.000000){\circle{7.957747}}
\put(39.440616, 30.894129){\circle{6.366198}}
\put(34.976943, 36.988860){\circle{1.591549}}
\end{picture} & \setlength{\unitlength}{1.524000mm}
\begin{picture}(50.000000, 50.000000)
\put(0, 0){\line(0,1){50.000000}}
\put(0, 50.000000){\line(1,0){50.000000}}
\put(50.000000, 50.000000){\line(0,-1){50.000000}}
\put(50.000000, 0){\line(-1,0){50.000000}}
\put(25.000000, 25.000000){\circle{7.957747}}
\put(37.923575, 29.508224){\circle{4.774648}}
\put(34.662533, 34.694242){\circle{1.591549}}
\put(32.180328, 36.652714){\circle{1.591549}}
\end{picture} \\
\MySubfigureFragments{\MyShortNameG}{\MyShortNameGAJB}{\MyShortNameGCLC} & 
\MySubfigureFragments{\MyShortNameH}{\MyShortNameHAJB}{\MyShortNameHCLC}\\
%\MyShortNameG notional diagram & \MyShortNameG notional diagram \\
\end{tabular}
\MyCaption{Notional diagrams for test cases \MyShortNameE, \MyShortNameF, \MyShortNameG  and \MyShortNameH.}
{\MyCircleExplanation}{tbl:circles.2}
\end{table}
\begin{table}
\centering
\begin{tabular}{c c}
\setlength{\unitlength}{1.524000mm}
\begin{picture}(50.000000, 50.000000)
\put(0, 0){\line(0,1){50.000000}}
\put(0, 50.000000){\line(1,0){50.000000}}
\put(50.000000, 50.000000){\line(0,-1){50.000000}}
\put(50.000000, 0){\line(-1,0){50.000000}}
\put(25.000000, 25.000000){\circle{7.957747}}
\put(33.877362, 25.792622){\circle{0.795775}}
\put(33.596522, 27.352790){\circle{0.795775}}
\put(33.043726, 28.838526){\circle{0.795775}}
\put(32.236462, 30.202829){\circle{0.795775}}
\put(31.200269, 31.402537){\circle{0.795775}}
\put(29.967926, 32.399697){\circle{0.795775}}
\put(28.578421, 33.162764){\circle{0.795775}}
\put(27.075710, 33.667597){\circle{0.795775}}
\put(25.507333, 33.898226){\circle{0.795775}}
\put(23.922906, 33.847354){\circle{0.795775}}
\end{picture} & \setlength{\unitlength}{1.524000mm}
\begin{picture}(50.000000, 50.000000)
\put(0, 0){\line(0,1){50.000000}}
\put(0, 50.000000){\line(1,0){50.000000}}
\put(50.000000, 50.000000){\line(0,-1){50.000000}}
\put(50.000000, 0){\line(-1,0){50.000000}}
\put(25.000000, 25.000000){\circle{3.183099}}
\put(31.375130, 27.897786){\circle{3.183099}}
\put(27.008603, 31.708574){\circle{3.183099}}
\put(21.266320, 30.924448){\circle{3.183099}}
\put(18.081352, 26.082481){\circle{3.183099}}
\end{picture} \\
\MySubfigureFragments{\MyShortNameI}{\MyShortNameIAJB}{\MyShortNameICLC} & 
\MySubfigureFragments{\MyShortNameJ}{\MyShortNameJAJB}{\MyShortNameJCLC}\\
%\MyShortNameI notional diagram & \MyShortNameJ notional diagram \\
\setlength{\unitlength}{1.524000mm}
\begin{picture}(50.000000, 50.000000)
\put(0, 0){\line(0,1){50.000000}}
\put(0, 50.000000){\line(1,0){50.000000}}
\put(50.000000, 50.000000){\line(0,-1){50.000000}}
\put(50.000000, 0){\line(-1,0){50.000000}}
\put(25.000000, 25.000000){\circle{2.546479}}
\put(29.950865, 27.184205){\circle{2.387324}}
\put(26.851471, 30.084671){\circle{2.228169}}
\put(22.859887, 29.970084){\circle{2.069014}}
\put(20.285132, 27.655531){\circle{1.591549}}
\put(19.591482, 24.827506){\circle{1.432394}}
\put(20.308694, 22.303063){\circle{1.273240}}
\put(21.878291, 20.579960){\circle{1.114085}}
\put(23.427958, 19.822114){\circle{0.636620}}
\put(24.515548, 19.610461){\circle{0.477465}}
\put(25.150220, 19.590817){\circle{0.159155}}
\end{picture} & \setlength{\unitlength}{1.524000mm}
\begin{picture}(50.000000, 50.000000)
\put(0, 0){\line(0,1){50.000000}}
\put(0, 50.000000){\line(1,0){50.000000}}
\put(50.000000, 50.000000){\line(0,-1){50.000000}}
\put(50.000000, 0){\line(-1,0){50.000000}}
\put(25.000000, 25.000000){\circle{1.591549}}
\put(28.187565, 26.448893){\circle{1.591549}}
\put(26.004301, 28.354287){\circle{1.591549}}
\put(23.133160, 27.962224){\circle{1.591549}}
\put(21.540676, 25.541240){\circle{1.591549}}
\put(22.317592, 22.749544){\circle{1.591549}}
\put(24.931774, 21.499256){\circle{1.591549}}
\put(27.592686, 22.646738){\circle{1.591549}}
\put(30.374759, 25.627521){\circle{1.591549}}
\put(28.656300, 28.989147){\circle{1.591549}}
\end{picture} \\
\MySubfigureFragments{\MyShortNameK}{\MyShortNameKAJB}{\MyShortNameKCLC} & 
\MySubfigureFragments{\MyShortNameL}{\MyShortNameLAJB}{\MyShortNameLCLC}\\
%\MyShortNameK notional diagram & \MyShortNameL notional diagram \\
\end{tabular}
\MyCaption{Notional diagrams for test cases \MyShortNameI, \MyShortNameJ, \MyShortNameK  and \MyShortNameL.}{\MyCircleExplanation}{tbl:circles.3}
\end{table}
\begin{table}
\centering
\begin{tabular}{c c}
\setlength{\unitlength}{1.524000mm}
\begin{picture}(50.000000, 50.000000)
\put(0, 0){\line(0,1){50.000000}}
\put(0, 50.000000){\line(1,0){50.000000}}
\put(50.000000, 50.000000){\line(0,-1){50.000000}}
\put(50.000000, 0){\line(-1,0){50.000000}}
\put(25.000000, 25.000000){\circle{1.591549}}
\put(28.038207, 26.314609){\circle{1.432394}}
\put(26.121730, 28.114582){\circle{1.432394}}
\put(23.497675, 27.949902){\circle{1.432394}}
\put(21.821274, 25.924447){\circle{1.432394}}
\put(22.149988, 23.315859){\circle{1.432394}}
\put(24.276467, 21.769613){\circle{1.432394}}
\put(26.859344, 22.261070){\circle{1.432394}}
\put(28.269362, 24.480222){\circle{1.432394}}
\put(28.975968, 28.079854){\circle{1.432394}}
\put(26.126449, 29.901523){\circle{1.432394}}
\end{picture} & \setlength{\unitlength}{1.524000mm}
\begin{picture}(50.000000, 50.000000)
\put(0, 0){\line(0,1){50.000000}}
\put(0, 50.000000){\line(1,0){50.000000}}
\put(50.000000, 50.000000){\line(0,-1){50.000000}}
\put(50.000000, 0){\line(-1,0){50.000000}}
\put(25.000000, 25.000000){\circle{1.591549}}
\put(28.187565, 26.448893){\circle{1.591549}}
\put(26.004301, 28.354287){\circle{1.591549}}
\put(23.133160, 27.962224){\circle{1.591549}}
\put(21.540676, 25.541240){\circle{1.591549}}
\put(22.317592, 22.749544){\circle{1.591549}}
\put(24.931774, 21.499256){\circle{1.591549}}
\put(27.592686, 22.646738){\circle{1.591549}}
\put(30.374759, 25.627521){\circle{1.591549}}
\put(28.656300, 28.989147){\circle{1.591549}}
\put(25.914357, 30.333458){\circle{1.591549}}
\put(22.881218, 29.979215){\circle{1.591549}}
\put(20.522850, 28.039235){\circle{1.591549}}
\put(19.590326, 25.131346){\circle{1.591549}}
\put(20.380628, 22.181627){\circle{1.591549}}
\put(22.642068, 20.129479){\circle{1.591549}}
\put(25.654441, 19.628452){\circle{1.591549}}
\put(28.458394, 20.838109){\circle{1.591549}}
\put(30.160947, 23.373210){\circle{1.591549}}
\put(32.062200, 26.929830){\circle{1.591549}}
\put(30.271884, 30.079975){\circle{1.591549}}
\put(27.687067, 31.810182){\circle{1.591549}}
\put(24.617220, 32.311114){\circle{1.591549}}
\put(21.616467, 31.492350){\circle{1.591549}}
\put(19.226460, 29.501682){\circle{1.591549}}
\put(17.878608, 26.698436){\circle{1.591549}}
\put(17.816206, 23.588614){\circle{1.591549}}
\put(19.050518, 20.733553){\circle{1.591549}}
\put(21.358743, 18.648610){\circle{1.591549}}
\put(24.324235, 17.710127){\circle{1.591549}}
\put(27.411706, 18.087506){\circle{1.591549}}
\put(30.063851, 19.712627){\circle{1.591549}}
\put(31.801944, 22.292147){\circle{1.591549}}
\put(34.219792, 25.454481){\circle{1.591549}}
\put(33.374997, 28.882080){\circle{1.591549}}
\put(31.591448, 31.462501){\circle{1.591549}}
\put(29.046763, 33.296676){\circle{1.591549}}
\put(26.034784, 34.172804){\circle{1.591549}}
\put(22.903314, 33.989718){\circle{1.591549}}
\put(20.013956, 32.768557){\circle{1.591549}}
\put(17.700354, 30.650335){\circle{1.591549}}
\put(16.229667, 27.879649){\circle{1.591549}}
\put(15.771721, 24.776440){\circle{1.591549}}
\put(16.379396, 21.699047){\circle{1.591549}}
\put(17.982521, 19.002825){\circle{1.591549}}
\put(20.395979, 16.999118){\circle{1.591549}}
\put(23.341079, 15.919301){\circle{1.591549}}
\put(26.477741, 15.888063){\circle{1.591549}}
\put(29.443762, 16.909012){\circle{1.591549}}
\put(31.896647, 18.864255){\circle{1.591549}}
\put(33.553153, 21.528013){\circle{1.591549}}
\put(34.221996, 24.592694){\circle{1.591549}}
\put(35.652002, 28.263941){\circle{1.591549}}
\put(34.131798, 31.381905){\circle{1.591549}}
\put(31.979593, 33.683532){\circle{1.591549}}
\put(29.269020, 35.290477){\circle{1.591549}}
\put(26.216926, 36.074184){\circle{1.591549}}
\put(23.067477, 35.971955){\circle{1.591549}}
\put(20.072631, 34.991971){\circle{1.591549}}
\put(17.471974, 33.212629){\circle{1.591549}}
\put(15.473559, 30.776276){\circle{1.591549}}
\put(14.237259, 27.877822){\circle{1.591549}}
\put(13.861979, 24.749141){\circle{1.591549}}
\put(14.377740, 21.640530){\circle{1.591549}}
\put(15.743282, 18.800676){\circle{1.591549}}
\put(17.849362, 16.456768){\circle{1.591549}}
\put(20.527492, 14.796318){\circle{1.591549}}
\put(23.563424, 13.952163){\circle{1.591549}}
\put(26.714281, 13.991835){\circle{1.591549}}
\put(29.727996, 14.912160){\circle{1.591549}}
\put(32.363471, 16.639513){\circle{1.591549}}
\put(34.409869, 19.035704){\circle{1.591549}}
\put(35.703477, 21.909039){\circle{1.591549}}
\put(38.050659, 25.034734){\circle{1.591549}}
\put(37.592020, 28.429568){\circle{1.591549}}
\put(36.398747, 31.355271){\circle{1.591549}}
\put(34.537318, 33.908450){\circle{1.591549}}
\put(32.116843, 35.939445){\circle{1.591549}}
\put(29.279203, 37.329207){\circle{1.591549}}
\put(26.190730, 37.996271){\circle{1.591549}}
\put(23.032460, 37.901539){\circle{1.591549}}
\put(19.989521, 37.050561){\circle{1.591549}}
\put(17.240280, 35.493219){\circle{1.591549}}
\put(14.945887, 33.320801){\circle{1.591549}}
\put(13.240833, 30.660644){\circle{1.591549}}
\put(12.225061, 27.668680){\circle{1.591549}}
\put(11.958114, 24.520287){\circle{1.591549}}
\put(12.455639, 21.400013){\circle{1.591549}}
\put(13.688471, 18.490758){\circle{1.591549}}
\put(15.584348, 15.963053){\circle{1.591549}}
\put(18.032139, 13.965064){\circle{1.591549}}
\put(20.888362, 12.613905){\circle{1.591549}}
\put(23.985596, 11.988778){\circle{1.591549}}
\put(27.142290, 12.126325){\circle{1.591549}}
\put(30.173411, 13.018484){\circle{1.591549}}
\put(32.901284, 14.612959){\circle{1.591549}}
\put(35.166010, 16.816287){\circle{1.591549}}
\put(36.834839, 19.499318){\circle{1.591549}}
\put(37.809949, 22.504780){\circle{1.591549}}
\put(39.941624, 25.752577){\circle{1.591549}}
\end{picture} \\
\MySubfigureFragments{\MyShortNameM}{\MyShortNameMAJB}{\MyShortNameMCLC} & 
\MySubfigureFragments{\MyShortNameN}{\MyShortNameNAJB}{\MyShortNameNCLC}\\
%\MyShortNameM notional diagram & \MyShortNameN notional diagram \\
\end{tabular}
\MyCaption{Notional diagrams for test cases \MyShortNameM  and \MyShortNameN.}{\MyCircleExplanation}{tbl:circles.4}
\end{table}

\subsection{Comparison equations}
Now that we have the basic definitions and constraints out of the way, we can begin to look at how AJB's \MyLargeS and \MyLittleS will
be evaluated.  A set of equations was selected  that seemed like they might be of use.  The set includes:
\begin{enumerate}
\item The median value of all the fragments, except the $LCC$.
\item The average size of all the fragments, except the $LCC$.
\item The standard deviation of all the fragments, except the $LCC$.
\item The harmonic mean of all the fragments, except the $LCC$.
\item The geometric mean of all the fragments, except the $LCC$.
\item A variation on the information retrieval (IR) metric \MyEquationInline{fScoreLHS} 
\MyEquationReference{equ:fscore} (a 2 value harmonic mean).  We selected \MyEquationInline{fScoreLHS}
because it had been used in other applications and we thought that it might be useful.  
In the IR world, \MyEquationInline{fScoreLHS} traditionally
operates on the values of \textit{precision} and \textit{recall}. 
For the purposes of analysis \MyLargeS was treated as \textit{precision} and \MyLittleS was
treated as \textit{recall}.
\MyEquationLabeled{fScore}{equ:fscore}
\item A generalized  \MyEquationInline{fScoreBetaLHS} \MyEquationReference{equ:fscorebeta} metric that incorporates a value $\beta$ 
that is used to weight \textit{precision} relative to
\textit{recall}.
\MyEquationLabeled{fScoreBeta}{equ:fscorebeta}
\item A simple arithmetic mean of \MyLargeS and \MyLittleS.
\item A geometric mean of \MyLargeS and \MyLittleS \MyEquationReference{equ:geomean}.
\MyEquationLabeled{geometricMean}{equ:geomean}
\item A quadratic mean of \MyLargeS and \MyLittleS \MyEquationReference{equ:quadmean}.
\MyEquationLabeled{quadraticMean}{equ:quadmean}
\item Ratio of \MyLittleS to \MyLargeS.
\item \MyLargeS raised to the \MyLittleS power.
\item \MyLittleS raised to the \MyLargeS power.
\end{enumerate}

In equations \ref{equ:fscore} through \ref{equ:quadmean}, $x_{1} = S$ and $x_{2}=s$.

\section{Analysis}\label{sec:analysis}
The interaction between \MyLargeS, $LLC$ and \MyLittleS is of interest and is summarized in Table \ref{tbl:s2}.  Various cells in Table \ref{tbl:s2} are have different colors and color is significant.
Cells that are filled with \color{\MyMath} \MyMath \color{black}~ violate some basic mathematical operation.
Cells that are filled with \color{\MyLCC} \MyLCC  \color{black}~ violate some some logical restriction on $LCC$ \MyConstraintReference{equ:lcc}.
Cells that are filled with \color{\MyM} \MyM \color{black}~ violate some logical restriction on $m$ \MyConstraintReference{equ:m}.  There may be cases when a  combination of $j$, $m$, \MyLargeS, $LLC$ and \MyLittleS
violates more than one constraint, in those cases the fill color will be chosen at random.  Cells that are not filled, do not violate any constraints.
There is a limited range of values for $m$ and \MyCloseLCC~ that do not violate some sort of mathematical or
logical constraint when attempting to compute \MyLittleS.  These limits are in keeping with the values
computed in AJB's paper.

The test cases from Table \ref {tbl:cases} were subjected to a series of mathematical investigations looking to
identify and quantify a metric that was near 0 for ``undamaged'' graphs and near 1 for
``damaged'' ones.  Table \ref{tbl:case01} shows the various mean and standard
deviation values for the test cases.  These approaches produced values that had
no discernible relationship to their state of damage.
Table \ref{tbl:case02} showed some useful information, but each of these more
sophisticated approaches, had some sort of ``hump'' or ``swale'' in the computed values.
Values produced by using these approaches would initially trend in the right direction
(from  low  to high) as the case numbers increased, but then the values would change direction
and start to go the other way.  Some of the exponentiation cases, created values that were too
large for the computer to handle reasonably.  While these computational limitations could
be overcome, there does not seem to be any reason to expend the effort to do so when
the data that was available was not well behaved.
None of the approaches in Tables \ref{tbl:case01} and
\ref{tbl:case02} showed the desired property of continuous directed change.

The investigation into a unit-less metric for assessing the ``damage'' inflicted upon a graph by fragmentation, led to writing an R script
that could produce three different types of graphs; random, small world and scale free.  These graph types were selected
because they are felt to represent approximately the extremes of the fundamental graph types.

The R script takes as an a argument the fragments that make up the test case \MyTableReference{tbl:cases}. Two graphs
are created based on the fragments.  The first graph is a simple connected graph whose size
is equal to the sum of the fragments.  The second graph is a simple disconnected graph whose
size is equal to the sum of the  fragments.

The average inverse path length (AIPL) \MyEquationReference{equ:02c} for the two graphs is
computed and then the ratio of the AIPLs is reported.  The hoped for behavior (a value near zero when the
graph is not too damaged, and near unity when severely damage) is exhibited by the ratio of the AIPLs \MyTableReference{tbl:case03}.

The ratio of the AIPLs metric for the test cases does range from 1.0 to 0.0
\MyTableReference{tbl:case03} fitting our intuition.  Now the question becomes, does that
metric continue (within reasonable bounds) as the size of the graph changes, this
is in keeping with the desirable behavior of the metric as listed in section \ref{sec:alt}.
The base size of the graph was increased by factors of 2, 4, 8 and 10,
the ratio was computed and reported \MyTableReference{tbl:case04}.  Data in Table \ref{tbl:case04} 
shows the metric starts at 1.0 for a non-fragmented graph and decreases towards 0.00 as the
graph becomes more fragmented and \MyCloseLCC becomes smaller.  Data in the table for
totally fragmented graphs does not reach 0.000 as the graph becomes larger possibly
because the round offs when computing all the paths and their inverses start to accumulate.
Where the expected value should be 0.000, it is in fact 0.0.  Computing all shortest paths
in a graph using the Floyd-Warshall algorithm can take $\Theta(V^{3})$ time \cite{cormenintroduction},
 so larger
graphs were not fully analyzed.

\begin{table}
\centering
\begin{tabular}{|c|c|c|c|c|c|c|}
\MyHline
\mc{2}{|c|}{\mr{2}{*}{}} &\mc{5}{c|}{$ \mid LCC \mid$} \\
\cline{3-7}
\mc{2}{|c|}{} & 1 & $\frac{n}{2}$ & $\frac{n}{j}$ & $n - 1$  & $n$ \\
\hline
\mr{6}{*}{$m$}
    & 1
    & $\frac{n-1}{1-1}=undef$
    \cellcolor{\MyMath}
    & $\frac{n - \frac{n}{2}}{1-1} = undef$
    \cellcolor{\MyMath}
    & $\frac{n-\frac{n}{j}}{1-1}=undef$
    \cellcolor{\MyMath}
    & $\frac{n-(n-1)}{1-1}=undef$
    \cellcolor{\MyMath}
    & $\frac{n-n}{1-1}=undef$
    \cellcolor{\MyMath}
\\
\cline{2-7}
    & $2$
    & $ \frac{n-1}{2-1} = n-1 \MyCRef{cont:5}$
    & $ \frac{n-\frac{n}{2}}{2-1} = \frac{n}{2} $
    & $ \frac{n-\frac{n}{j}}{2-1}=n(1-\frac{1}{j}) $
    & $ \frac{n-(n-1)}{2-1} = 1 $
    & $ \frac{n-n}{2-1} = 0 $
    \cellcolor{\MyM}
    \\
\cline{2-7}
    & $\frac{n}{2}$
    & $ \approx 2 \MyERef{equ:3-3}\MyCRef{cont:1}$
    & $ \approx 1 \MyERef{equ:3-4}$
    & $ \approx 2-\frac{2}{j}\MyERef{equ:3-5}$
    & $ \frac{2}{n}\MyERef{equ:3-6}\MyCRef{cont:4}$
    & $\frac{n-n}{\frac{n}{2}-1} = 0 $
    \cellcolor{\MyM}
    \\
\cline{2-7}
    & $\frac{n}{j}$
    & $\approx j \MyERef{equ:4-3}\MyCRef{cont:1}$
    & $\approx \frac{j}{2} \MyERef{equ:4-4}$
    & $\approx j - 1 \MyERef{equ:4-5}$
    & $=\frac{j}{n} \MyERef{equ:4-6}\MyCRef{cont:4}$
    & $\frac{n-n}{\frac{n}{j}-1}=0$
    \cellcolor{\MyM}
\\
\cline{2-7}
    & $n - 1$
    & $\approx 1 \MyERef{equ:5-3}\MyCRef{cont:1}$
    & $ \approx \frac{1}{2}\MyERef{equ:5-4}\MyCRef{cont:2}$
    & $\approx 1-\frac{1}{j}\MyERef{equ:5-5}\MyCRef{cont:2}$
    & $ \approx \frac{1}{n}\MyERef{equ:5-6}\MyCRef{cont:4}$
    & $\frac{n-n}{(n-1)-1}=0$
    \cellcolor{\MyM}
\\
\cline{2-7}
    & $n$
    & $\frac{n-1}{n-1}= 1$
    & $\approx \frac{1}{2}\MyERef{equ:6-4}\MyCRef{cont:2}$
    & $\approx 1 - \frac{1}{j}\MyERef{equ:6-5}\MyCRef{cont:2}$
    & $ \approx \frac{1}{n}\MyERef{equ:6-6}\MyCRef{cont:4}$
    & $\frac{n-n}{n-1}=0$
    \cellcolor{\MyM}
\\
\MyHline

\end{tabular}

\MyCaption{Analysis of \MyLittleS based on possible values of \MyCloseLCC~ and $m$.}{\AverageNonLCC is the average size of all fragments
in the graph, less the $LCC$.  The table summarizes the lower limit on \MyLittleS based on the maximum number of
fragments $m$ there can be in the graph based on \MyCloseLCC. Where the value in the cell is not obvious (i.e.,
how it was derived, what assumptions were made, etc.), the (E\#) refers to a set of equations 
that show how the value was obtained.  In some cells there is a constraint logical violation.  These constraint
violations are shown as (C\#).}{tbl:s2}
\end{table}

\begin{table}
\centering
\setlength{\extrarowheight}{3pt}
\begin{tabular} {|c|*{8}{r|}}
\MyHline
\multicolumn{1}{|c}{\textbf{}} &
\multicolumn{1}{|c}{\textbf{}} &
\multicolumn{1}{|c}{\textbf{}} &
\multicolumn{1}{|c}{\textbf{}} &
\multicolumn{1}{|c}{\textbf{}} &
\multicolumn{1}{|c}{\textbf{}} &
\multicolumn{1}{|c}{\textbf{Standard}} &
\multicolumn{1}{|c}{\textbf{Harmonic}} &
\multicolumn{1}{|c|}{\textbf{Geometric}} \\
\multicolumn{1}{|c}{\textbf{Name}} &
\multicolumn{1}{|c}{\textbf{S}} &
\multicolumn{1}{|c}{\textbf{s}} &
\multicolumn{1}{|c}{\textbf{m}} &
\multicolumn{1}{|c}{\textbf{Median}} &
\multicolumn{1}{|c}{\textbf{Mean}} &
\multicolumn{1}{|c}{\textbf{Deviation}} &
\multicolumn{1}{|c}{\textbf{Mean}} &
\multicolumn{1}{|c|}{\textbf{Mean}} \\
\hline
\MyShortNameA & 100 & NaN & 1 & 100.00 & 100.00 & NA & 100.00 & 100.00 \\ 
\MyShortNameB & 90 & 10 & 2 & 50.00 & 50.00 & 56.57 & 18.00 & 30.00 \\ 
\MyShortNameC & 90 & 1 & 11 & 1.00 & 9.09 & 26.83 & 1.10 & 1.51 \\ 
\MyShortNameD & 80 & 2 & 11 & 2.00 & 9.09 & 23.52 & 2.19 & 2.80 \\ 
\MyShortNameE & 50 & 50 & 2 & 50.00 & 50.00 & 0.00 & 50.00 & 50.00 \\ 
\MyShortNameF & 50 & 25 & 3 & 49.00 & 33.33 & 28.01 & 2.88 & 13.48 \\ 
\MyShortNameG & 50 & 25 & 3 & 40.00 & 33.33 & 20.82 & 20.69 & 27.14 \\ 
\MyShortNameH & 50 & 17 & 4 & 20.00 & 25.00 & 19.15 & 15.79 & 19.68 \\ 
\MyShortNameI & 50 & 5 & 11 & 5.00 & 9.09 & 13.57 & 5.45 & 6.16 \\ 
\MyShortNameJ & 20 & 20 & 5 & 20.00 & 20.00 & 0.00 & 20.00 & 20.00 \\ 
\MyShortNameK & 16 & 8 & 11 & 9.00 & 9.09 & 5.07 & 4.70 & 7.19 \\ 
\MyShortNameL & 10 & 10 & 10 & 10.00 & 10.00 & 0.00 & 10.00 & 10.00 \\ 
\MyShortNameM & 10 & 9 & 11 & 9.00 & 9.09 & 0.30 & 9.08 & 9.09 \\ 
\MyShortNameN & 1 & 1 & 100 & 1.00 & 1.00 & 0.00 & 1.00 & 1.00 \\ 
\MyHline
\end{tabular}
\MyCaption {Simple and standard statistical approaches applied to $S$ and the set of all fragments less the $LCC$.}
{The hoped for behavior of the metrics is to be a ``good'' value (approximately 0) for the low numbered cases and a ``bad'' value (approximately 1) for the high numbered cases.  The simple statistical approaches did not produce the type of hoped for behavior.}
{tbl:case01}
\end{table}

\begin{table}
\centering
\setlength{\extrarowheight}{3pt}
\begin{tabular} {|c|*{8}{r|}}
\MyHline
\multicolumn{1}{|c}{\textbf{}} &
\multicolumn{1}{|c}{\textbf{\MyEquationInline{fScoreLHS}}} &
\multicolumn{1}{|c}{\textbf{\MyEquationInline{fScoreBetaLHS}}} &
\multicolumn{1}{|c}{\textbf{Arithmetic}} &
\multicolumn{1}{|c}{\textbf{Geometric}} &
\multicolumn{1}{|c}{\textbf{Quadratic}} &
\multicolumn{1}{|c}{\textbf{Ratio}} &
\multicolumn{1}{|c}{\textbf{}} &
\multicolumn{1}{|c|}{\textbf{}} \\
\multicolumn{1}{|c}{\textbf{Name}} &
\multicolumn{1}{|c}{\textbf{}} &
\multicolumn{1}{|c}{\textbf{\ensuremath{\beta = 0.5}}} &
\multicolumn{1}{|c}{\textbf{Mean}} &
\multicolumn{1}{|c}{\textbf{Mean}} &
\multicolumn{1}{|c}{\textbf{Mean}} &
\multicolumn{1}{|c}{\textbf{(s/S)}} &
\multicolumn{1}{|c}{\textbf{\ensuremath{\log{(S^{s})}}}} &
\multicolumn{1}{|c|}{\textbf{\ensuremath{\log{(s^{S})}}}} \\
\hline
\MyShortNameA & NaN & NaN & 100.00 & 100.00 & 100.00 & NaN  & NaN  & NaN \\ 
\MyShortNameB & 18.00 & 34.62 & 50.00 & 30.00 & 64.03 & 0.11  & 45.00  & 207.23 \\ 
\MyShortNameC & 1.98 & 4.79 & 9.09 & 1.51 & 27.15 & 0.01  & 4.50  & 0.00 \\ 
\MyShortNameD & 3.90 & 9.09 & 9.09 & 2.80 & 24.20 & 0.03  & 8.76  & 55.45 \\ 
\MyShortNameE & 50.00 & 50.00 & 50.00 & 50.00 & 50.00 & 1.00  & 195.60  & 195.60 \\ 
\MyShortNameF & 33.33 & 41.67 & 33.33 & 13.48 & 40.42 & 0.50  & 97.80  & 160.94 \\ 
\MyShortNameG & 33.33 & 41.67 & 33.33 & 27.14 & 37.42 & 0.50  & 97.80  & 160.94 \\ 
\MyShortNameH & 25.00 & 35.71 & 25.00 & 19.68 & 30.00 & 0.33  & 65.20  & 140.67 \\ 
\MyShortNameI & 9.09 & 17.86 & 9.09 & 6.16 & 15.81 & 0.10  & 19.56  & 80.47 \\ 
\MyShortNameJ & 20.00 & 20.00 & 20.00 & 20.00 & 20.00 & 1.00  & 59.91  & 59.91 \\ 
\MyShortNameK & 11.02 & 13.55 & 9.09 & 7.19 & 10.30 & 0.53  & 23.29  & 34.05 \\ 
\MyShortNameL & 10.00 & 10.00 & 10.00 & 10.00 & 10.00 & 1.00  & 23.03  & 23.03 \\ 
\MyShortNameM & 9.47 & 9.78 & 9.09 & 9.09 & 9.10 & 0.90  & 20.72  & 21.97 \\ 
\MyShortNameN & 1.00 & 1.00 & 1.00 & 1.00 & 1.00 & 1.00  & 0.00  & 0.00 \\ 
\MyHline
\end{tabular}
\MyCaption {Slightly more sophisticated  statistical approaches applied to $S$ and the set of  all fragments less the $LCC$.}
{Equations  \ref{equ:fscore} through \ref{equ:quadmean} were applied to $S$ and $s$ where $x_{1} = S$ and $x_{2}=s$.  For all the test cases, the computed values had a ``hump'' and a ``swale'' minimizing their utility as a metric for the ``fitness'' or ``damage'' of a graph. The hoped for behavior of the metric is to be a ``good'' value (something approaching 0.0) for the low numbered cases and a ``bad'' value (something approaching 1.0) for the high numbered cases.  The ratio of $s$ and $S$ are not particularly usable, and most of the $s$ and $S$ exponentiations result in hugely large numbers that do not appear  to be very enlightening.}
{tbl:case02}
\end{table}

\begin{table}
\setlength{\extrarowheight}{3pt}
\centering
\begin{tabular} {|c|r|r|r|*{3}{c|}}
\MyHline
& \multicolumn{3}{c|}{\textbf{Albert, Jeong and Barab{\'a}si}}&\multicolumn{3}{c|}{\MyEquationInline{damageLHS}} \\
\multicolumn{1}{|c}{\textbf{}} &
\multicolumn{1}{|c}{\textbf{}} &
\multicolumn{1}{|c}{\textbf{}} &
\multicolumn{1}{|c}{\textbf{}} &
\multicolumn{1}{|c}{\textbf{}} &
\multicolumn{1}{|c}{\textbf{Small}} &
\multicolumn{1}{|c|}{\textbf{Scale}} \\
\multicolumn{1}{|c}{\textbf{Name}} &
\multicolumn{1}{|c}{\textbf{S}} &
\multicolumn{1}{|c}{\textbf{s}} &
\multicolumn{1}{|c}{\textbf{m}} &
\multicolumn{1}{|c}{\textbf{Random}} &
\multicolumn{1}{|c}{\textbf{World}} &
\multicolumn{1}{|c|}{\textbf{Free}} \\
\hline
\MyShortNameA & 100 & NaN & 1 & 0.000 & 0.000 & 0.000 \\ 
\MyShortNameB & 90 & 10 & 2 & 0.181 & 0.106 & 0.140 \\ 
\MyShortNameC & 90 & 1 & 11 & 0.191 & 0.135 & 0.159 \\ 
\MyShortNameD & 80 & 2 & 11 & 0.357 & 0.265 & 0.308 \\ 
\MyShortNameE & 50 & 50 & 2 & 0.506 & 0.275 & 0.387 \\ 
\MyShortNameF & 50 & 25 & 3 & 0.515 & 0.285 & 0.395 \\ 
\MyShortNameG & 50 & 25 & 3 & 0.585 & 0.345 & 0.459 \\ 
\MyShortNameH & 50 & 17 & 4 & 0.645 & 0.407 & 0.520 \\ 
\MyShortNameI & 50 & 5 & 11 & 0.733 & 0.573 & 0.638 \\ 
\MyShortNameJ & 20 & 20 & 5 & 0.803 & 0.536 & 0.658 \\ 
\MyShortNameK & 16 & 8 & 11 & 0.890 & 0.692 & 0.778 \\ 
\MyShortNameL & 10 & 10 & 10 & 0.907 & 0.712 & 0.807 \\ 
\MyShortNameM & 10 & 9 & 11 & 0.917 & 0.741 & 0.822 \\ 
\MyShortNameN & 1 & 1 & 100 & 1.000 & 1.000 & 1.000 \\ 
\MyHline
\end{tabular}
\MyCaption{Application of proposed damage metric to the test case graphs.}
{The hoped for behavior of the metrics is to be a ``good'' value (approximately 0) for the low numbered cases and a ``bad'' value (approximately 1) for the high numbered cases.  AJB in \cite{Albert2000} based their analysis on graph information that they obtained on Internet and HTTP connectivity.  Based on the statistics for those graphs, they constructed exponential (random degree distribution) and scale-free graphs with the same notional properties.}
{tbl:case03}
\end{table}

\begin{table}
\centering
\setlength{\extrarowheight}{3pt}
\begin{tabular} {|c|*{7}{r|}}
\MyHline
\multicolumn{1}{|c}{\textbf{}} &
\multicolumn{3}{|c}{\textbf{200 nodes}} &
\multicolumn{3}{|c|}{\textbf{400 nodes}} \\
\hline
\multicolumn{1}{|c}{\textbf{Base}} &
\multicolumn{1}{|c}{\textbf{Ran-}} &
\multicolumn{1}{|c}{\textbf{Small}} &
\multicolumn{1}{|c}{\textbf{Scale}} &
\multicolumn{1}{|c}{\textbf{Ran-}} &
\multicolumn{1}{|c}{\textbf{Small}} &
\multicolumn{1}{|c|}{\textbf{Scale}} \\
\multicolumn{1}{|c}{\textbf{Case}} &
\multicolumn{1}{|c}{\textbf{dom}} &
\multicolumn{1}{|c}{\textbf{World}} &
\multicolumn{1}{|c}{\textbf{Free}} &
\multicolumn{1}{|c}{\textbf{dom}} &
\multicolumn{1}{|c}{\textbf{World}} &
\multicolumn{1}{|c|}{\textbf{Free}} \\
\hline
\MyShortNameA & 0.000 & 0.000 & 0.000 & 0.000 & 0.000 & 0.000\\ 
\MyShortNameB & 0.181 & 0.080 & 0.149 & 0.181 & 0.131 & 0.154\\ 
\MyShortNameC & 0.190 & 0.115 & 0.167 & 0.190 & 0.167 & 0.168\\ 
\MyShortNameD & 0.359 & 0.202 & 0.311 & 0.357 & 0.227 & 0.320\\ 
\MyShortNameE & 0.505 & 0.194 & 0.409 & 0.501 & 0.206 & 0.444\\ 
\MyShortNameF & 0.514 & 0.205 & 0.417 & 0.511 & 0.215 & 0.454\\ 
\MyShortNameG & 0.584 & 0.266 & 0.482 & 0.582 & 0.250 & 0.516\\ 
\MyShortNameH & 0.644 & 0.334 & 0.540 & 0.642 & 0.317 & 0.573\\ 
\MyShortNameI & 0.729 & 0.481 & 0.647 & 0.726 & 0.455 & 0.666\\ 
\MyShortNameJ & 0.804 & 0.469 & 0.682 & 0.802 & 0.418 & 0.719\\ 
\MyShortNameK & 0.886 & 0.608 & 0.779 & 0.886 & 0.563 & 0.807\\ 
\MyShortNameL & 0.902 & 0.626 & 0.798 & 0.902 & 0.578 & 0.823\\ 
\MyShortNameM & 0.911 & 0.649 & 0.812 & 0.911 & 0.597 & 0.830\\ 
\MyShortNameN & 0.994 & 0.974 & 0.975 & 0.993 & 0.939 & 0.970\\ 
\MyHline
\multicolumn{1}{|c}{\textbf{}} &
\multicolumn{3}{|c}{\textbf{800 nodes}} &
\multicolumn{3}{|c|}{\textbf{1000 nodes}} \\
\hline
\multicolumn{1}{|c}{\textbf{Base}} &
\multicolumn{1}{|c}{\textbf{Ran-}} &
\multicolumn{1}{|c}{\textbf{Small}} &
\multicolumn{1}{|c}{\textbf{Scale}} &
\multicolumn{1}{|c}{\textbf{Ran-}} &
\multicolumn{1}{|c}{\textbf{Small}} &
\multicolumn{1}{|c|}{\textbf{Scale}} \\
\multicolumn{1}{|c}{\textbf{Case}} &
\multicolumn{1}{|c}{\textbf{dom}} &
\multicolumn{1}{|c}{\textbf{World}} &
\multicolumn{1}{|c}{\textbf{Free}} &
\multicolumn{1}{|c}{\textbf{dom}} &
\multicolumn{1}{|c}{\textbf{World}} &
\multicolumn{1}{|c|}{\textbf{Free}} \\
\hline
\MyShortNameA & 0.000 & 0.000 & 0.000 & 0.000 & 0.000 & 0.000\\ 
\MyShortNameB & 0.180 & 0.044 & 0.161 & 0.180 & 0.124 & 0.161\\ 
\MyShortNameC & 0.189 & 0.073 & 0.174 & 0.189 & 0.148 & 0.173\\ 
\MyShortNameD & 0.356 & 0.228 & 0.328 & 0.356 & 0.307 & 0.330\\ 
\MyShortNameE & 0.501 & 0.350 & 0.434 & 0.501 & 0.330 & 0.444\\ 
\MyShortNameF & 0.510 & 0.362 & 0.444 & 0.510 & 0.333 & 0.454\\ 
\MyShortNameG & 0.581 & 0.395 & 0.512 & 0.581 & 0.416 & 0.520\\ 
\MyShortNameH & 0.641 & 0.436 & 0.575 & 0.641 & 0.456 & 0.582\\ 
\MyShortNameI & 0.726 & 0.538 & 0.667 & 0.726 & 0.541 & 0.675\\ 
\MyShortNameJ & 0.801 & 0.495 & 0.733 & 0.801 & 0.574 & 0.741\\ 
\MyShortNameK & 0.885 & 0.609 & 0.824 & 0.885 & 0.637 & 0.829\\ 
\MyShortNameL & 0.901 & 0.621 & 0.841 & 0.901 & 0.656 & 0.847\\ 
\MyShortNameM & 0.910 & 0.650 & 0.852 & 0.910 & 0.668 & 0.856\\ 
\MyShortNameN & 0.991 & 0.907 & 0.970 & 0.991 & 0.901 & 0.970\\ 
\MyHline
\end{tabular}
\MyCaption {Results of testing the proposed metric on larger graphs.}
{The base case of a 100 node graph was increased by factors of 2, 4,  8 and 10 to ensure that the metric  continued to perform correctly.  In all cases, the ratio worked intuitively, starting at 0.0 for a non-fragmented graph and increasing towards 1.0 for a totally fragmented graph.  Some of the fully fragmented graphs, did not reach 1.000 possibly due to round off errors in the  computations.  Those fully fragmented graphs that did not reach 1.000 did reach 1.0 as expected.}
{tbl:case04}
\end{table}

\section{Conclusion}\label{sec:conclusion}
Considerable time was spent examining the equation Albert, Jeong and Barab{\'a}si \AverageNonLCC
from \cite{Albert2000} to see how it could be used to quantify the ``damage'' to a
graph when the graph becomes fragmented.  This investigation was spurred on by the
equation's use in \cite{Albert2000,brandes2005network} and the
 belief that there was more information there that could be of use.
The equation was analyzed and limits (both mathematical and logical)
were identified.  These limits fit nicely with the graphs in both references.

Because of the limitations experienced using the tuple (\MyLittleS, $m$, \MyLargeS) from AJB
 and the desire to have a unit-less
metric that reflects the efficiency of the graph; a different approach was identified.
Netotea and Pongor in \cite{netotea2006evolution} and Crucitti et al. in
\cite{crucitti2004error} proposed
the use of the average inverse path length (AIPL) as a way of quantifying the efficiency of a graph. 
 We used the equations from Crucitti  to compute the AIPL of
a connected  graph that is equal to the sum of all
the fragments and the original   disconnected graph consisting of the fragments.  
A ratio was computed using these  AIPLs.  This ratio has the desired effect of being: (1) unit-less,
(2) independent of graph size, and (3) does not require a priori knowledge of
the graph.

The ratio of the average inverse path lengths of a connected and a disconnected graph
can be used as a metric about the health of a graph.  The damage (i.e., the converse of health) of a fragmented graph can be
computed using \MyEquationInline{damage-02.tex}.

\section{Acknowledgment}\label{sec:acknowledgement}
 This work supported in part by the NSF, Project 370161.

\bibliographystyle{abbrv}
\bibliography{master}
\newpage
\appendix
\newpage
\section{Comparison of connected and disconnected  graph metrics}\label{sec:metrics}
Within this paper the following terms and ideas are used:

\begin{enumerate}
\item A graph \MyEquationInline{graph} is an ordered pair of disjoint sets
(\MyEquationInline{vertexLHS},\MyEquationInline{edgeLHS}) such that \MyEquationInline{edgeLHS} is a subset of $\MyEquationInline{vertexLHS}^{2}$ of the unordered pairs of
\MyEquationInline{vertexLHS} \cite{bollobas1998modern}.
\item The terms \emph{vertex} and \emph{node} are used interchangeably and mean the
same thing.

\item The term \emph{connected} means that there is a series \emph{edges} between
any arbitrary nodes source \emph{s} and terminus \emph{t} that can be used to get from
node \emph{s} to \emph{t}\cite{bollobas1998modern}. 
 A graph is \emph{disconnected} when nodes \emph{s} and \emph{t}
cannot be reached by any series of edges.

\item The term \emph{directed} means that the edge connecting nodes \emph{s} and \emph{t} 
is unidirectional.   \emph{t} is an immediate neighbor to \emph{s} because they are
separated by one edge and it takes more than one edge for \emph{t} to reach \emph{s}.

\item The term \emph{undirected} means that the edge connecting nodes \emph{s} and \emph{t} 
is bidirectional.     \emph{t} is an immediate neighbor to \emph{s} because they are
separated by one edge and the same edge connects  \emph{t} to \emph{s}.

\item The term \emph{simple} means that there is only one \emph{edge} between
any adjacent nodes.

\item The terms \emph{fragment}, \emph{cluster} or \emph{component} are used
interchangeably and mean a set of \emph{nodes} (there may be only 1 node) that
are connected to each other.  A graph \emph{G} may have more than one \emph{component}.

\item The difference between a \emph{graph} and a \emph{network} is the assignment
of different \emph{weights} to each \emph{edge} in the graph.  By default, all
\emph{edges} in a \emph{graph} have a weight of 1.  While, \emph{edges} in a \emph{network}
may have different weights.

\item A node could have an edge that started and ended at the same source node.  These edges are
called \emph{self loops}.
\end{enumerate}

The graphs in this paper are: \emph{undirected}, \emph{simple}, \emph{self loops} are not permitted 
and may have more than one \emph{component}.
\subsection{Connected graph metrics} \label{sec:connected}
Here we review  a collection of characteristic metrics for connected graphs.
In many cases the characteristic does not have meaning, or a computable value when the
graph is not connected.
\begin{description}
\item Path length \cite{bollobas1998modern}. \\
The number of edges in a path $P$ from a starting node $u$ to terminating node $v$.
 \MyEquationLabeled{distanceEdge}{equ:00c}
\item Average path length (APL)\cite{brinkmeier211}.\\
 The average of all shortest path lengths  between nodes $u$ and $v$. 
The  lower an APL, the fewer edges on average there are between nodes.
 \MyEquationLabeled{averageDistance}{equ:01c}

\MyIgnore{\item Average inverse path length (AIPL) \cite{holme2002attack}.\\
 The average of the inverse  of all shortest paths 
between all nodes $u$ and $v$.  AIPL is also known as average
inverse shortest path (AISP) \cite{beygelzimer2005improving} and average inverse shortest path length (AISPL) \cite{venuturumilli2006obtaining}.
\MyEquationLabeled{averageDistanceInverse}{equ:02c}
If a path does not exist between nodes $u$ and $v$   then by definition the path's length is infinite $\infty$.
}
\item Centrality, betweenness of an edge \cite{koschutzkicentrality}. \\
The proportion of shortest paths between nodes $s$ and $t$ that use edge $e$.
\MyEquationLabeled{betweennessEdge}{equ:03c}

\item Centrality, betweenness of an edge relative to all edges in a graph.\\
The edge that has the highest centrality of all edges is the edge that is
most used by all shortest paths in the graph.
 \MyEquationLabeled{betweennessGlobalEdge}{equ:14c}

\item Centrality, betweenness of a vertex \cite{koschutzkicentrality}. \\
The proportion of shortest paths between nodes $s$ and $t$ that use vertex $v$.
 \MyEquationLabeled{betweennessVertex}{equ:04c}

\item Centrality, betweenness of a vertex  relative to all vertices in a graph.\\
The vertex that has the highest centrality of all vertices is the vertex that is used
by the most shortest paths in the graph.
 \MyEquationLabeled{betweennessGlobalVertex}{equ:15c}

\MyIgnore{
\item Centrality, closeness of a vertex \cite{csardi2006igraph}. \\
 How close (fewest
number of edges) $u$ is to all other vertices.
\MyEquationLabeled{closenessNormalized}{equ:05c}

\item Centrality, degree. \\
The number of edges incident to a vertex.  \MyEquationLabeled{degreeUndirectedDegree}{equ:06c}
} % End of \MyIgnore

\item Clustering coefficient \cite{brinkmeier211,watts:collective_dynamics}.\\
The likelihood that two neighbors of $v$ are connected.
\MyEquationLabeled{clusteringCoefficient}{equ:07c}

\item Degree of a node.\\
The number of edges incident to a node.
\MyEquationLabeled{degree}{equ:12c}

\item Diameter of a graph \cite{brinkmeier211}.\\
The  maximal shortest path between any vertices $u$ and $v$.
\MyEquationLabeled{diameter}{equ:08c}

\item Eccentricity of a node \cite{brinkmeier211,west2001introduction}.\\
 The maximal distance between vertex $u$ and any other vertex  $v$.
\MyEquationLabeled{eccentricity}{equ:09c}

\item Eccentricity of a graph. \\
The maximal eccentricity of all nodes $u$ in $G$.
\MyEquationLabeled{eccentricityGlobal}{equ:11c}

\item Radius of a graph \cite{brinkmeier211,west2001introduction}.\\
The minimal eccentricity of all vertices in $G$.
\MyEquationLabeled{radius}{equ:10c}

\item Triangles based on a node \cite{brinkmeier211}.\\
The number of subgraphs of the graph $G$ that have exactly three nodes and three edges and one
of the nodes is $v$.
\MyEquationLabeled{triangleLocal}{equ:13c}

\end{description}

Equations \ref{equ:01c}, \ref{equ:08c} and \ref{equ:09c} are directly related to the
length of the path between nodes $u$ and $v$ \MyEquationReference{equ:00c}. Equations \ref{equ:03c},
\ref{equ:14c}, \ref{equ:04c}, \ref{equ:15c}, \ref{equ:07c}, \ref{equ:09c} and \ref{equ:10c} are
indirected related to the path length.

\subsection{Disconnected graph metrics}\label{sec:disconnected}
Here we review  a  collection of characteristic metrics for disconnected graphs.
In many cases the connected graph characteristic does not have meaning, or is not  computable
 when the graph is disconnected.
\begin{description}
\item Constrained average path length (CAPL).\\
 The average of all shortest path lengths  between nodes $u$ and $v$, 
given that there is a path between  $u$ and $v$.
The  lower an CAPL, the fewer edges on average there are between nodes.
 \MyEquationLabeled{averageDistanceConnected}{equ:01d}

\item Average inverse path length (AIPL) \cite{holme2002attack}.\\
 The average of the inverse  of all shortest paths 
between all nodes $u$ and $v$.  AIPL is also known as average
inverse shortest path (AISP) \cite{beygelzimer2005improving} and average inverse shortest path length (AISPL) \cite{venuturumilli2006obtaining}.
\MyEquationLabeled{averageDistanceInverse}{equ:02c}
If a path does not exist between nodes $u$ and $v$   then by definition the path's length is infinite $\infty$.

\end{description}

Equation \ref{equ:01d} is an constrained APL as
compared to a un-constrained  APL \MyEquationReference{equ:01c} that restricts the
path lengths between nodes to those whose path length is not $\infty$.  Equation \ref{equ:02c} at first appears to be
dependent on a path length, but in fact, it does not.  If a path does not exist between nodes $u$ and $v$
then, by definition, the path length is infinite $\infty$.  Any number divided by $\infty$ is defined
to be 0.

\subsection{The effect of directivity and self loops}
Many of the graph metric equations use the number of edges in the
graph, but often the authors do not specify how the edges are 
selected or limited.  Table \ref{tbl:edges:comparison} identifies
how many edges can be used based on two criteria; whether or not the
edges are directed or whether or not the graph permits edges back to the
originating vertex.  Based on these restrictions, the number of edges
can range from \ensuremath{\frac{n*(n-1)}{2}}~ to~ \ensuremath{n*(n+1)}.

\begin{table}
\centering
\newcommand{\LocalEdgeCount}[0]{\ensuremath{\mathclose \mid E_{max} \mathclose \mid}~}
\newcolumntype{C}{>{\centering\arraybackslash\hspace{0pt}} m{2cm} } 
\begin{tabular}{|cc|c|c|}
\MyHline
&&\multicolumn{2}{c|}{\textbf{Are   directed edges permitted?}} \\
&& \textbf{Yes} & \textbf{No} \\
\hline
\multirow {4}{*}{\rotatebox{90}{\textbf{Self loops permitted?}}} &
\rotatebox{90}{\textbf{~~~~~~~~~~~~~~~~~~~~~~~~~~~~~~~~~~~Yes}} &
\includegraphics{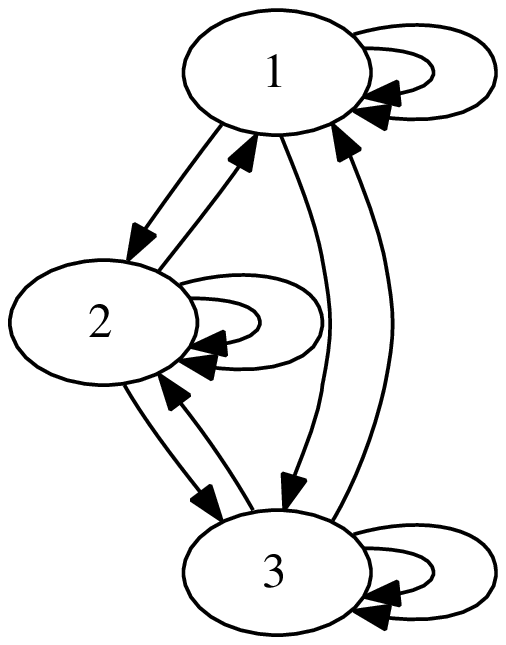} & \includegraphics{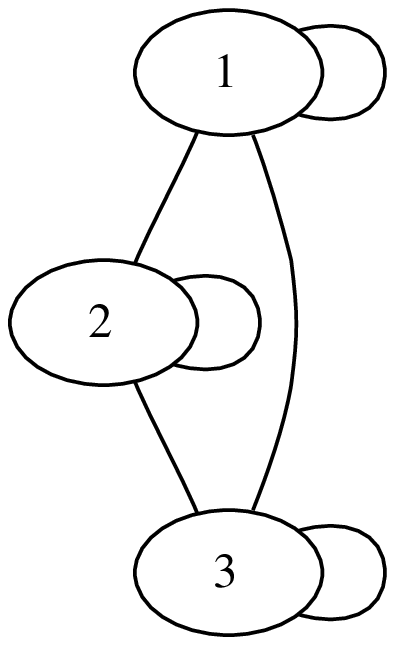} \\
 & & \ensuremath{\LocalEdgeCount = n*(n+1)=12} &\ensuremath{\LocalEdgeCount = \frac{n*(n+1)}{2}=6} \\
\cline{2-4}
  & \rotatebox{90}{\textbf{~~~~~~~~~~~~~~~~~~~~~~~~~~~~~~~~~~~No}} &
 \includegraphics{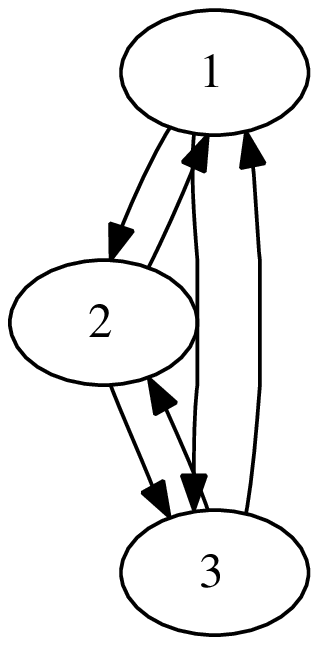} & \includegraphics{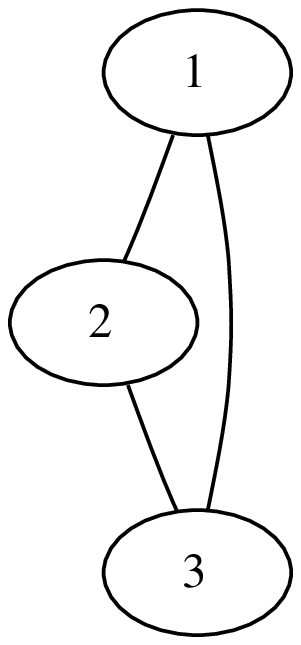} \\
 & & \ensuremath{\LocalEdgeCount = n*(n-1)=6} &\ensuremath{\LocalEdgeCount = \frac{n*(n-1)}{2}=3} \\
\MyHline
\end{tabular}
\MyCaption{Maximum number of edges based on directivity and self loops.}
{A sample three node graph is used to illustrate the maximum number of edges 
a graph can have based on whether edges are directed or not and whether
the graph permits edges that originate and return to the same node.  The number 
of edges that can be used various graph theoretical computations can range
from \ensuremath{\frac{n*(n-1)}{2}}~ to~ \ensuremath{n*(n+1)}.  The apparently 
redundant double edges when  directed edges are allowed and self loops are permitted
reflect that there is two-way communication.  In effect, the node is ``talking'' to
itself.
}
{tbl:edges:comparison}
\end{table}

\MyIgnore{
CLC CLC CLC CLC CLC CLC CLC CLC CLC CLC CLC CLC CLC CLC CLC 

\newcolumntype{x}[1]{>{\centering\arraybackslash\hspace{0pt}}m{#1}}
\begin{table}[ht]\footnotesize
\centering
\begin{tabular*}{0.95\textwidth }{@{\extracolsep{\fill}}|llllllllllllllll|x{0.15\textwidth}|}
\hline
\multicolumn{16}{|c|}{Small text} & Long long long long long text \\ \hline 
1 & 5 & 6 & 7 & 8 & 9 & 10 & 11 & 12 & 13 & 16 & 19 & 20 & 21 & 22 & 23 & 0.9636149 \\ \hline 
\end{tabular*}
\caption{Caption}
\label{tab:Example}
\end{table}
}

\newpage
\section{Derivation of various Albert, Jeong and Barab{\'a}si related estimations}\label{sec:derivation}
Often papers have only the solution to a problem or perhaps only the first and last steps.  What follows is
a collection of all the equations and their derivations for the solutions in Table \ref{tbl:s2}.

Table \ref{tbl:s2} is repeated here for convenience.  In most cases this is basic algebra and the equations are
here because sometimes it is hard to remember how an answer was derived when only an answer is given.
\begin{table}[ht]
\centering
\begin{tabular}{|c|c|c|c|c|c|c|}
\MyHline
\mc{2}{|c|}{\mr{2}{*}{}} &\mc{5}{c|}{$ \mid LCC \mid$} \\
\cline{3-7}
\mc{2}{|c|}{} & 1 & $\frac{n}{2}$ & $\frac{n}{j}$ & $n - 1$  & $n$ \\
\hline
\mr{6}{*}{$m$}
    & 1
    & $\frac{n-1}{1-1}=undef$
    \cellcolor{\MyMath}
    & $\frac{n - \frac{n}{2}}{1-1} = undef$
    \cellcolor{\MyMath}
    & $\frac{n-\frac{n}{j}}{1-1}=undef$
    \cellcolor{\MyMath}
    & $\frac{n-(n-1)}{1-1}=undef$
    \cellcolor{\MyMath}
    & $\frac{n-n}{1-1}=undef$
    \cellcolor{\MyMath}
\\
\cline{2-7}
    & $2$
    & $ \frac{n-1}{2-1} = n-1 \MyCRef{cont:5}$
    & $ \frac{n-\frac{n}{2}}{2-1} = \frac{n}{2} $
    & $ \frac{n-\frac{n}{j}}{2-1}=n(1-\frac{1}{j}) $
    & $ \frac{n-(n-1)}{2-1} = 1 $
    & $ \frac{n-n}{2-1} = 0 $
    \cellcolor{\MyM}
    \\
\cline{2-7}
    & $\frac{n}{2}$
    & $ \approx 2 \MyERef{equ:3-3}\MyCRef{cont:1}$
    & $ \approx 1 \MyERef{equ:3-4}$
    & $ \approx 2-\frac{2}{j}\MyERef{equ:3-5}$
    & $ \frac{2}{n}\MyERef{equ:3-6}\MyCRef{cont:4}$
    & $\frac{n-n}{\frac{n}{2}-1} = 0 $
    \cellcolor{\MyM}
    \\
\cline{2-7}
    & $\frac{n}{j}$
    & $\approx j \MyERef{equ:4-3}\MyCRef{cont:1}$
    & $\approx \frac{j}{2} \MyERef{equ:4-4}$
    & $\approx j - 1 \MyERef{equ:4-5}$
    & $=\frac{j}{n} \MyERef{equ:4-6}\MyCRef{cont:4}$
    & $\frac{n-n}{\frac{n}{j}-1}=0$
    \cellcolor{\MyM}
\\
\cline{2-7}
    & $n - 1$
    & $\approx 1 \MyERef{equ:5-3}\MyCRef{cont:1}$
    & $ \approx \frac{1}{2}\MyERef{equ:5-4}\MyCRef{cont:2}$
    & $\approx 1-\frac{1}{j}\MyERef{equ:5-5}\MyCRef{cont:2}$
    & $ \approx \frac{1}{n}\MyERef{equ:5-6}\MyCRef{cont:4}$
    & $\frac{n-n}{(n-1)-1}=0$
    \cellcolor{\MyM}
\\
\cline{2-7}
    & $n$
    & $\frac{n-1}{n-1}= 1$
    & $\approx \frac{1}{2}\MyERef{equ:6-4}\MyCRef{cont:2}$
    & $\approx 1 - \frac{1}{j}\MyERef{equ:6-5}\MyCRef{cont:2}$
    & $ \approx \frac{1}{n}\MyERef{equ:6-6}\MyCRef{cont:4}$
    & $\frac{n-n}{n-1}=0$
    \cellcolor{\MyM}
\\
\MyHline

\end{tabular}

\end{table}

\MyExplanation{\frac{n}{2}}{1}{equ:3-3}
\begin{eqnarray}
  \frac{n-1}{\frac{n}{2}-1} & \approx & \frac{n}{\frac{n}{2}} \nonumber \\
  & \approx & 2    \label{equ:3-3}
\end{eqnarray}

\MyExplanation{\frac{n}{2}}{\frac{n}{2}}{equ:3-4}
\begin{eqnarray}
\frac{n-\frac{n}{2}}{\frac{n}{2}-1} &\approx &\frac{\frac{n}{2}}{\frac{n}{2}} \nonumber \\
& \approx & 1   \label{equ:3-4}
\end{eqnarray}

\MyExplanation{\frac{n}{2}}{\frac{n}{j}}{equ:3-5}
\begin{eqnarray}
\frac{n-\frac{n}{j}}{\frac{n}{2}-1} &\approx &\frac{n(1-\frac{1}{j})2}{n} \nonumber \\
& \approx& 2-\frac{2}{j}   \label{equ:3-5}
\end{eqnarray}

\MyExplanation{\frac{n}{2}}{n-1}{equ:3-6}
\begin{eqnarray}
\frac{n-(n-1)}{\frac{n}{2}-1} & \approx & \frac {1}{\frac{n}{2}} \nonumber \\
& \approx & \frac {2}{n}   \label{equ:3-6}
\end{eqnarray}

\MyExplanation{\frac{n}{j}}{1}{equ:4-3}
\begin{eqnarray}
\frac{n-1}{\frac{n}{j}-1} & \approx & \frac {n}{\frac{n}{j}} \nonumber \\
& \approx & j   \label{equ:4-3}
\end{eqnarray}

\MyExplanation{\frac{n}{j}}{\frac{n}{2}}{equ:4-4}
\begin{eqnarray}
\frac{n - \frac{n}{2}}{\frac{n}{j}-1} &\approx &\frac{n(1-\frac{1}{2})j}{n} \nonumber \\
& \approx & \frac {j}{2}   \label{equ:4-4}
\end{eqnarray}

\MyExplanation{\frac{n}{j}}{\frac{n}{j}}{equ:4-5}
\begin{eqnarray}
\frac{n-\frac{n}{j}}{\frac{n}{j}-1} & \approx  & \frac{n-\frac{n}{j}} {\frac{n}{j}} \nonumber \\
& \approx & \frac{j(n-\frac{n}{j})}{n} \nonumber \\
& \approx & \frac{jn(1-\frac{1}{j})}{n} \nonumber \\
& \approx & j(1-\frac{1}{j}) \nonumber \\
& \approx &  j - 1  \label{equ:4-5}
\end{eqnarray}

\MyExplanation{\frac{n}{j}}{n-1}{equ:4-6}
\begin{eqnarray}
\frac{n-(n-1)}{\frac{n}{j}-1} & = & \frac {1}{\frac{n}{j}} \nonumber \\
& = & \frac{j}{n}   \label{equ:4-6}
\end{eqnarray}

\MyExplanation{n-1}{1}{equ:5-3}
\begin{eqnarray}
\frac{n-1}{(n-1)-1} & = & \frac{n-1}{n-2}  \nonumber \\
&\approx & \frac{n-1}{n-1}  \nonumber \\
&\approx & 1   \label{equ:5-3}
\end{eqnarray}

\MyExplanation{n-1}{\frac{n}{2}}{equ:5-4}
\begin{eqnarray}
\frac{n - \frac{n}{2}}{(n-1)-1} & = & \frac{\frac{n}{2}}{n-2}  \nonumber \\
& \approx &\frac{\frac{n}{2}}{n} \nonumber \\
& \approx & \frac{1}{2}   \label{equ:5-4}
\end{eqnarray}

\MyExplanation{n-1}{\frac{n}{j}}{equ:5-5}
\begin{eqnarray}
\frac{n-\frac{n}{j}}{(n-1)-1} & = & \frac {n-\frac{n}{j}}{n-2}  \nonumber \\
& \approx & \frac{n(1 - \frac{1}{j})}{n} \nonumber \\
& \approx & 1 - \frac {1}{j}   \label{equ:5-5}
\end{eqnarray}

\MyExplanation{n-1}{n-1}{equ:5-6}
\begin{eqnarray}
\frac{n-(n-1)}{(n-1)-1} & = & \frac{1}{n-2}  \nonumber \\
& \approx & \frac{1}{n}   \label{equ:5-6}
\end{eqnarray}

\MyExplanation{n}{\frac{n}{2}}{equ:6-4}
\begin{eqnarray}
\frac{n - \frac{n}{2}}{n-1} & \approx & \frac {n(1 - \frac{1}{2})}{n} \nonumber \\
& \approx & 1 - \frac {1}{2}  \label{equ:6-4}
\end{eqnarray}

\MyExplanation{n}{\frac{n}{j}}{equ:6-5}
\begin{eqnarray}
\frac{n-\frac{n}{j}}{n-1} & \approx & \frac {n(1 - \frac{1}{j})}{n}  \nonumber \\
& \approx & 1 - \frac {1}{j} \label{equ:6-5}
\end{eqnarray}

\MyExplanation{n}{n-1}{equ:6-6}
\begin{eqnarray}
\frac{n-(n-1)}{n-1} & = & \frac {1}{n-1}  \nonumber \\
& \approx & \frac {1}{n}   \label{equ:6-6}
\end{eqnarray}

\newpage
\section{Graph attack profiles}\label{sec:attackProfiles}
\subsection{Comparison of errors and attacks}
Errors and attacks remove components from a system.  The distinguishing characteristic between the two
types of losses is how components are selected.  This characteristic can be explained by using a computer network as a
graph.  The network is a graph where vertices are represented by routers, switches and computers.  While edges
are represented by the connections between the vertices, either wired or wireless connections.

The loss of a router through hardware failure, or mis-configuration,
or the severing of the communications links to the router can be considered to be accidental.
An error is the accidental loss of a component from a system.
The simultaneous loss of a set of routers, perhaps without a readily apparent reason, could be considered to
be an attack.  An attack is the deliberate loss of components, or a component from a system.

The survivability of a graph to error or attack depends  on the underlying structure of the graph (for example scale-free or exponential).
Scale-free graphs are very robust in the face of random failures, but are very susceptible to attacks \cite{Albert2000}.  Where
exponential graphs have just the opposite behavior.
\subsection{Selection of graph component to attack}
Ultimately there are only two graph components that an attacker can attack, edges or vertices.  The selection of which
of these components to attack has to be based on some metric rather than random selection.  Holme and Kim \cite{holme2002attack}
looked at how an attacker could maximize the damage to a graph by one of two approaches.  The approaches being:
\begin{enumerate}
\item To remove the vertex with the highest initial degree (ID)
 \MyEquation{degreeUndirectedDegree}
\item Or, the vertex with the highest in-betweenness
centrality (IB) \MyEquation{betweennessVertex}
\end{enumerate}
Their idea about betweenness can be extended to include removing the edge with the highest in-betweenness centrality
\MyEquation{betweennessEdge}.

Lee et al. in \cite{lee2006rnt} put forth failures in a network as being either \textbf{node}, \textbf{link}, or \textbf{path} related.
Their \textbf{node} corresponds to our \textbf{vertex}.  Their \textbf{link} to our \textbf{edge}.  And, their \textbf{path} to our
\textbf{betweenness}.  The betweenness of a component is a measurement of the component's contribution to all the shortest paths
\MyEquationInline{pathLHS} in the graph.  The higher the betweenness value, the more shortest paths use that component.

In the following subsections, we will use a sample graph to show the effects of an attacker's limited knowledge of the global graph
on which component to remove.

\subsubsection{Size of subgraph to evaluate}
An attacker has to select a graph component to attack, and identifying which component to remove is
based on the attacker's knowledge of some portion of the graph.  The attacker's knowledge can
range from a single component to complete knowledge of the graph.  One approach to gaining knowledge of a graph's organization
is to identify  a vertex and then determine those vertices that are at a path length distance of 1 edge from
the initial vertex.  This process is repeated again and again until the attacker decides to stop increasing the
path length \MyFigureReference{fig:neighbor}.

In Figure \ref{fig:neighbor}, vertex 5 is the source vertex and is colored red.  The path length is initially
set to 1 and the attacker now knows about the vertex set \{4, 5, 6, 8, 9\} \MyFigureReference{fig:neighbor:1}.  All
attacker discovered vertices are colored pink.  As the
path length increases from 2 \MyFigureReference{fig:neighbor:2} to 4 \MyFigureReference{fig:neighbor:4}, more and more
of the global graph becomes known.  As readers, we know what the global graph looks like because we have an omnipotent view point.
The attacker does not enjoy this view and must blindly continue to work outwards from his initial vertex.
The attacker must expend time and energy to increase his knowledge of the graph, until at some point
he will have spent ``enough'' and believes that sending additional time will not be worth the effort.

The attacker uses this limited local knowledge of the global graph to select the
component whose removal  will cause the greatest damage to the graph.  If the path length is increased
enough, the entire graph will be discovered. Barab{\'a}si hypothesized that the entire INTERNET
could be discovered with a path length of 19 \cite{barabsi:world_wide_web_forms_a_large_directed_graph}.  The
resources for attempting to conduct such a discovery may be too large to be practical.

\begin{figure}
\centering
\MyLocalSubfigure{Path length = 1, discovered diameter = 2 }{fig:neighbor:1}{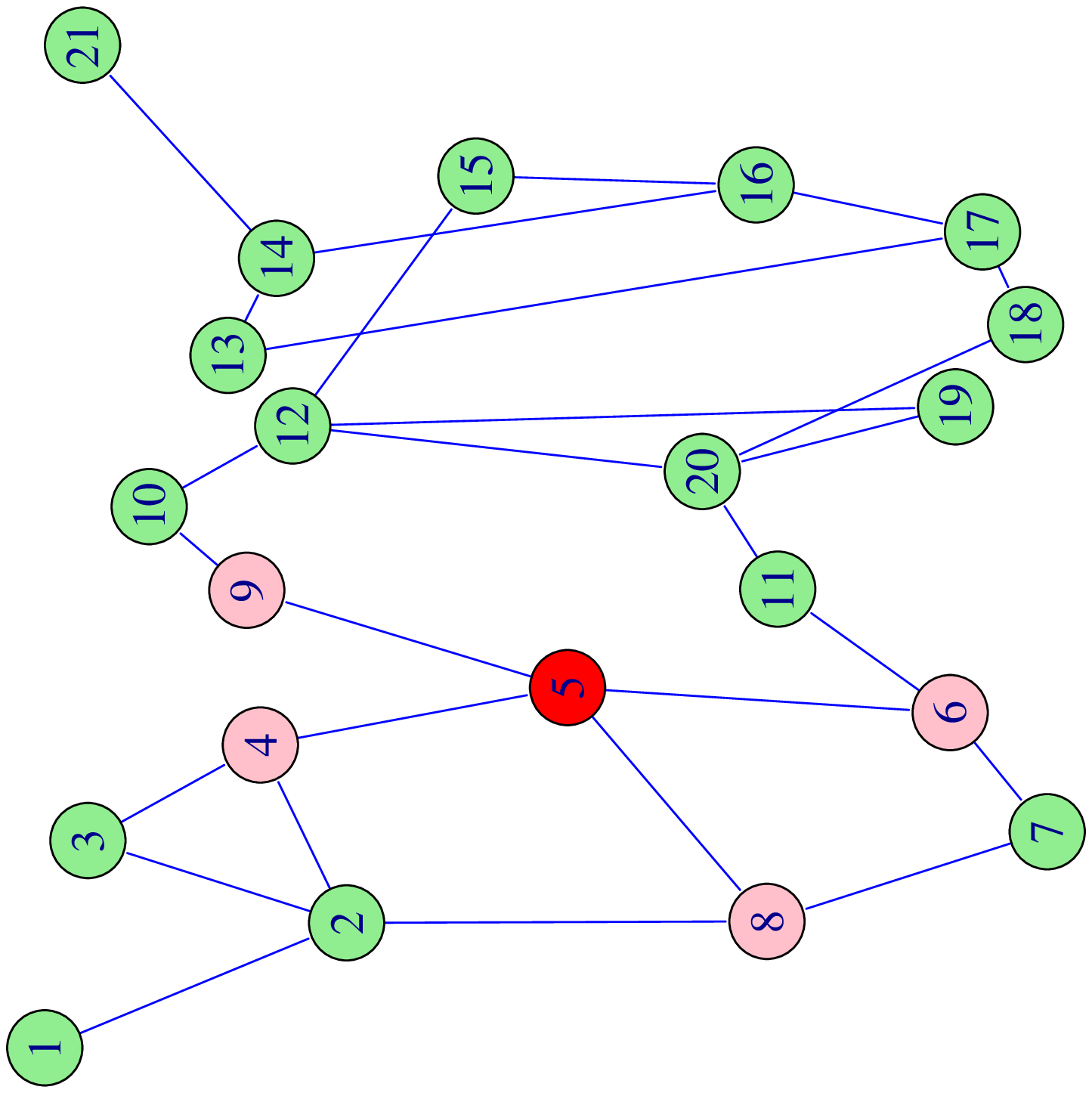}
\MyLocalSubfigure{Path length = 2, discovered diameter = 4}{fig:neighbor:2}{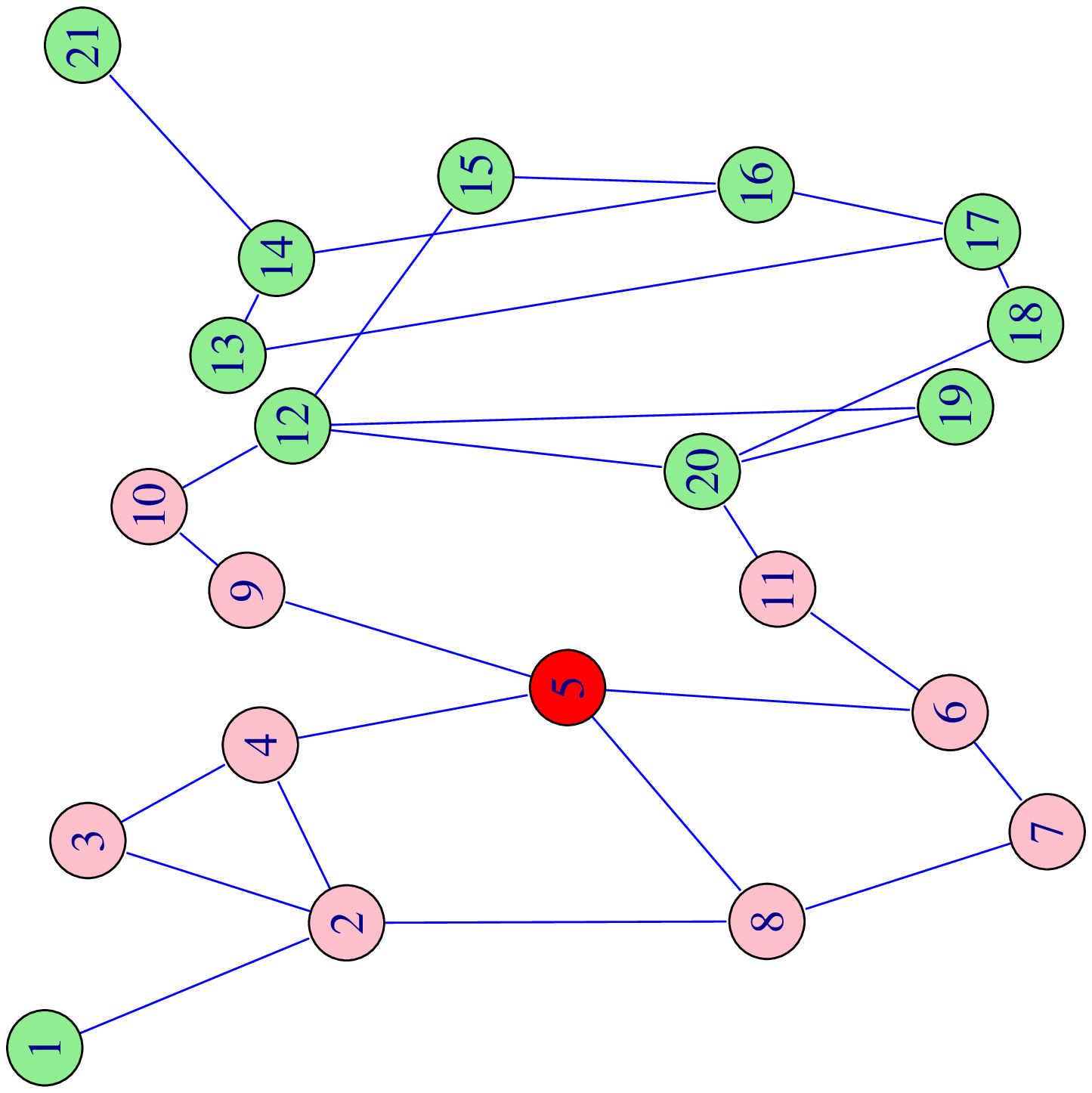}
\MyLocalSubfigure{Path length = 3, discovered diameter = 6}{fig:neighbor:3}{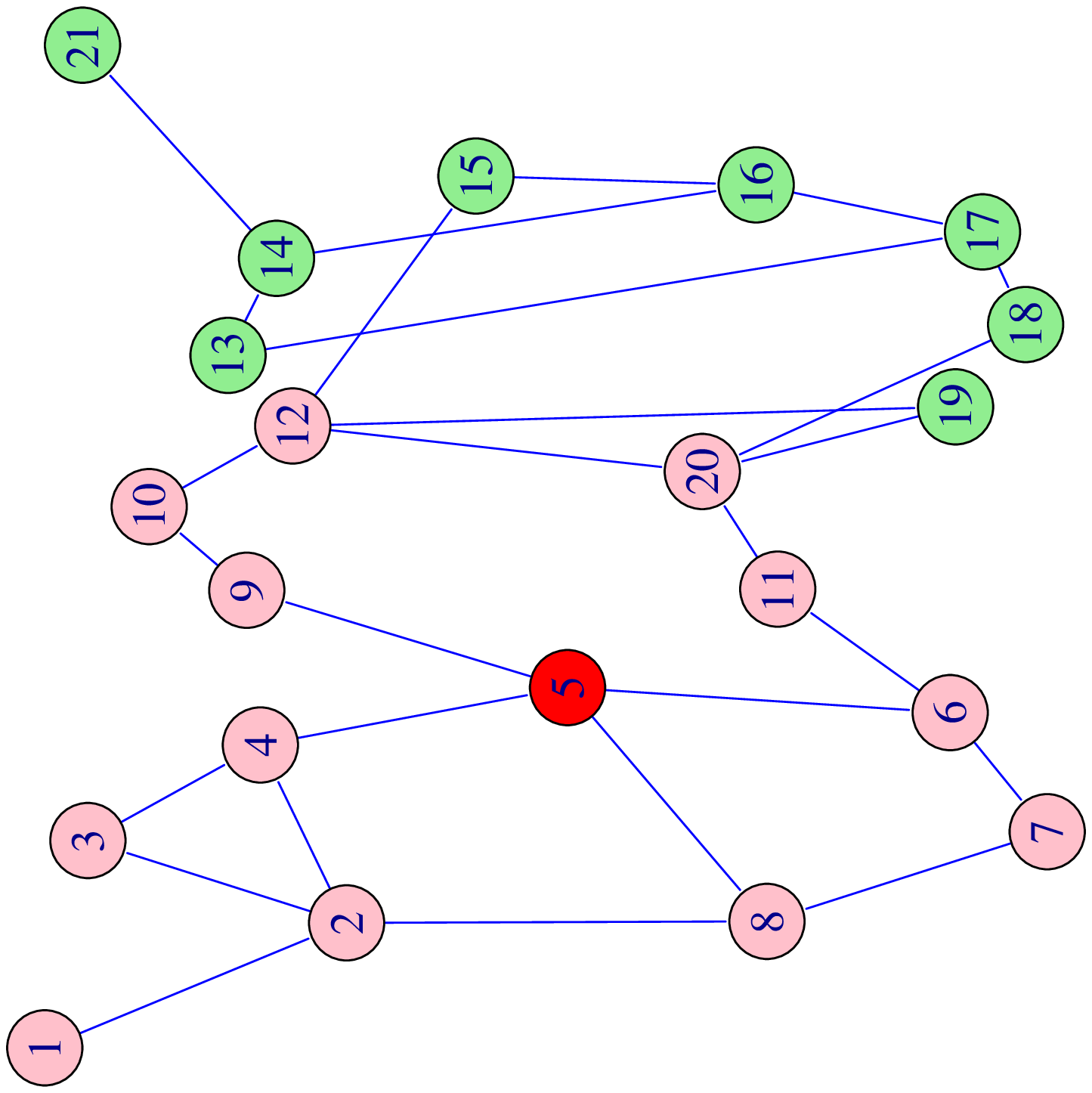}
\MyLocalSubfigure{Path length = 4, discovered diameter = 7}{fig:neighbor:4}{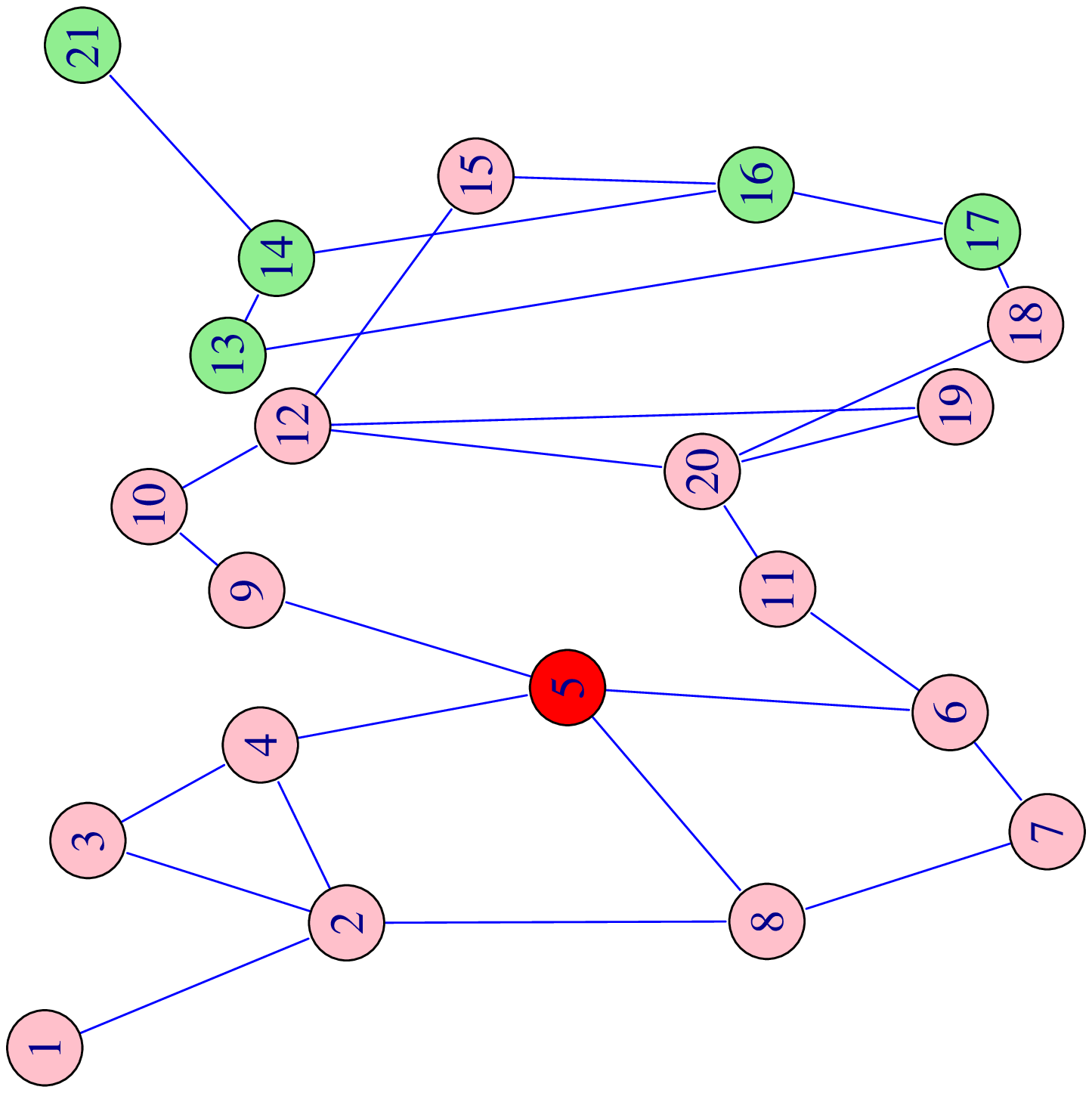}
\MyCaption{The effects of different path lengths starting from a fixed vertex in discovering the global graph.}
{Vertex 5 is the center vertex.  Each sub-figure shows the subgraph that is discovered based on the path length
from the center vertex as the path length increments from 1 to 4. The diameter of the discovered subgraph is at most twice the
path length.  As the path length increases, more and more of the global graph is discovered.}
{fig:neighbor}
\end{figure}

\subsubsection{Edge selection}\label{sec:edgeSelection}
The selection of an edge to remove from the graph is based on how much of the graph that the attacker has
discovered.  As the discovered graph becomes larger and larger (as measured by the path length from a initial/central)
vertex to the rest of the graph \MyFigureReference{fig:neighbor}, the more accurate the computed value betweenness value
of the edge is to the edge's betweenness value for the entire graph.  The edge betweenness value for all edges in the
global graph and for the discovered subgraph is shown in Table \ref{tbl:neighborhood:edge}.
In the table, the first two columns are the vertices that are connected by an edge.  The third column is the edge betweenness
for that edge based on the global graph.  The remaining columns show the edge betweenness value as the path length from
the central vertex gets longer and longer.  In those cases where the discovered subgraph has not discovered a particular
vertex in the global graph, the edge betweenness value is marked with a --- indicating no value possible.  It is interesting to
see how the value of an edge changes as the size of the graph changes.  In most  cases the value of an edge decreases as graph size
increases. 
\begin{table}
 \centering
\setlength{\extrarowheight}{3pt}
\begin{tabular} {|*{7}{c|}}
\MyHline
\multicolumn{1}{|p{0.75in}}{\textbf{Source node}} &
\multicolumn{1}{|p{0.75in}}{\textbf{Dest. node}} &
\multicolumn{1}{|p{0.75in}}{\textbf{Edge Betweenness}} &
\multicolumn{1}{|p{0.75in}}{\textbf{Path length 1}} &
\multicolumn{1}{|p{0.75in}}{\textbf{Path length 2}} &
\multicolumn{1}{|p{0.75in}}{\textbf{Path length 3}} &
\multicolumn{1}{|p{0.75in}|}{\textbf{Path length 4}} \\
\hline
1
&
2
&
20.00
&
---
&
---
&
12.00
&
15.00
\\
2
&
3
&
4.00
&
---
&
3.00
&
4.00
&
4.00
\\
2
&
4
&
15.63
&
---
&
4.00
&
8.83
&
11.30
\\
2
&
8
&
24.03
&
---
&
7.67
&
14.83
&
18.37
\\
3
&
4
&
16.00
&
---
&
6.00
&
8.00
&
11.00
\\
\rowcolor{gray!25}
4
&
5
&
43.97
&
4.00
&
13.33
&
21.17
&
29.63
\\
\rowcolor{gray!25}
5
&
6
&
36.47
&
4.00
&
14.17
&
19.33
&
26.80
\\
\rowcolor{gray!25}
5
&
8
&
26.80
&
4.00
&
8.50
&
13.83
&
18.47
\\
\rowcolor{gray!25}
5
&
9
&
49.63
&
4.00
&
16.00
&
23.00
&
31.63
\\
6
&
7
&
21.57
&
---
&
5.50
&
9.33
&
14.23
\\
6
&
11
&
56.37
&
---
&
9.00
&
19.00
&
34.37
\\
7
&
8
&
13.90
&
---
&
6.83
&
9.67
&
11.57
\\
9
&
10
&
51.63
&
---
&
9.00
&
17.00
&
28.63
\\
10
&
12
&
53.63
&
---
&
---
&
11.00
&
25.63
\\
11
&
20
&
58.37
&
---
&
---
&
13.00
&
31.37
\\
12
&
15
&
52.67
&
---
&
---
&
---
&
15.00
\\
12
&
19
&
9.63
&
---
&
---
&
---
&
6.63
\\
12
&
20
&
21.67
&
---
&
---
&
7.00
&
15.00
\\
13
&
14
&
8.67
&
---
&
---
&
---
&
---
\\
13
&
17
&
20.67
&
---
&
---
&
---
&
---
\\
14
&
16
&
33.33
&
---
&
---
&
---
&
---
\\
14
&
21
&
20.00
&
---
&
---
&
---
&
---
\\
15
&
16
&
43.67
&
---
&
---
&
---
&
---
\\
16
&
17
&
13.67
&
---
&
---
&
---
&
---
\\
17
&
18
&
37.33
&
---
&
---
&
---
&
---
\\
18
&
20
&
46.33
&
---
&
---
&
---
&
15.00
\\
19
&
20
&
10.37
&
---
&
---
&
---
&
8.37
\\
\MyHline
\end{tabular}

\MyCaption{Comparing the betweenness of edges based on the neighborhood discovered from a central vertex.}
{The size of the neighborhood increases from 1 to 4 based around vertex 5 \MyFigureReference{fig:neighbor}.
As the size of the neighborhood gets closer and closer to the global graph,  the betweenness values
get closer and closer to the global values.  Those edges that have not been discovered because they belong
to a portion of the global graph that has not  been discovered are marked with a ---.}
{tbl:neighborhood:edge}
\end{table}
\subsubsection{Vertex selection}

The selection of a vertex to remove from the graph is based on how much of the graph that the attacker has
discovered.  As the discovered graph becomes larger and larger (as measured by the path length from a initial/central)
vertex to the rest of the graph \MyFigureReference{fig:neighbor}, the more accurate the computed value betweenness value
of the vertex is to the vertex's betweenness value for the entire graph.  The betweenness value for all vertices in the
global graph and for the discovered subgraph is shown in Table \ref{tbl:neighborhood:vertex}.
In the table, the first column is the vertex number.  The second column is the
vertex's betweenness value based on the global graph.  The remaining columns show the vertex betweenness value as the path length from
the central vertex gets longer and longer.  In those cases where the discovered subgraph has not discovered a particular
vertex in the global graph, the vertex betweenness value is marked with a --- indicating no value possible.  It is interesting to
see how the value of an vertex changes as the size of the graph changes.  In most  cases the value of an vertex decreases as graph size
increases.  One notable exception is the vertex 2.  As the graph size increases, that vertex's
betweenness increase and decreases and yet in the global graph, its value is less than in some of the subgraphs.

\begin{table}
 \centering
\setlength{\extrarowheight}{3pt}
\begin{tabular} {|*{6}{c|}}
\MyHline
\multicolumn{1}{|p{0.75in}}{\textbf{Node}} &
\multicolumn{1}{|p{0.75in}}{\textbf{Vertex Betweenness}} &
\multicolumn{1}{|p{0.75in}}{\textbf{Path length 1}} &
\multicolumn{1}{|p{0.75in}}{\textbf{Path length 2}} &
\multicolumn{1}{|p{0.75in}}{\textbf{Path length 3}} &
\multicolumn{1}{|p{0.75in}|}{\textbf{Path length 4}} \\
\hline
1
&
0.00
&
---
&
---
&
0.00
&
0.00
\\
2
&
0.32
&
---
&
0.13
&
0.42
&
0.37
\\
3
&
0.00
&
---
&
0.00
&
0.00
&
0.00
\\
4
&
0.41
&
0.00
&
0.33
&
0.40
&
0.40
\\
\rowcolor{gray!25}
5
&
1.00
&
1.00
&
1.00
&
1.00
&
1.00
\\
6
&
0.69
&
0.00
&
0.46
&
0.55
&
0.66
\\
7
&
0.11
&
---
&
0.08
&
0.11
&
0.12
\\
8
&
0.33
&
0.00
&
0.33
&
0.40
&
0.36
\\
9
&
0.59
&
0.00
&
0.37
&
0.43
&
0.49
\\
10
&
0.62
&
---
&
0.00
&
0.24
&
0.43
\\
11
&
0.69
&
---
&
0.00
&
0.31
&
0.55
\\
12
&
0.86
&
---
&
---
&
0.09
&
0.52
\\
13
&
0.07
&
---
&
---
&
---
&
---
\\
14
&
0.31
&
---
&
---
&
---
&
---
\\
15
&
0.56
&
---
&
---
&
---
&
0.00
\\
16
&
0.52
&
---
&
---
&
---
&
---
\\
17
&
0.38
&
---
&
---
&
---
&
---
\\
18
&
0.47
&
---
&
---
&
---
&
0.00
\\
19
&
0.00
&
---
&
---
&
---
&
0.00
\\
20
&
0.85
&
---
&
---
&
0.12
&
0.60
\\
21
&
0.00
&
---
&
---
&
---
&
---
\\
\MyHline
\end{tabular}

\MyCaption{Comparing the betweenness of vertices based on the neighborhood discovered from a central vertex.}
{The size of the neighborhood increases from 1 to 4 based around  vertex 5 \MyFigureReference{fig:neighbor}.
As the size of the neighborhood get closer and closer to the global graph,  the betweenness values
get closer and closer to the global values.  Those vertices that have not been discovered because they belong
to a portion of the global graph that has not been discovered are marked with a ---.  The betweenness values have been
normalized to the range (0,1) to allow comparisons across different sized graphs.}
{tbl:neighborhood:vertex}
\end{table}
\subsubsection{Degree selection}
 Discovering the degree of a node is based on the idea that the nodes exchange messages between themselves and
that the attacker can intercept these messages.  As the attacker intercepts more and more messages; a node's neighbors
(a.k.a., degree) can be determined.  The degree of a node can be used as a criterion to determine if the node is worthy of
attack.

The degrees for the discovered graph based on differing path lengths is shown in Table \ref{tbl:neighborhood:degree}.  The first column
is the vertex number.  The second column is the vertex's global degree.  The remaining columns show the degree of the each
of the discovered vertices as the path length increases.  If the vertex has not been discovered based on a particular path length
 then the marker --- is used to indicate that no data is available.  It is interesting to note that the degree of a vertex
always increases as the path length increases until the global degree value is reached.  Once the global value is reached, it remains constant.
\begin{table}
 \centering
\setlength{\extrarowheight}{3pt}
\begin{tabular} {|*{6}{c|}}
\MyHline
\multicolumn{1}{|c}{\textbf{Vertex}} &
\multicolumn{1}{|c}{\textbf{Degree}} &
\multicolumn{1}{|c}{\textbf{Path length 1}} &
\multicolumn{1}{|c}{\textbf{Path length 2}} &
\multicolumn{1}{|c}{\textbf{Path length 3}} &
\multicolumn{1}{|c|}{\textbf{Path length 4}} \\
\hline
1
&
1
&
---
&
---
&
1
&
1
\\
2
&
4
&
---
&
3
&
4
&
4
\\
3
&
2
&
---
&
2
&
2
&
2
\\
4
&
3
&
1
&
3
&
3
&
3
\\
\rowcolor{gray!25}
5
&
4
&
4
&
4
&
4
&
4
\\
6
&
3
&
1
&
3
&
3
&
3
\\
7
&
2
&
---
&
2
&
2
&
2
\\
8
&
3
&
1
&
3
&
3
&
3
\\
9
&
2
&
1
&
2
&
2
&
2
\\
10
&
2
&
---
&
1
&
2
&
2
\\
11
&
2
&
---
&
1
&
2
&
2
\\
12
&
4
&
---
&
---
&
2
&
4
\\
13
&
2
&
---
&
---
&
---
&
---
\\
14
&
3
&
---
&
---
&
---
&
---
\\
15
&
2
&
---
&
---
&
---
&
1
\\
16
&
3
&
---
&
---
&
---
&
---
\\
17
&
3
&
---
&
---
&
---
&
---
\\
18
&
2
&
---
&
---
&
---
&
1
\\
19
&
2
&
---
&
---
&
---
&
2
\\
20
&
4
&
---
&
---
&
2
&
4
\\
21
&
1
&
---
&
---
&
---
&
---
\\
\MyHline
\end{tabular}

\MyCaption{Comparing the degreeness of each vertex based on the neighborhood discovered from a central vertex.}
{The size of the neighborhood increases from 1 to 4 based around  vertex 5 \MyFigureReference{fig:neighbor}.
As the size of the neighborhood get closer and closer to the global graph,  the betweenness values
get closer and closer to the global values.}
{tbl:neighborhood:degree}
\end{table}
\subsection{Attack Profile Notation}
An attacker can target any graph component for removal based on the damage estimate or other criteria and
whether to use the highest, or lowest valued component based on those criteria.  We introduce
the notation \MyAttackNotation{C}{V} as a short hand way to identify a specific profile.  
The first subscript in \MyAttackNotation{C}{V} is the metric that is being used to select
a component $C\in\{E,V,D,*\}$ for \textit{edge, vertex, degree or any} respectively.
The second subscript is the value of the metric that is being used
$V\in\{L,M,H,R,*\}$ for \textit{low, medium, high, random or any} respectively.
The notation
\MyAttackNotation{D}{H} means that the attacker is using a profile that targets nodes based on their
degree $D$ and choose the highest $H$ valued one.
\subsection{Effectiveness of different attack profiles}
The damage to a graph by fragmentation can be calculated \MyEquationReference{equ:damage} using the fragmented graph and
approximating the graph without fragmentation.
\MyEquationLabeled{damage-02.tex}{equ:damage}
An unfragmented graph is created from the fragmented graph by adding an edge between
each of the highest degreed nodes of each fragment.  As each edge is added to coalesce the fragments into a larger and larger connected
component, the highest degreed node may change based on the order in which the fragments are coalesced.  Therefore the highest
degreed node in the coalescing component must be evaluated after each fragment addition.  At the end of the collation process,
there will be a single connected component containing the same number of nodes as the fragmented graph and one additional edge for
each of the original fragments.

As the original graph becomes more and more fragmented, its AIPL will decrease.  The AIPL of the unfragmented approximation will decrease and
the \MyEquationInline{damageLHS} will increase as well.  This behavior is readily apparent when edges are removed from the
original graph in order to create the fragments.  When vertices are removed, the behavior is similar, until the last vertex is removed.  
In the limiting case, 
AIPL of the fragmented graph with one fragment and one node in that  fragment, is the same as the AIPL of a connected component with one node.
Using Equation \ref{equ:damage} results in a value of 0 meaning that the graph is undamaged.
\subsubsection{Edge selection}
The attacker can compute the betweenness of any edge in the subgraph that he has discovered \MyTableReference{tbl:neighborhood:edge}.  Based on these
computed betweenness values, the attacker can select either the highest or lowest valued edge to remove.  After the removal
of this edge, the betweenness values can be recomputed for the newly modified subgraph and the process repeated again and again
until there are no edges left in the discovered graph (the discovered graph is totally destroyed).

Figures \ref{fig:delete:edge:low} and \ref{fig:delete:edge:high} show the effects of repeatedly applying attack \MyAttackNotation{E}{L} or
\MyAttackNotation{V}{L} profile to the discovered subgraph of path length 3.  In each figure, the betweenness value of each edge is written on the edge.  The
edge with the lowest \MyFigureReference{fig:delete:edge:low} or highest \MyFigureReference{fig:delete:edge:high} betweenness value
is highlighted in red, prior to it being removed.  After the removal of the edge, the betweenness values of all the
remaining edges is computed shown in the next subfigure, along with the next edge that has been selected for removal.  The four subfigures in Figures \ref{fig:delete:edge:low} and \ref{fig:delete:edge:high} show this process.  When two or more edges have the same betweenness value, the
selection of which edge to remove it totally random.

Attack profile \MyAttackNotation{E}{L}  tends to attack the periphery of the graph.  While profile \MyAttackNotation{E}{H} tends
to attack the core of the graph.  Either profile  will result in a fully disconnected graph with the same number of
removals, selecting the highest valued edge causes more damage quicker.

\begin{figure}
\centering
\MyLocalSubfigure{First lowest has been identified}{fig:delete:edge:low:0}{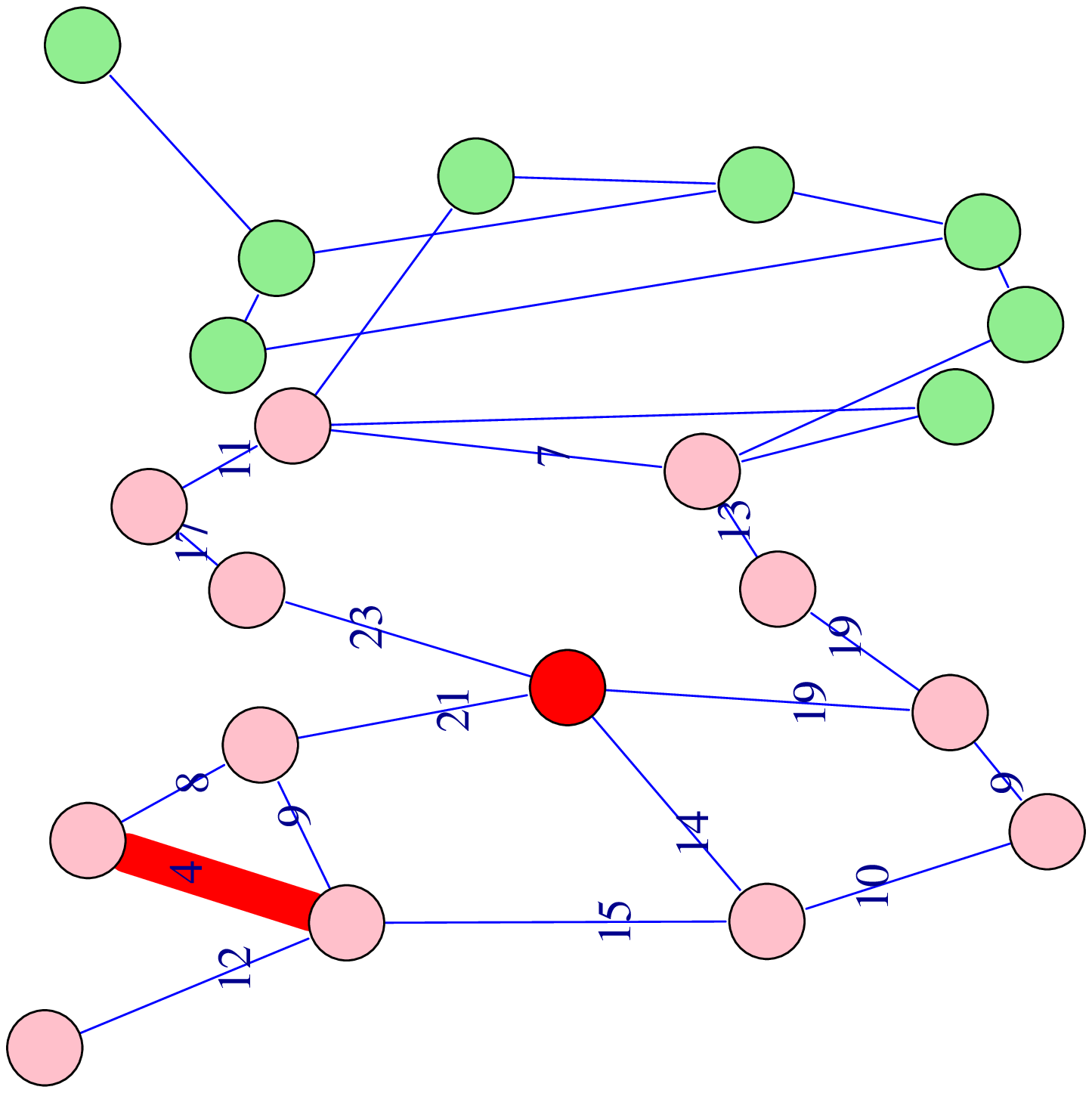}
\MyLocalSubfigure{Previous lowest has been removed, new lowest identified}{fig:delete:edge:low:1}{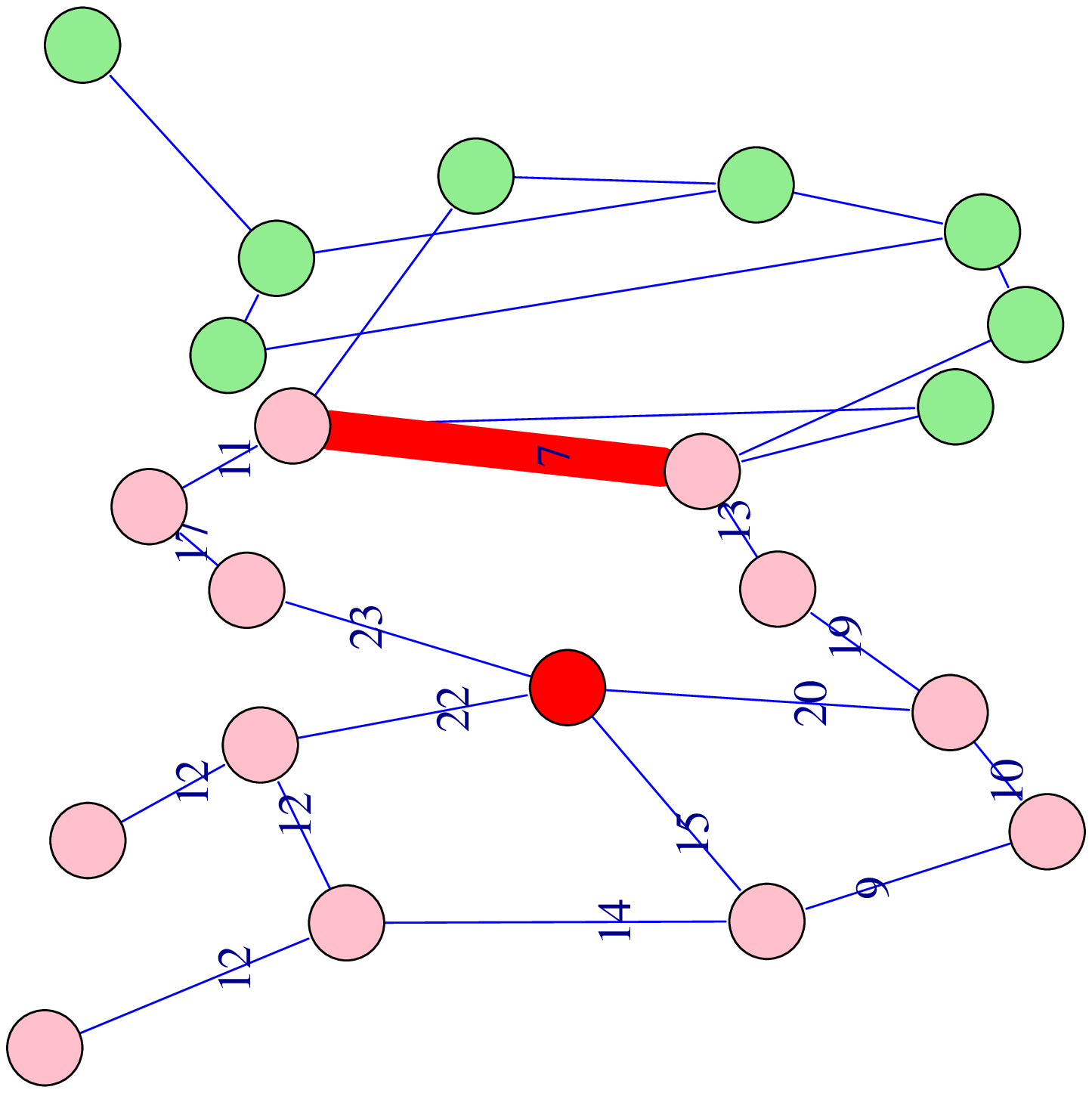}
\MyLocalSubfigure{Previous lowest has been removed, new lowest identified}{fig:delete:edge:low:2}{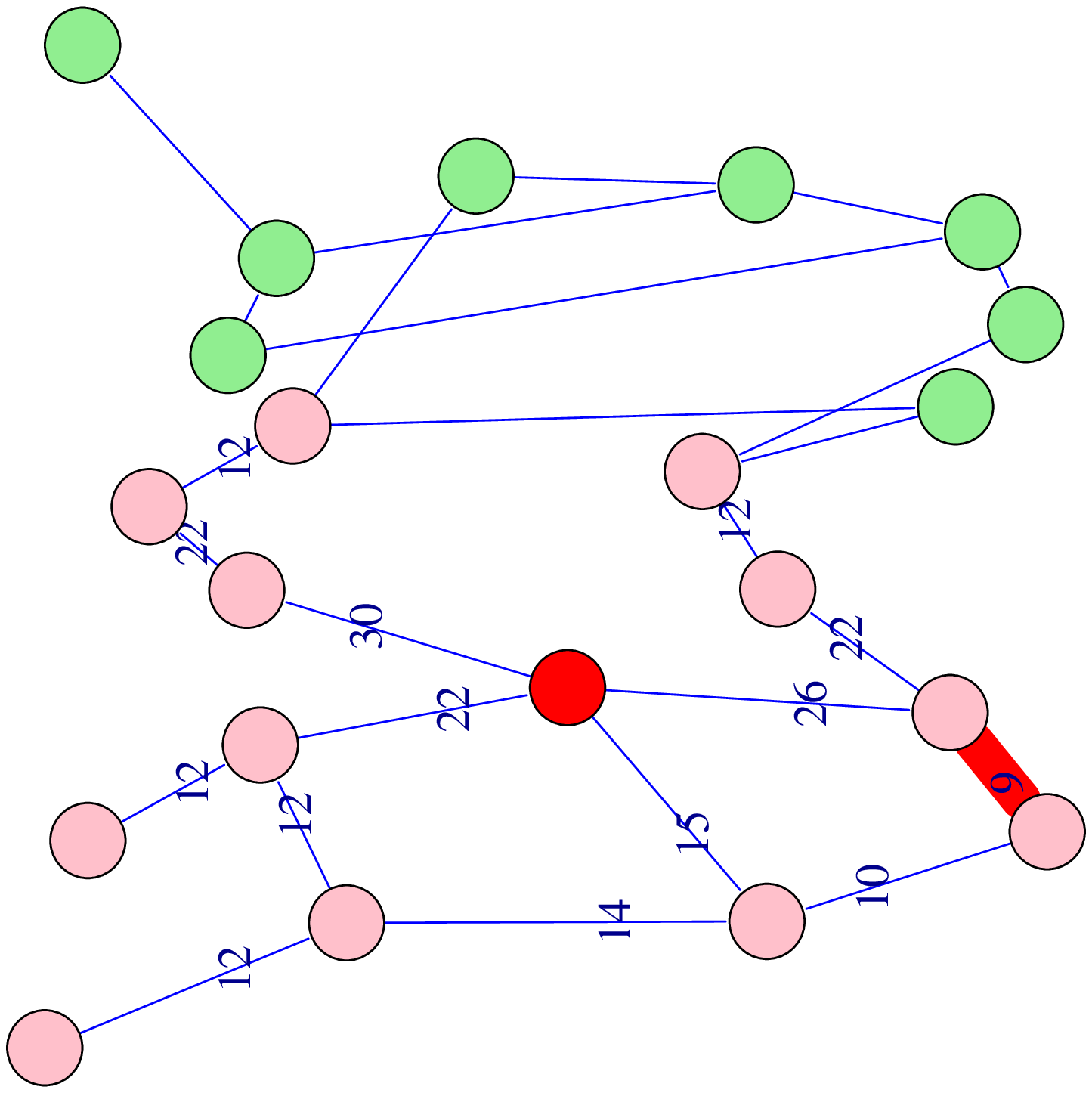}
\MyLocalSubfigure{Previous lowest has been removed, new lowest identified}{fig:delete:edge:low:3}{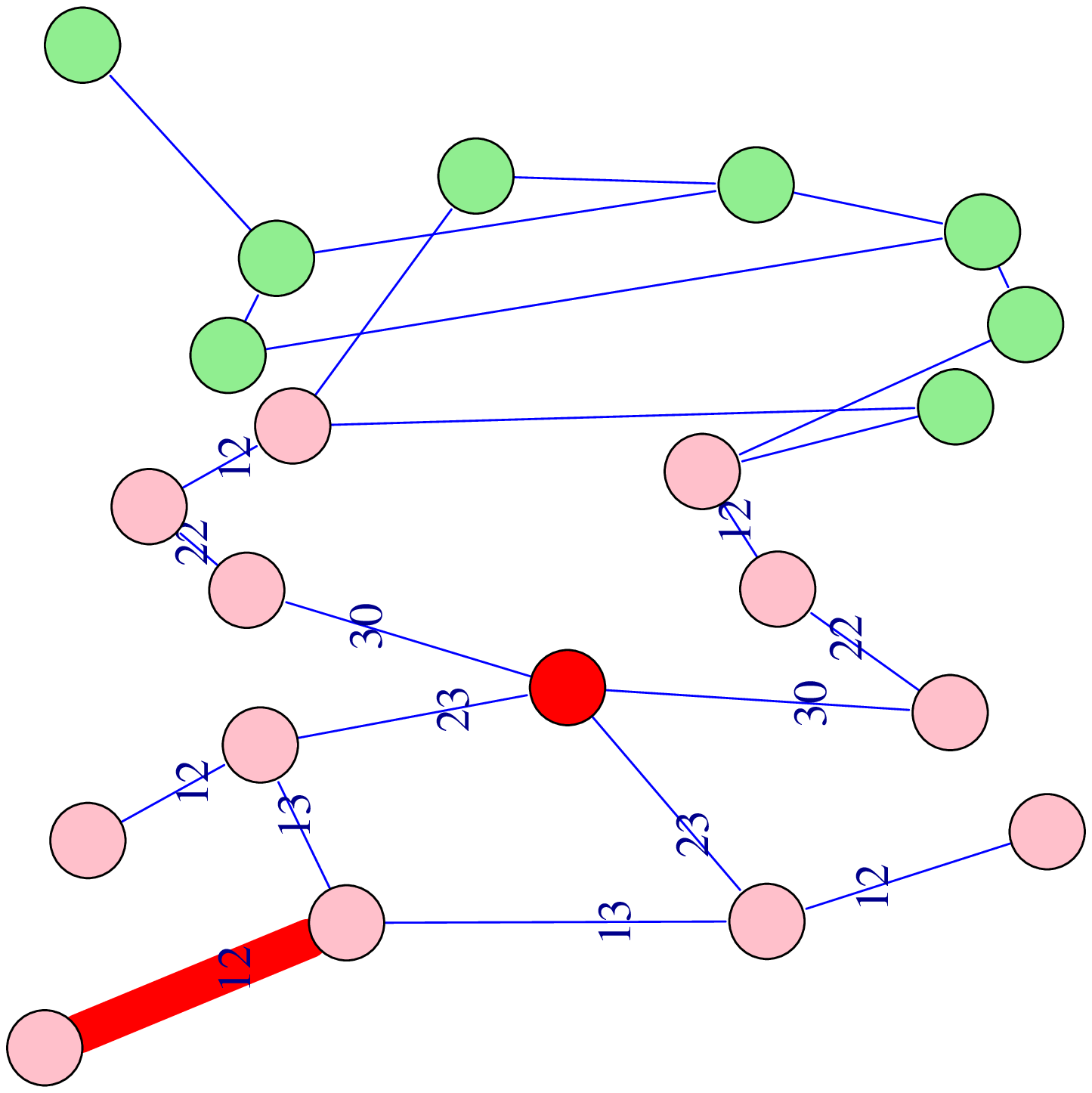}
\MyCaption{The effects of the \MyAttackNotation{E}{L} attack profile on the sample graph.}
{Vertex 5 is the center vertex and is marked in red.  The discovered graph is at a path length of 3 from the center vertex 
and is marked in pink.
  The edge with the lowest
betweenness value is marked in red.  After each deletion, all edge betweenness values are recomputed because the
graph has changed. Some of the edges are unlabeled because the attacker has not ``discovered'' them.}
{fig:delete:edge:low}
\end{figure}
\begin{figure}
\centering
\MyLocalSubfigure{First highest has been identified}{fig:delete:edge:high:0}{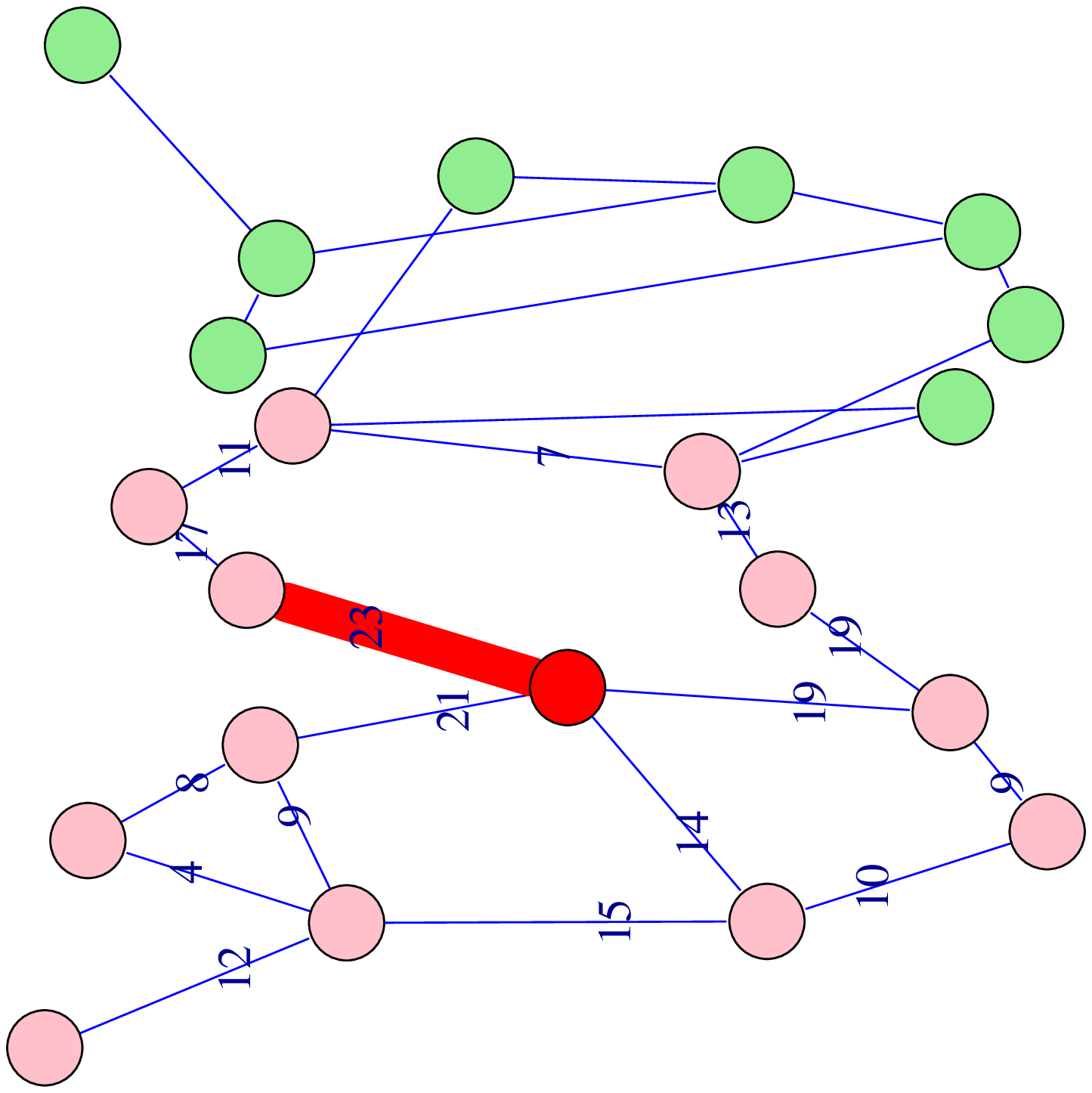}
\MyLocalSubfigure{Previous highest has been removed, new highest identified}{fig:delete:edge:high:1}{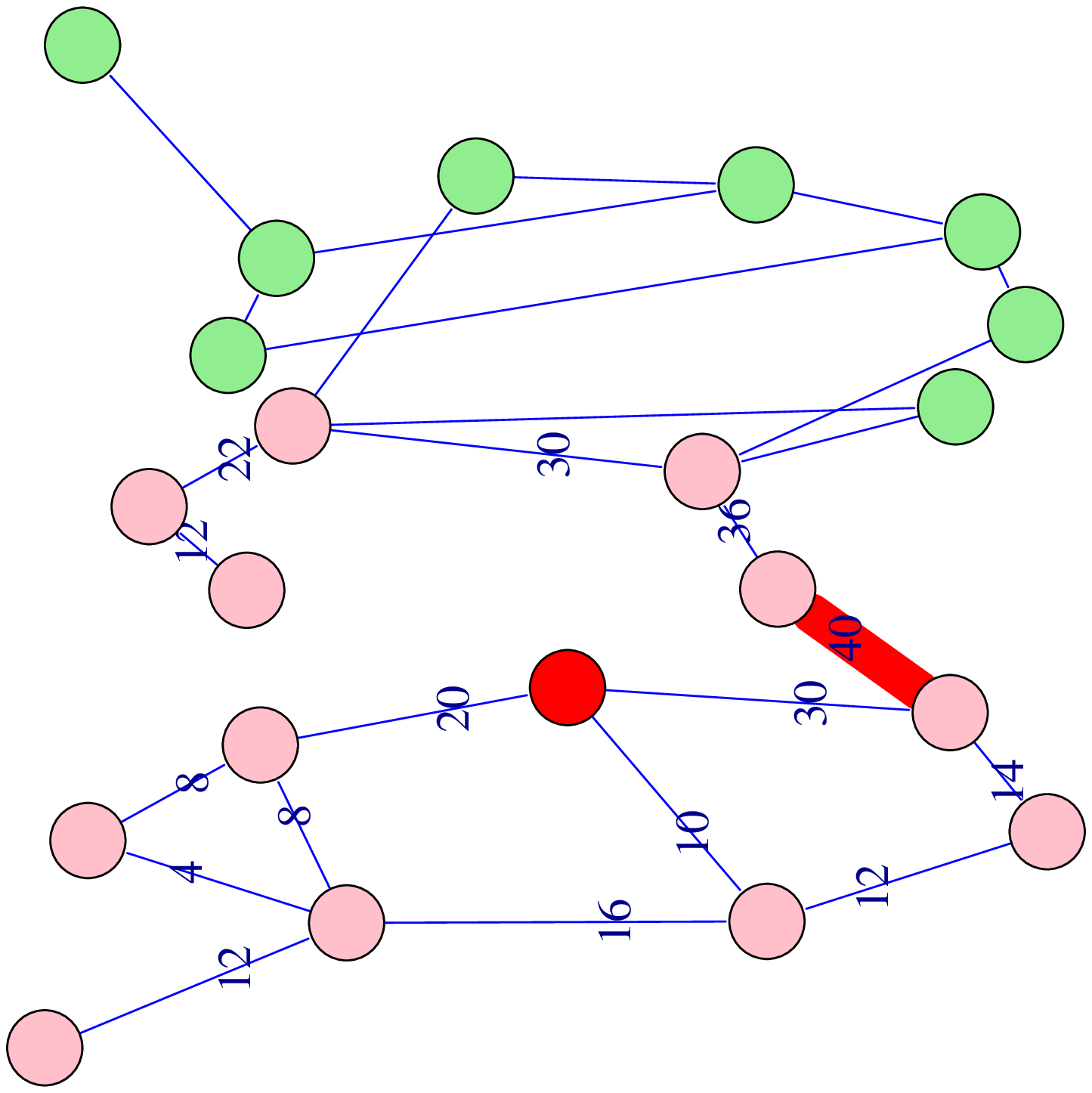}
\MyLocalSubfigure{Previous highest has been removed, new highest identified}{fig:delete:edge:high:2}{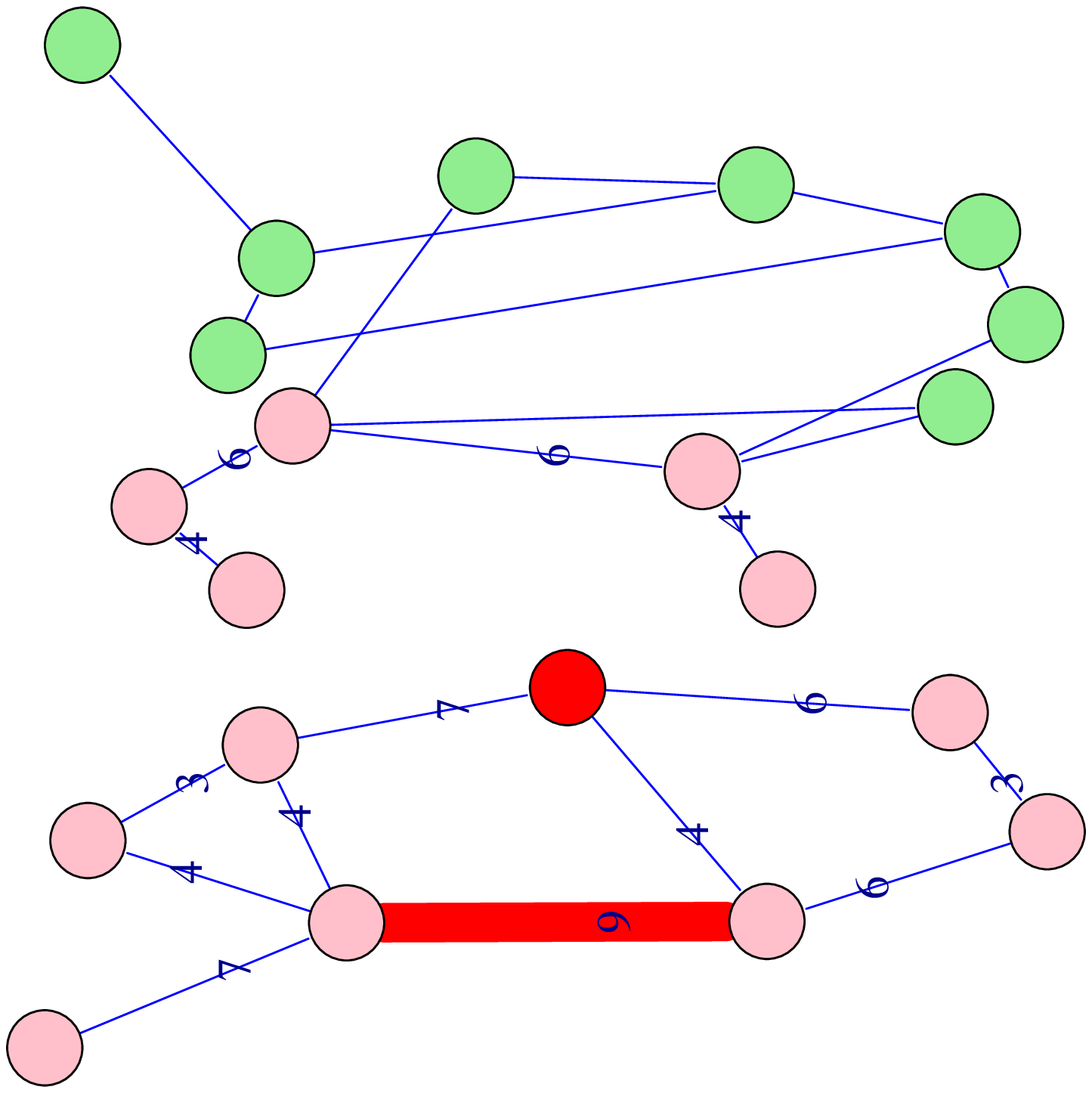}
\MyLocalSubfigure{Previous highest has been removed, new highest identified}{fig:delete:edge:high:3}{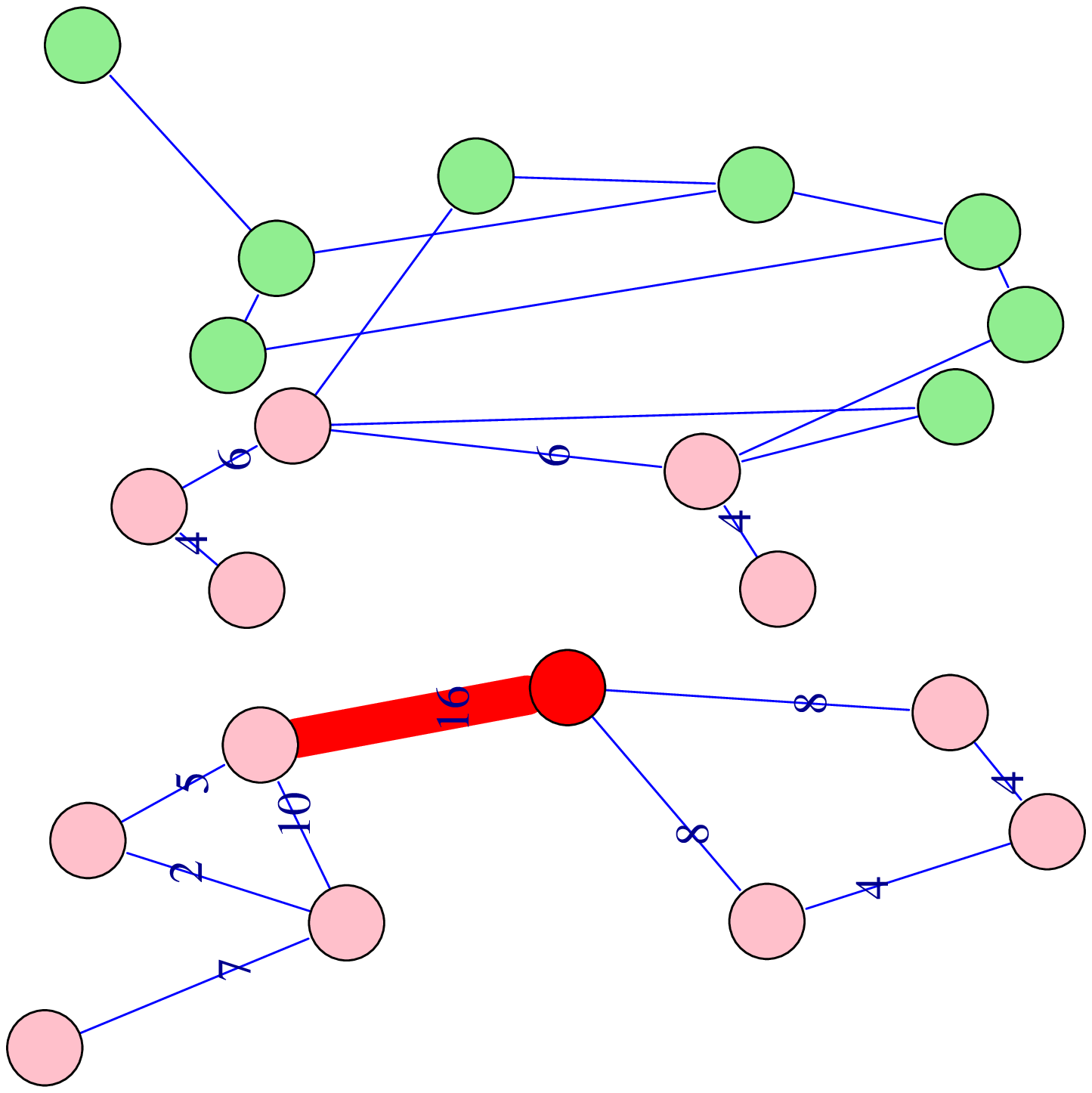}
\MyCaption{The effects of the \MyAttackNotation{E}{H} attack profile on the sample graph.}
{Vertex 5 is the center vertex and is marked in red.  The discovered graph is at a path length of 3 from the center vertex
and is marked in pink.  The edge with the highest
betweenness value is marked in red.  After each deletion, all edge betweenness values are recomputed because the
graph has changed.  Some of the edges are unlabeled because the attacker has not ``discovered'' them.}
{fig:delete:edge:high}
\end{figure}

Table \ref{tbl:damage:edge} lists the computed damage to the discovered subgraph after the removal of either the highest or lowest
betweenness valued edge.  Figure \ref{fig:damage:edge} shows the damage plotted against the deletion.
 There are 16 edges in the discovered subgraph and damage is total upon the removal of the last edge.
\begin{table}
 \centering
\setlength{\extrarowheight}{3pt}
\begin{tabular} {|*{5}{c|}}
\MyHline
\multicolumn{1}{|p{0.75in}}{\textbf{Deletion}} &
\multicolumn{1}{|p{0.75in}}{\textbf{Local damage due to \MyAttackNotation{E}{H}}} &
\multicolumn{1}{|p{0.75in}}{\textbf{Global damage due to local damage by \MyAttackNotation{E}{H}}} &
\multicolumn{1}{|p{0.75in}}{\textbf{Local damage due to \MyAttackNotation{E}{L}}} &
\multicolumn{1}{|p{0.75in}|}{\textbf{Global damage due to local damage by \MyAttackNotation{E}{L}}} \\
\hline
0
&
0.00
&
0.00
&
0.00
&
0.00
\\
1
&
0.10
&
0.06
&
0.02
&
0.01
\\
2
&
0.36
&
0.31
&
0.07
&
0.03
\\
3
&
0.41
&
0.33
&
0.10
&
0.05
\\
4
&
0.57
&
0.41
&
0.21
&
0.12
\\
5
&
0.65
&
0.50
&
0.23
&
0.13
\\
6
&
0.70
&
0.53
&
0.34
&
0.19
\\
7
&
0.72
&
0.54
&
0.44
&
0.26
\\
8
&
0.78
&
0.57
&
0.54
&
0.32
\\
9
&
0.82
&
0.62
&
0.62
&
0.38
\\
10
&
0.83
&
0.62
&
0.71
&
0.43
\\
11
&
0.87
&
0.64
&
0.77
&
0.48
\\
12
&
0.89
&
0.65
&
0.83
&
0.52
\\
13
&
0.92
&
0.67
&
0.89
&
0.57
\\
14
&
0.95
&
0.68
&
0.93
&
0.61
\\
15
&
0.97
&
0.69
&
0.97
&
0.65
\\
16
&
1.00
&
0.70
&
1.00
&
0.70
\\
\MyHline
\end{tabular}

\MyCaption{Damage to the discovered subgraph of path length 3 based on  \MyAttackNotation{E}{*} attack profiles.}
{The betweenness of each edge is recomputed after the
removal of either the highest or lowest betweenness valued edge.  The process is repeated again and again until all edges
are removed.}
{tbl:damage:edge}
\end{table}
\begin{figure}
 \centering
\includegraphics[width = 4.5in, angle = -90]{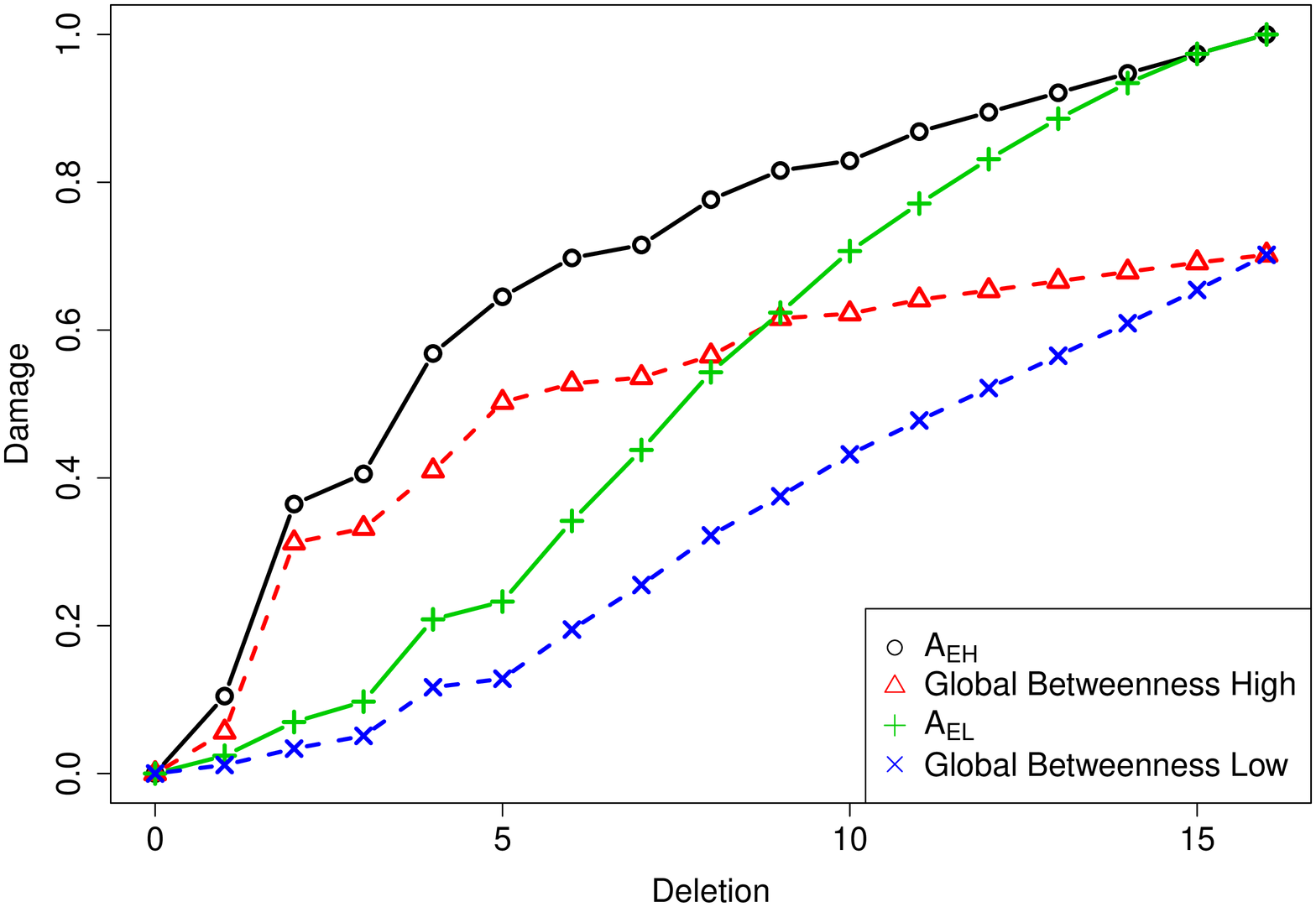}
\MyCaption{Damage to the discovered graph of path length 3 based on \MyAttackNotation{E}{*} attack profiles.}{
The ``local'' values are those that come from the discovered graph, while the 
global values are from the total graph.  Damage inflicted on the discovered graph when 
using the high edge betweenness value and the resulting impact on the total
graph are show in black and red respectively.  In a similar manner, damage caused by choosing the low 
betweenness is shown in the green and blue lines respectively.  The betweenness of each edge is recomputed after the
removal of either the highest or lowest betweenness valued edge.  The process is repeated again and again until all edges
are removed.}
{fig:damage:edge}
\end{figure}

\subsubsection{Vertex selection}
The attacker can compute the betweenness of any vertex in the subgraph that he has discovered \MyTableReference{tbl:neighborhood:vertex}.  Based on these
computed betweenness values, the attacker can select either the highest or lowest valued vertex to remove.  After the removal
of this vertex, the betweenness values can be recomputed for the newly modified subgraph and the process repeated again and again
until there are no vertices left in the discovered graph (the discovered graph is totally destroyed).

Figures \ref{fig:delete:vertex:low} and \ref{fig:delete:vertex:high} show the effects of repeatedly applying 
\MyAttackNotation{V}{L} or \MyAttackNotation{V}{L} profile to the
 discovered subgraph of path length 3.  In each figure, the betweenness value of each vertex is written in the vertex.  The
vertex with the lowest \MyFigureReference{fig:delete:vertex:low} or highest \MyFigureReference{fig:delete:vertex:high} betweenness value
is highlighted in yellow, prior to it being removed.  After the removal of the vertex, the betweenness values of all the
remaining vertices are computed and shown in the next subfigure, along with the next vertex that has been selected for removal.  
The four subfigures in Figures \ref{fig:delete:vertex:low} and \ref{fig:delete:vertex:high} show this process.  When two or more vertices have the same betweenness value, the
selection of which edge to remove it totally random.

Attack profile \MyAttackNotation{V}{L} tends to attack the periphery of the subgraph.  While attack 
profile \MyAttackNotation{V}{H} tends
to attack the core of the graph.  While both selection choices will result in a fully disconnected graph with the same number of
removals, selecting the highest valued vertex causes more damage quicker.

The betweenness computation, removal and damage computation process is shown in Table \ref{tbl:damage:vertex} and Figure \ref{fig:damage:vertex}.
The global high line in Figure \ref{fig:damage:vertex} goes flat after the fifth deletion 
while the global low line continues to increase.  This behavior is explained by looking
at Figures \ref{fig:delete:vertex:high:4} and \ref{fig:delete:vertex:low:4}.  By the fifth
high deletion, the discovered and global graphs are disconnected and further local
deletions do not affect the global graph.  In Figure \ref{fig:delete:vertex:high:4}, 
the discovered and global graphs are still connected and local deletions will affect the 
global graph.

\begin{figure}
\centering
\MyLocalSubfigure{First lowest has been identified}{fig:vertex:edge:low:0}{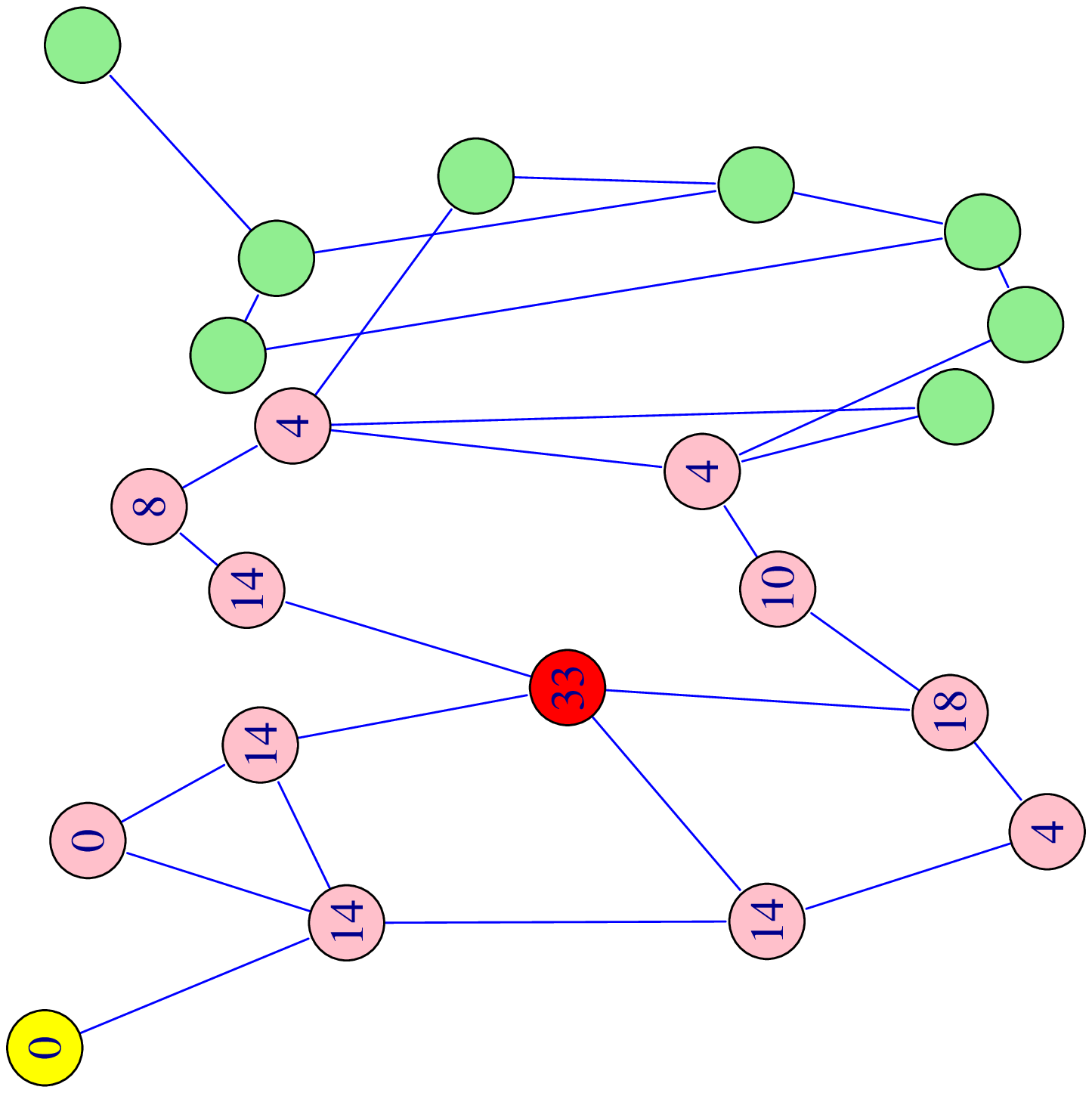}
\MyLocalSubfigure{Previous lowest has been removed, new lowest identified}{fig:vertex:edge:low:1}{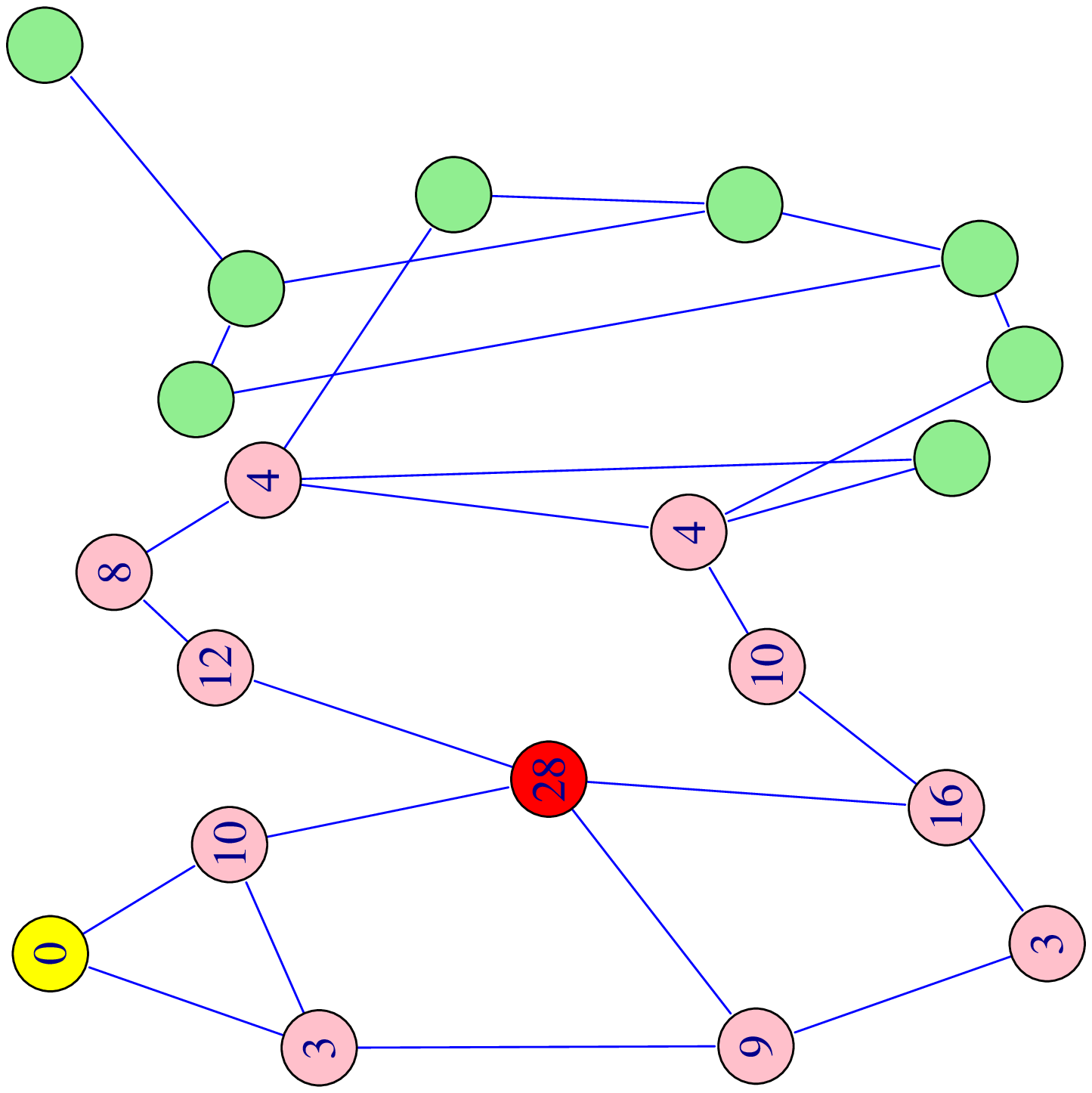}
\MyLocalSubfigure{Previous lowest has been removed, new lowest identified}{fig:vertex:edge:low:2}{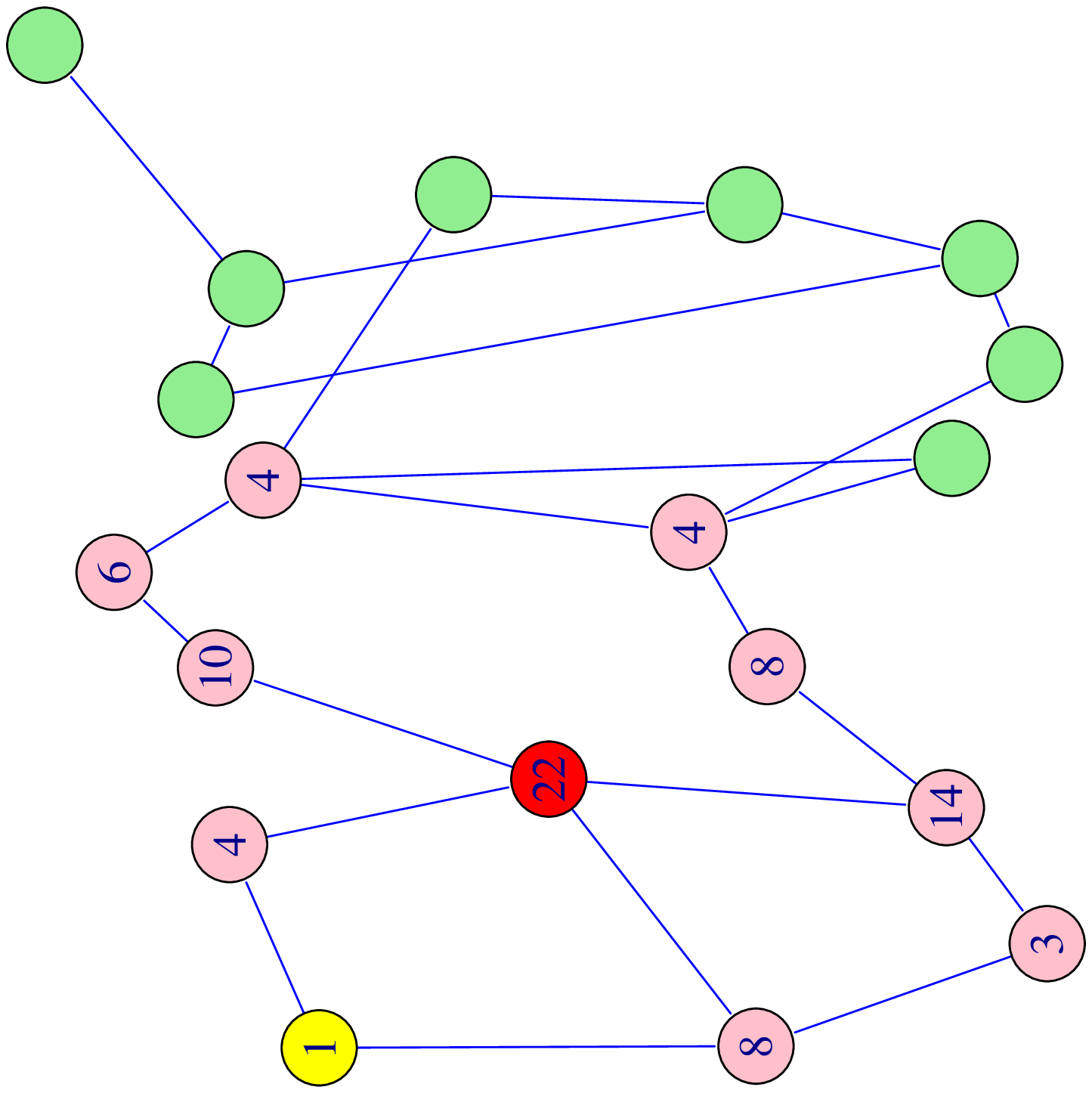}
\MyLocalSubfigure{Previous lowest has been removed, new lowest identified}{fig:vertex:edge:low:3}{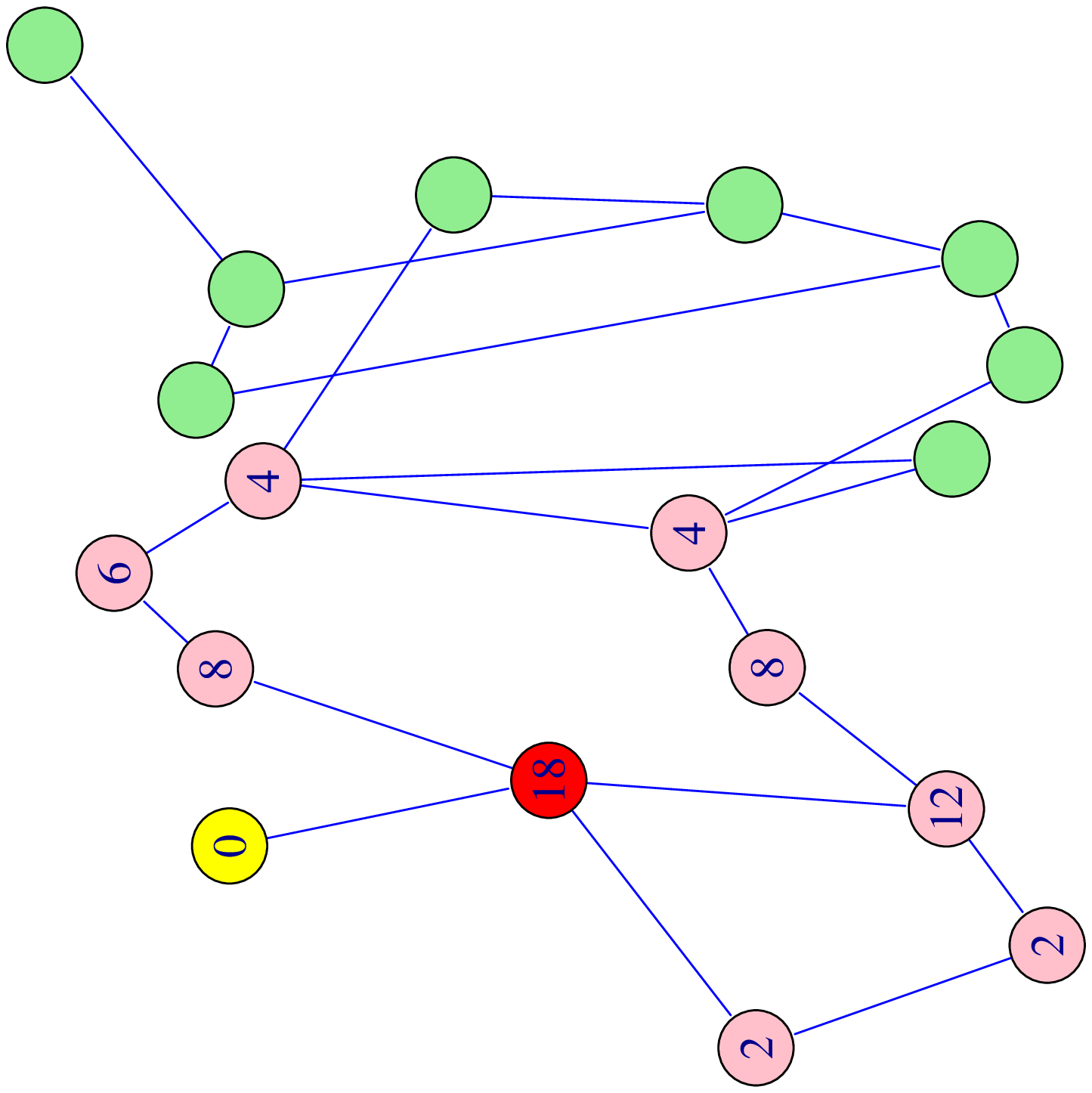}
\MyCaption{The effects of an \MyAttackNotation{V}{L} attack profile on the sample graph.}
{Vertex 5 is the center vertex and is shown in red.  The discovered graph, in pink is at a distance of 3 from the center vertex.
Each vertex is labeled with the number of shortest paths that go use that vertex.
The vertex with the lowest betweenness is drawn in yellow.  Each time, the lowest valued vertex is removed from the
discovered graph and all betweenness values for the discovered graph are recomputed.  If there is more
than one vertex with the same low value, one is selected at random for removal.
 Some of the vertices are unlabeled because the attacker has not ``discovered'' them.}
{fig:delete:vertex:low}
\end{figure}
\begin{figure}
\centering
\MyLocalSubfigure{First highest has been identified}{fig:delete:vertex:high:0}{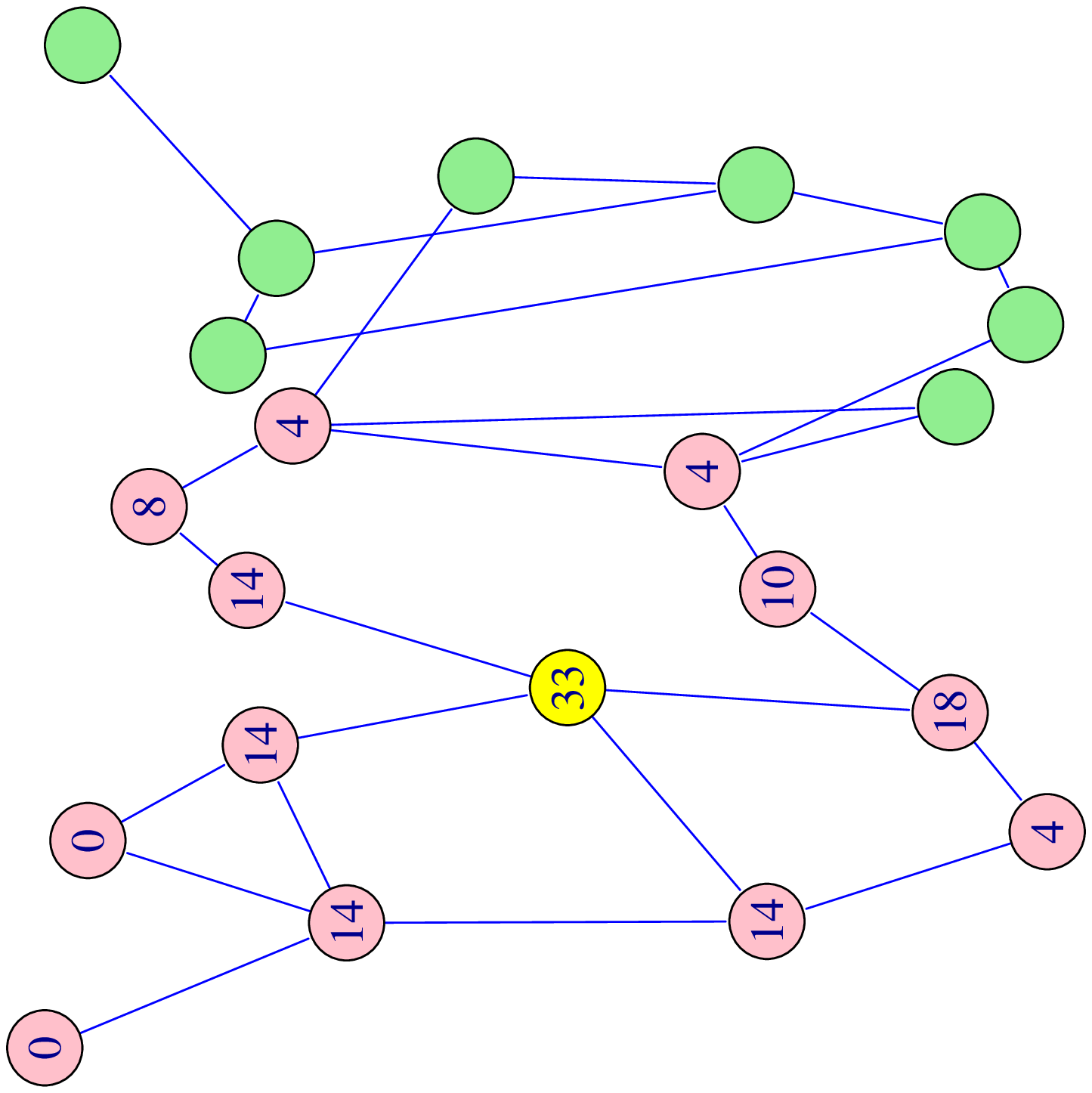}
\MyLocalSubfigure{Previous highest has been removed, new highest identified}{fig:delete:vertex:high:1}{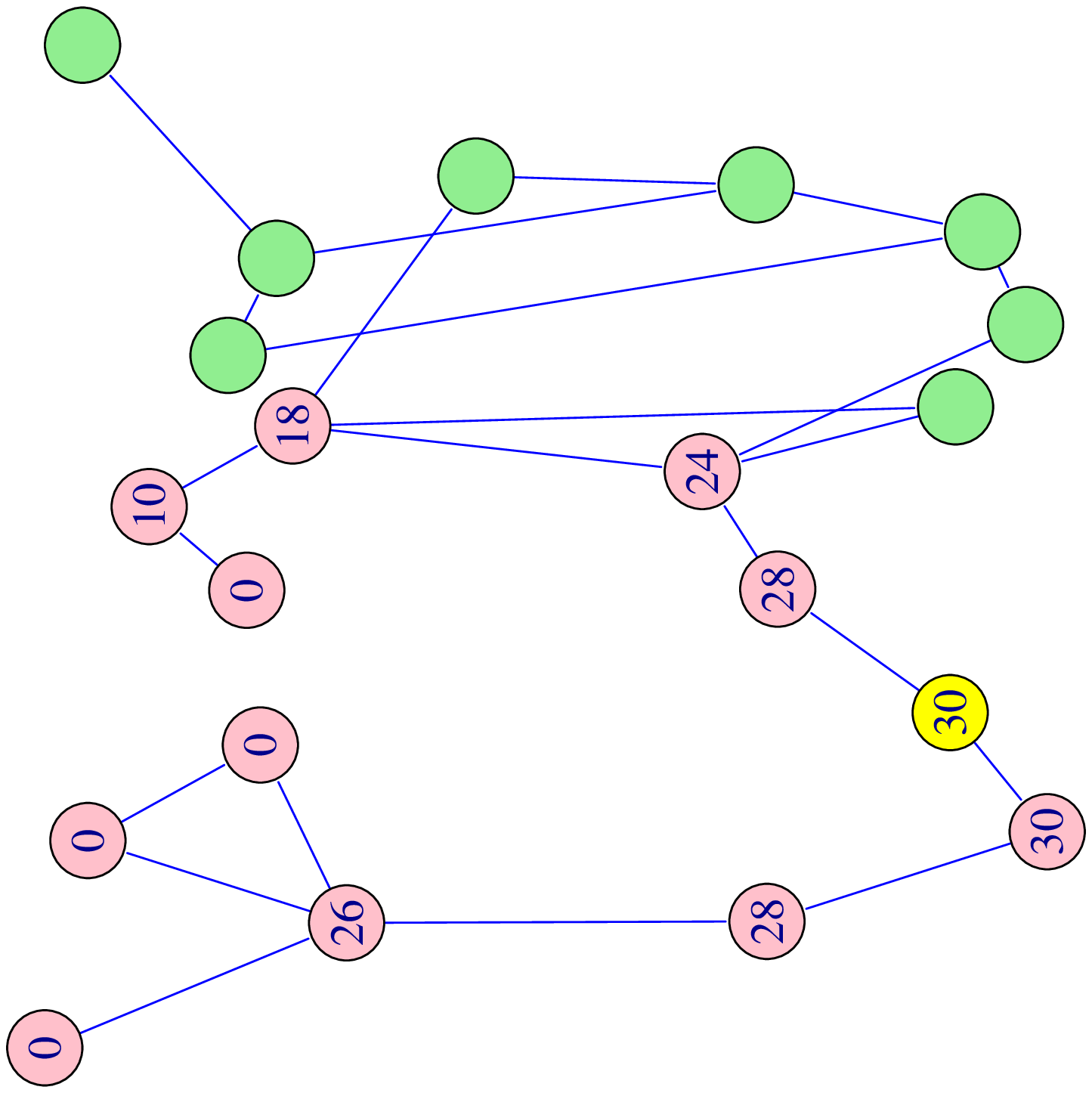}
\MyLocalSubfigure{Previous highest has been removed, new highest identified}{fig:delete:vertex:high:2}{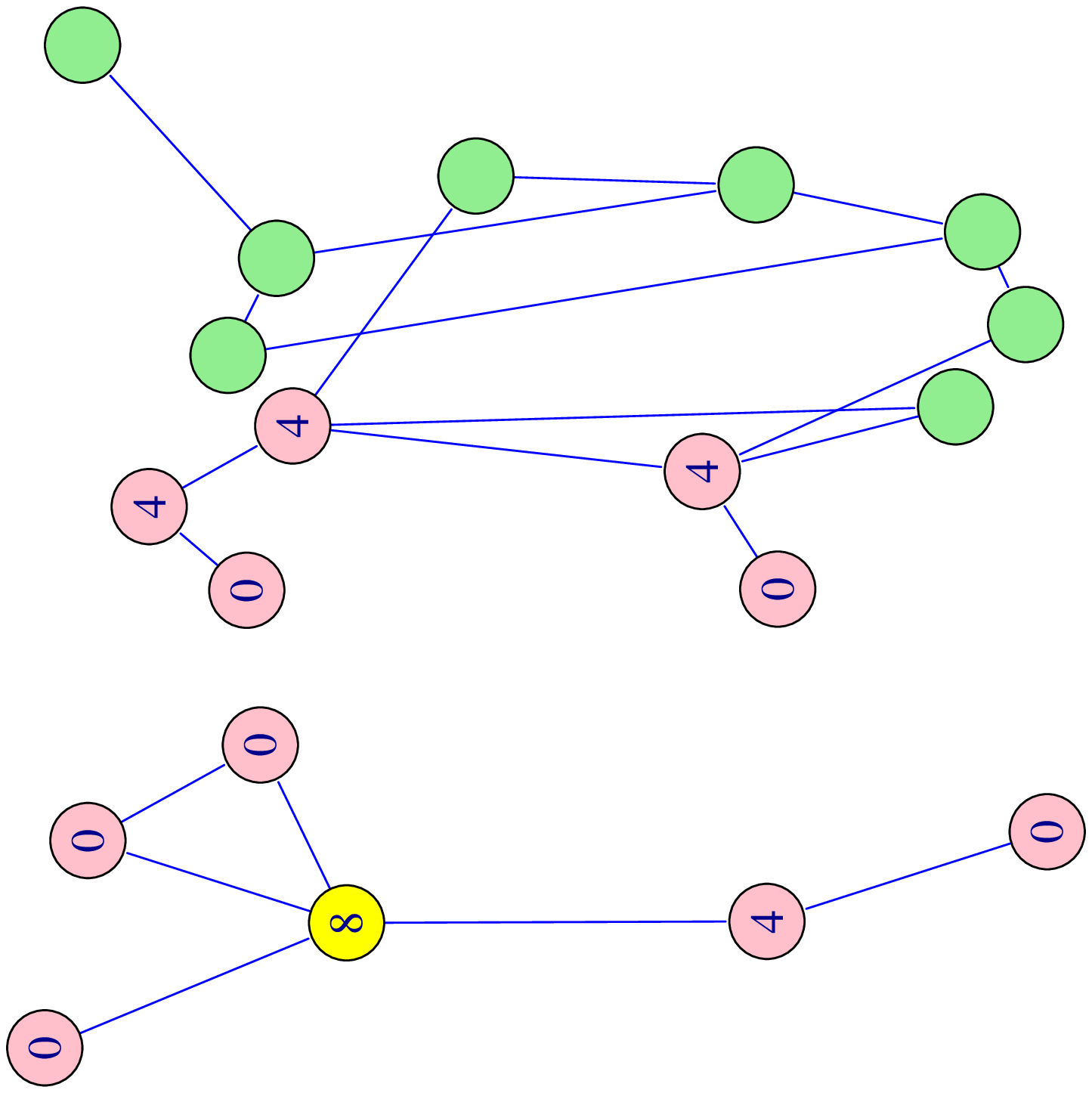}
\MyLocalSubfigure{Previous highest has been removed, new highest identified}{fig:delete:vertex:high:3}{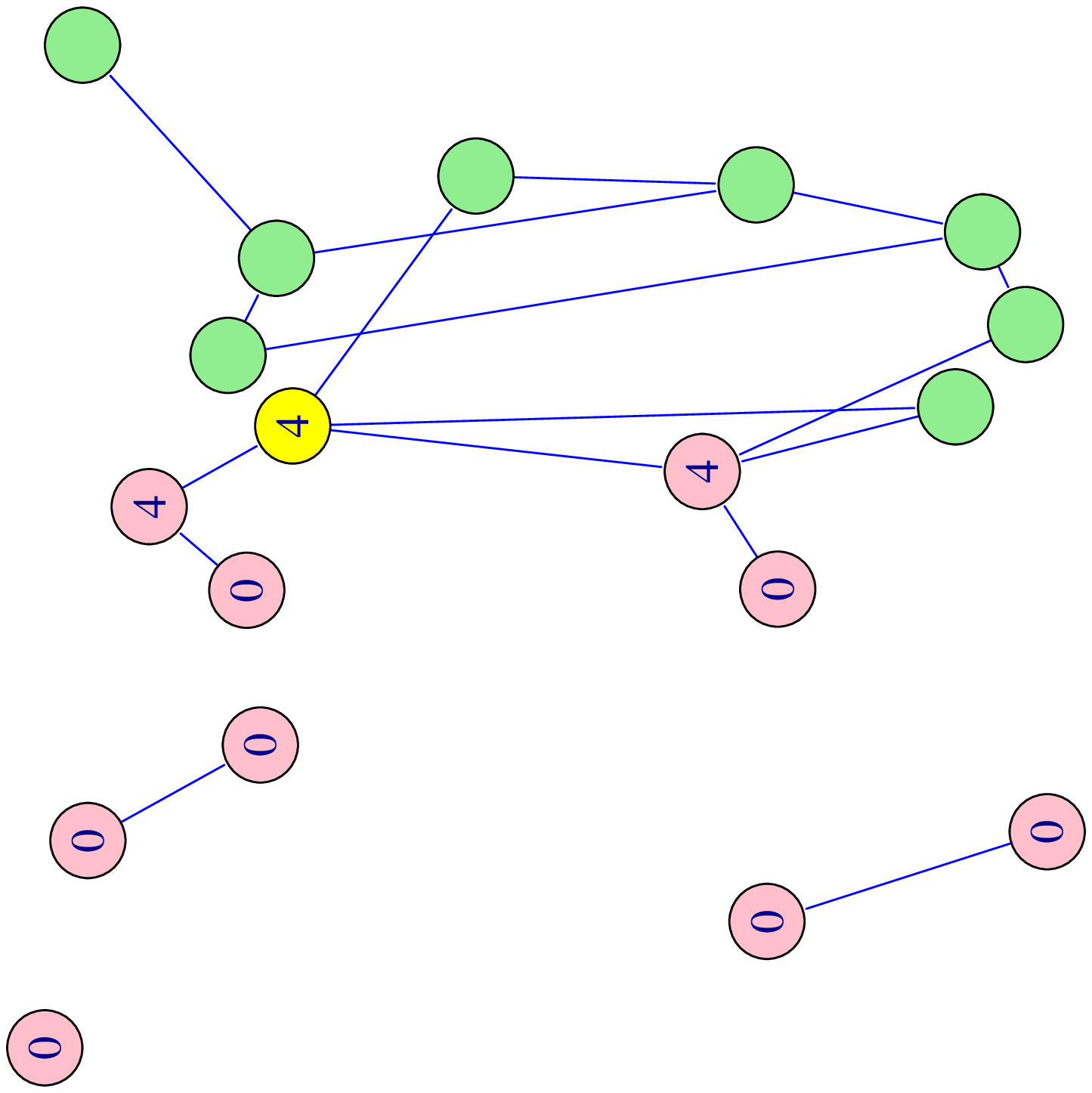}
\MyCaption{The effects of an \MyAttackNotation{V}{H} attack profile on the sample graph.}
{Vertex 5 is the center vertex.  The discovered graph is at a distance of 3 from the center vertex.
The vertex with the highest betweenness is drawn in yellow.  Each time, the highest valued vertex is removed from the
discovered graph and all betweenness values for the discovered graph are recomputed.  If there is more
than one vertex with the same high value, one is selected at random for removal.
 Some of the vertices are unlabeled because the attacker has not ``discovered'' them.}
{fig:delete:vertex:high}
\end{figure}
\begin{table}
 \centering
\setlength{\extrarowheight}{3pt}
\begin{tabular} {|*{5}{c|}}
\MyHline
\multicolumn{1}{|p{0.75in}}{\textbf{Deletion}} &
\multicolumn{1}{|p{0.75in}}{\textbf{Local damage due to \MyAttackNotation{V}{H}}} &
\multicolumn{1}{|p{0.75in}}{\textbf{Global damage due to local damage by \MyAttackNotation{V}{H}}} &
\multicolumn{1}{|p{0.75in}}{\textbf{Local damage due to \MyAttackNotation{V}{L}}} &
\multicolumn{1}{|p{0.75in}|}{\textbf{Global damage due to local damage by \MyAttackNotation{V}{L}}} \\
\hline
0
&
0.00
&
0.00
&
0.00
&
0.00
\\
1
&
0.29
&
0.17
&
0.12
&
0.07
\\
2
&
0.57
&
0.41
&
0.24
&
0.14
\\
3
&
0.78
&
0.51
&
0.36
&
0.22
\\
4
&
0.89
&
0.68
&
0.47
&
0.28
\\
5
&
0.89
&
0.68
&
0.58
&
0.35
\\
6
&
0.92
&
0.70
&
0.66
&
0.41
\\
7
&
0.92
&
0.70
&
0.77
&
0.48
\\
8
&
0.95
&
0.71
&
0.83
&
0.54
\\
9
&
0.95
&
0.71
&
0.89
&
0.59
\\
10
&
0.97
&
0.72
&
0.93
&
0.64
\\
11
&
0.97
&
0.72
&
0.97
&
0.69
\\
12
&
1.00
&
0.77
&
1.00
&
0.77
\\
\MyHline
\end{tabular}

\MyCaption{Damage to the discovered subgraph of path length 3 based on \MyAttackNotation{V}{*} attack profiles.}
{The betweenness of each vertex is recomputed after the
removal of either the highest or lowest betweenness valued vertex.  The process is repeated again and again until all vertices
are removed.}
{tbl:damage:vertex}
\end{table}
\begin{figure}
 \centering
\includegraphics[width = 4.5in, angle = -90]{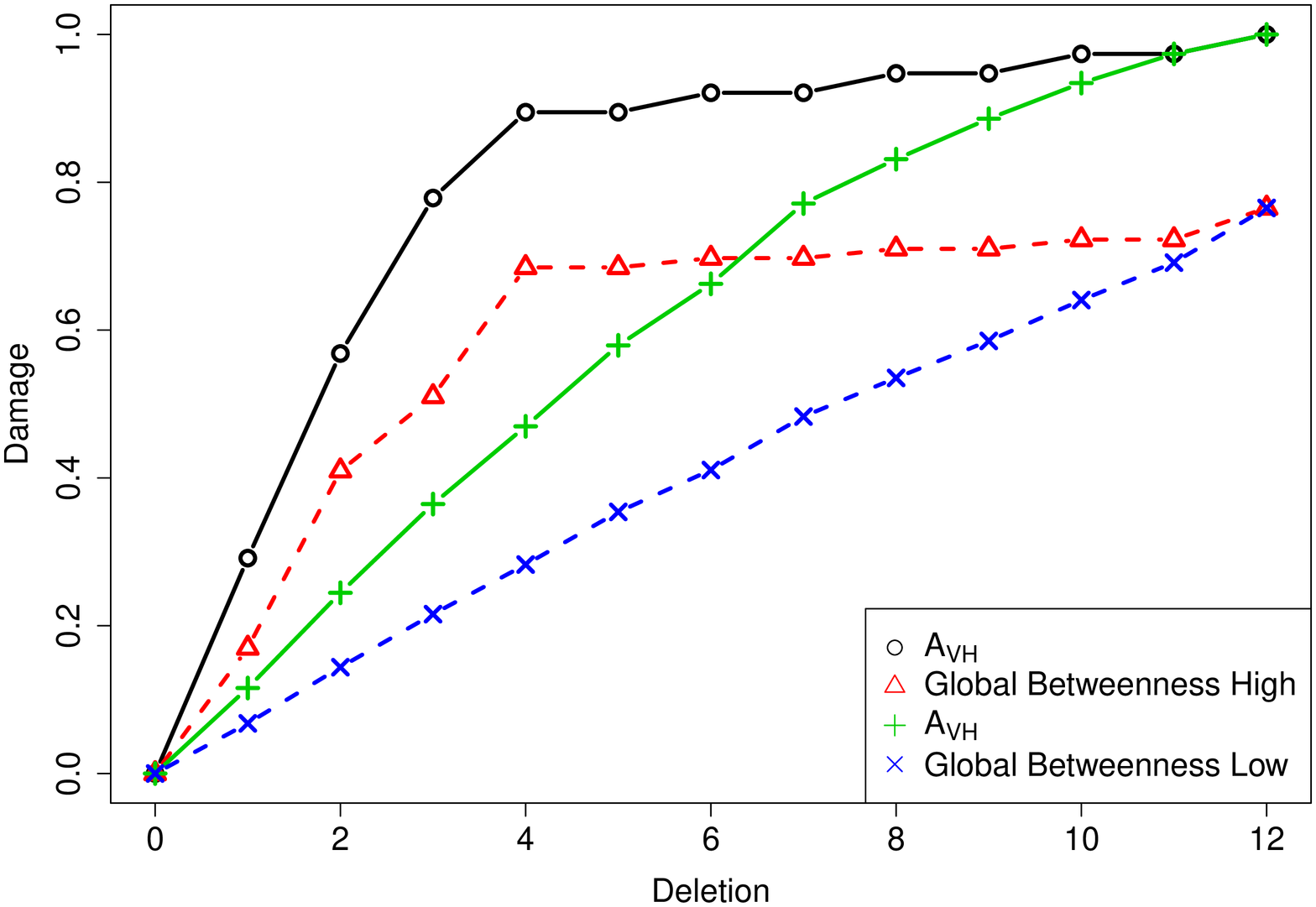}
\MyCaption{Damage to the discovered subgraph of path length 3 based on \MyAttackNotation{V}{*} attack profiles.}{
The ``local'' values are those that come from the discovered graph, while the 
global values are from the total graph.  Damage inflicted on the discovered graph when 
using the high vertex betweenness value and the resulting impact on the total
graph are show in black and red respectively.  In a similar manner, damage caused by choosing the low 
betweenness is shown in the green and blue lines respectively.
The betweenness of each vertex
 is recomputed after the
removal of either the highest or lowest betweenness valued vertex.  The process is repeated again and again until all vertices
are removed.  Damage to the global graph is flat from deletion 4 through 11, while the local damage increases due to the selection
of the particular high valued vertices to remove.  The low betweenness option does not show this type of behavior.
The system of graphs for high and low selection is shown in Figure \ref{fig:damage:vertex:explain}. }
{fig:damage:vertex}
\end{figure}

\begin{figure}
\centering
\MyLocalSubfigure{Results of \MyAttackNotation{V}{H}}{fig:delete:vertex:high:4}{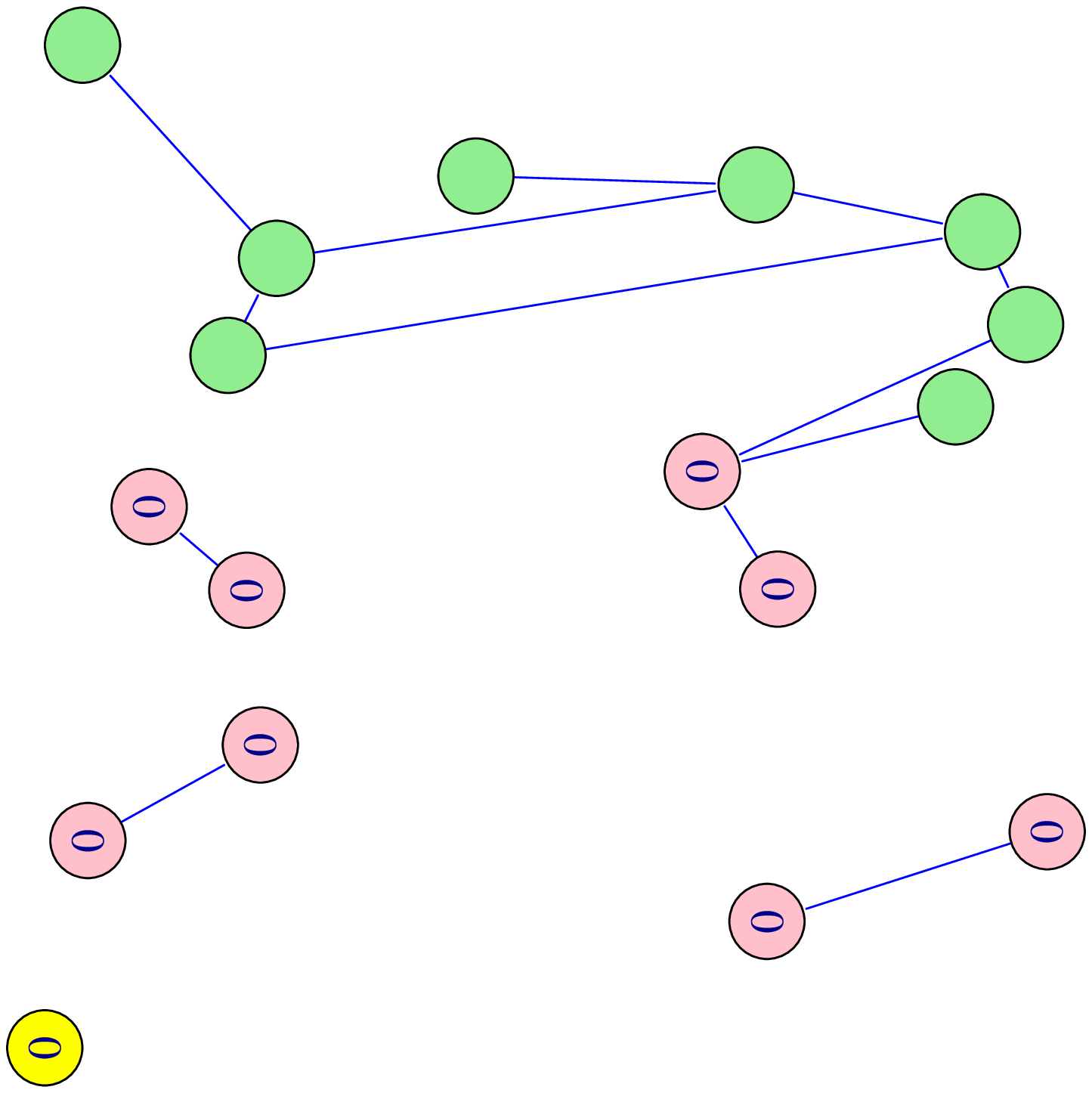}
\MyLocalSubfigure{Results of \MyAttackNotation{V}{L}}{fig:delete:vertex:low:4}{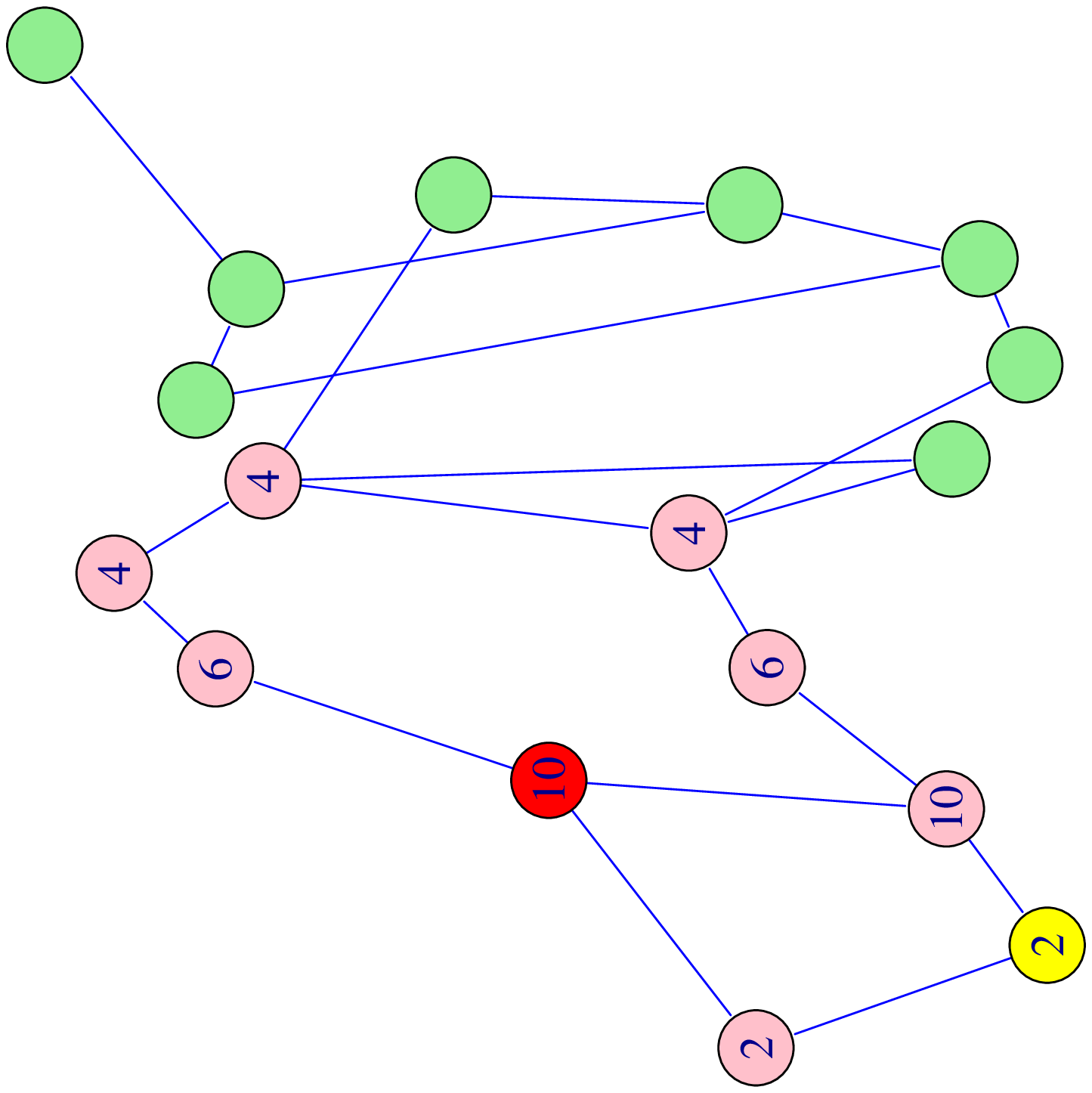}
\MyCaption{Markedly different graphs resulting from the differences in choosing \MyAttackNotation{V}{H} or 
\MyAttackNotation{V}{L} attack profiles.}
{Both subfigures show the sample graph after 4 deletions based on \MyAttackNotation{V}{H} or \MyAttackNotation{V}{L} 
attack profiles.
Continued deletions in the  discovered graph (in pink) in the high betweenness case \MyFigureReference{fig:delete:vertex:high:4},
will have only marginal effect on the global graph (the union of pink and green).  Deletions 
in the discovered graph in low betweenness case \MyFigureReference{fig:delete:vertex:low:4} will continue
to affect the union of the pink and the green nodes because the two graphs (pink and green) are still connected.
 Some of the vertices are unlabeled because the attacker has not ``discovered'' them.}
{fig:damage:vertex:explain}
\end{figure}

\subsubsection{Degree selection}
The attacker can compute the degreeness of any vertex in the subgraph that he has discovered \MyTableReference{tbl:neighborhood:degree}.  Based on these
values, the attacker can select either the highest or lowest valued vertex to remove.  After the removal
of this vertex, the degreeness values can be recomputed for the newly modified subgraph and the process repeated again and again
until there are no vertices left in the discovered graph (the discovered graph is totally destroyed).

Figures \ref{fig:delete:degree:low} and \ref{fig:delete:degree:high} show the effects of repeatedly applying attack \MyAttackNotation{D}{L} or
\MyAttackNotation{D}{L} profiles to the  
 discovered subgraph of path length 3.  In each figure, the degreeness value of each vertex is written in the vertex.  The
edge with the lowest \MyFigureReference{fig:delete:degree:low} or highest \MyFigureReference{fig:delete:degree:high} betweenness value
is highlighted in yellow, prior to it being removed.  After the removal of the vertex, the degreeness values of all the
remaining vertices are computed shown in the next subfigure, along with the next vertex that has been selected for removal.  The four subfigures in Figures \ref{fig:delete:degree:low} and \ref{fig:delete:degree:high} show this process.  When two or more vertices have the same degreeness value, the
selection of which edge to remove it totally random.

Attack profile \MyAttackNotation{D}{L} tends to attack the periphery of the subgraph.  While attack
profile \MyAttackNotation{D}{H} tends to attack the core of the graph.  
While both selection choices will result in a fully disconnected graph with the same number of
removals, selecting the highest valued vertex causes more damage quicker.

The betweenness computation, removal and damage computation process is shown in Table \ref{tbl:damage:degree} and Figure \ref{fig:damage:degree}.
The Global High line in Figure \ref{fig:damage:degree} goes flat after the fifth deletion while the Global Low line continues to 
increase.  This behavior is explained by looking at Figures \ref{fig:delete:degree:high:5} and \ref{fig:delete:degree:low:5}.  Using a 
\MyAttackNotation{D}{H} profile, the discovered and global graphs are disconnected and further local deletions do not affect the
global graph.  Using  \MyAttackNotation{D}{L} profile in Figure \ref{fig:delete:degree:low:5} results in the 
discovered and global graphs still being connected, so any deletions on the discovered graph affect the global graph.
the fifth deletion the discovered and global graphs are
\begin{figure}
\centering
\MyLocalSubfigure{First lowest has been identified}{fig:delete:degree:low:0}{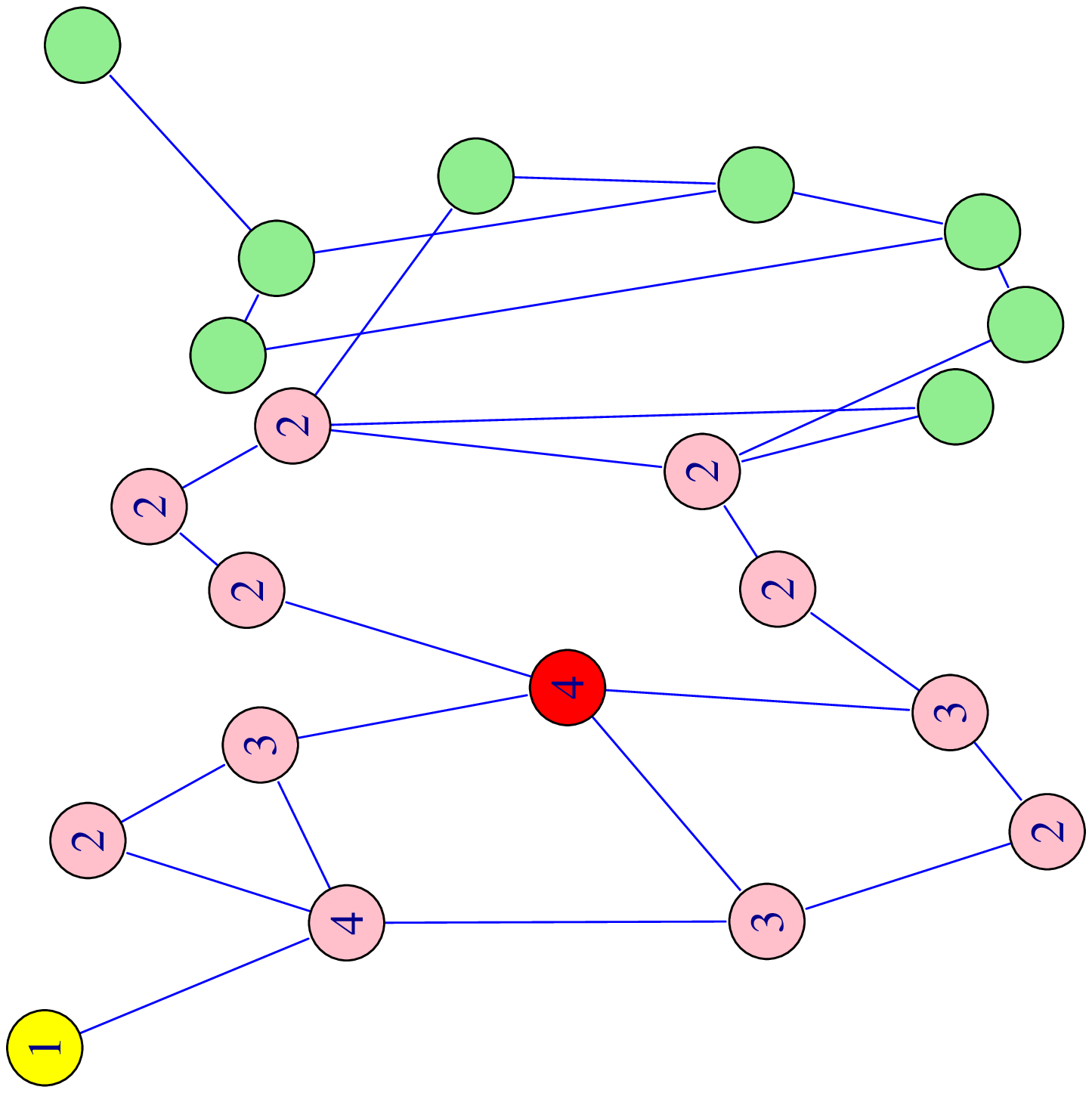}
\MyLocalSubfigure{Previous lowest has been removed, new lowest identified}{fig:delete:degree:low:1}{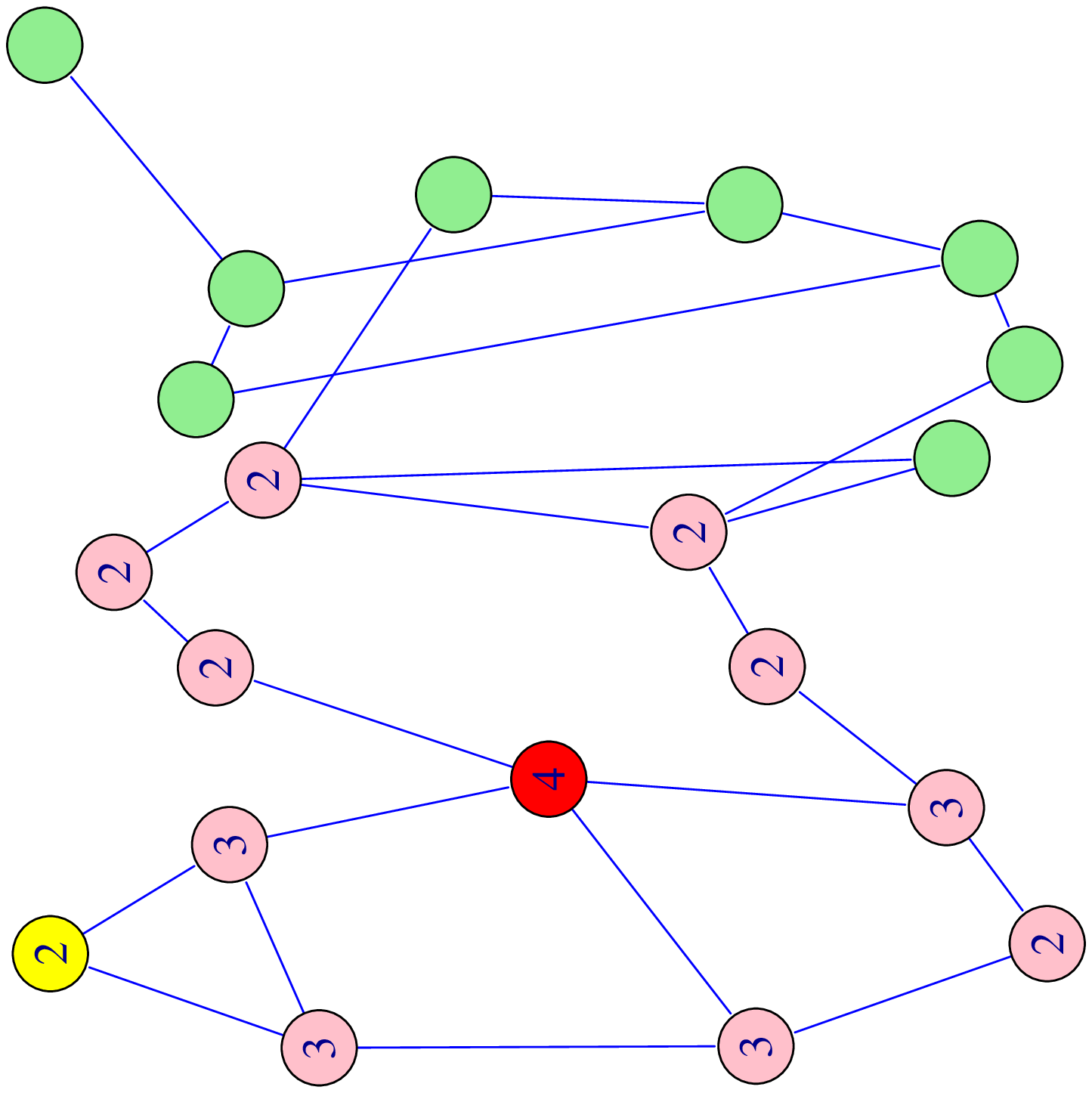}
\MyLocalSubfigure{Previous lowest has been removed, new lowest identified}{fig:delete:degree:low:2}{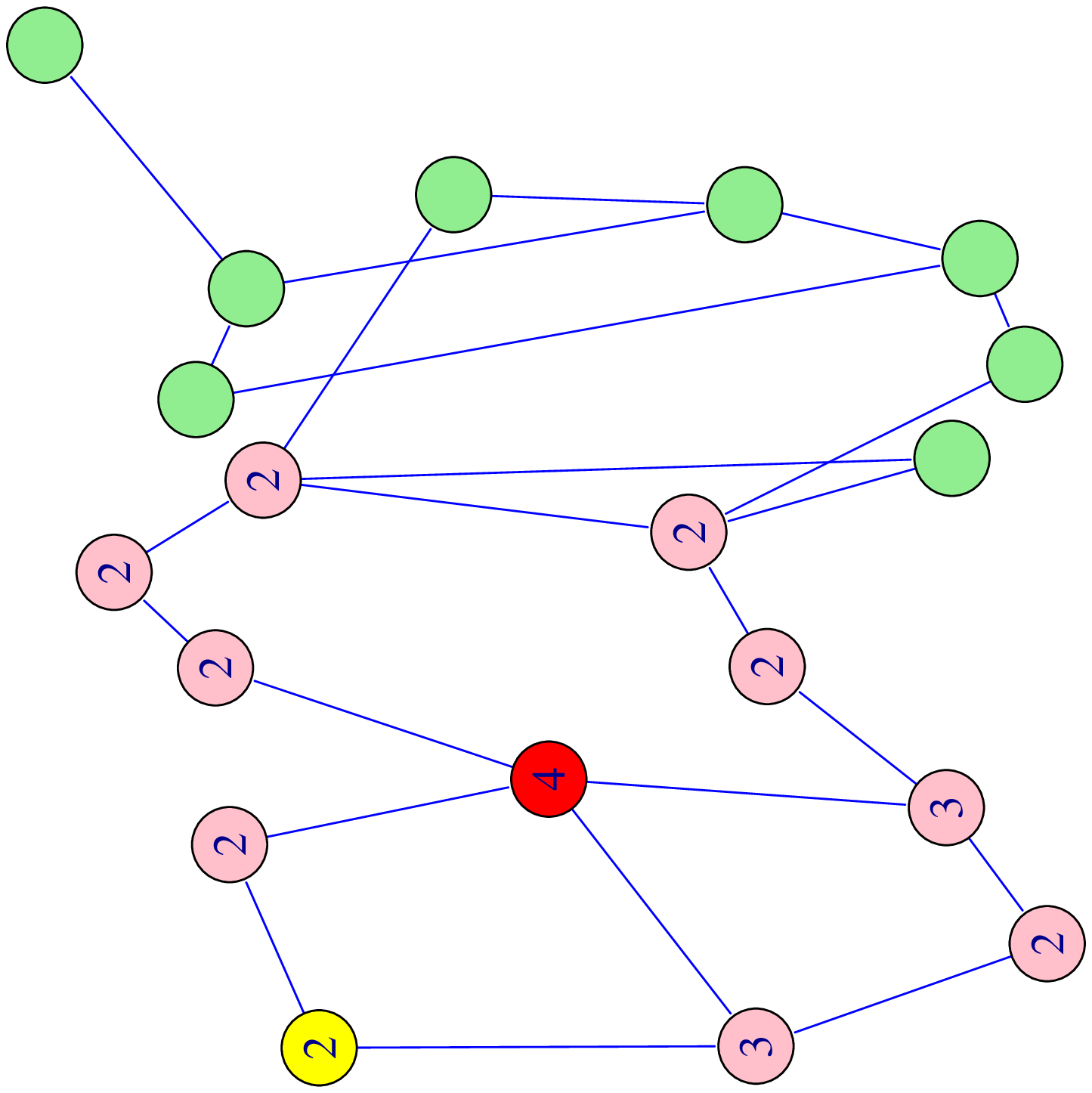}
\MyLocalSubfigure{Previous lowest has been removed, new lowest identified}{fig:delete:degree:low:3}{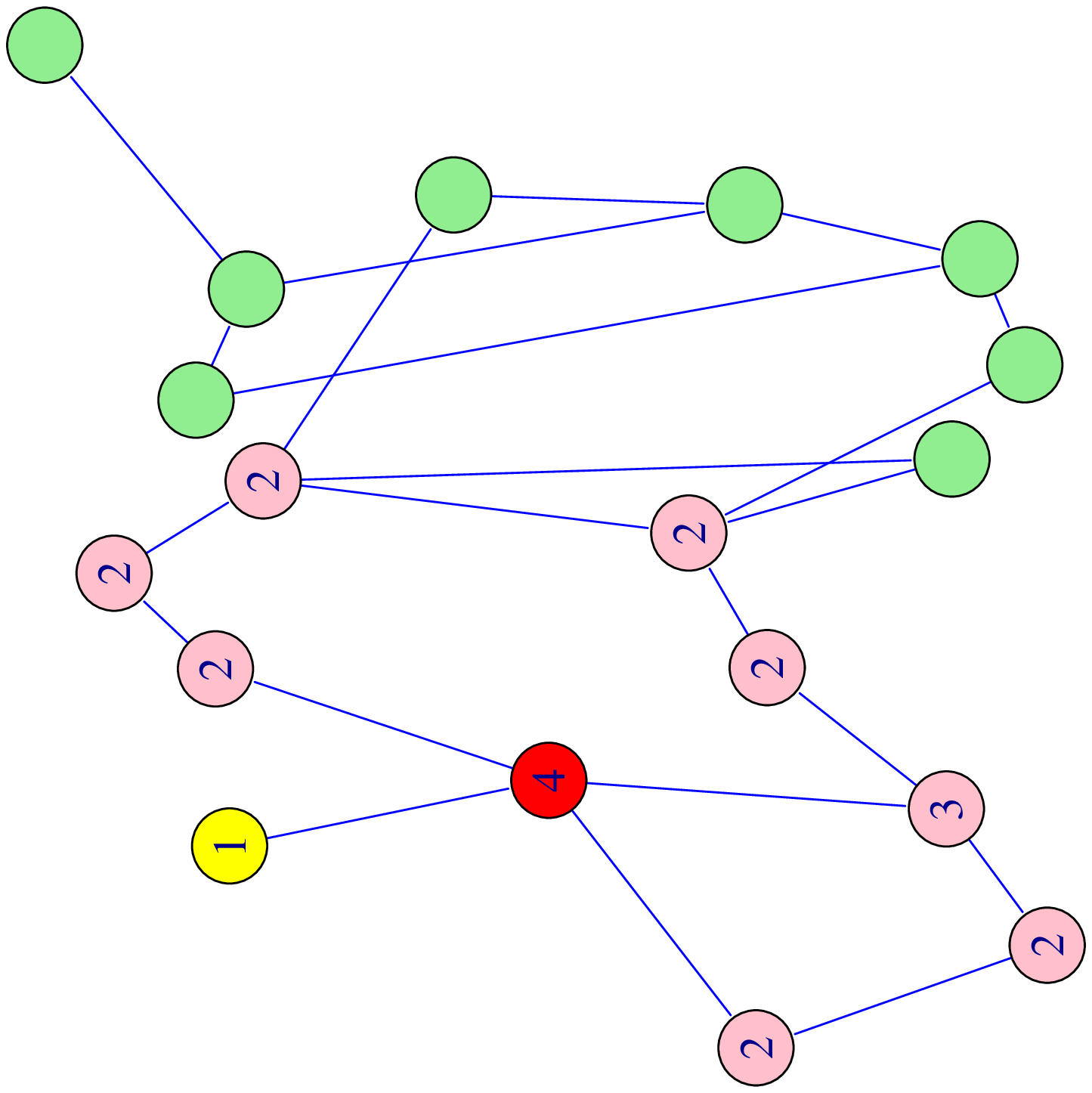}
\MyCaption{The effects of an \MyAttackNotation{D}{L} attack profile on the sample graph.}
{Vertex 5 (marked in red) is the center vertex.  The discovered graph is at a distance of 3 from the center vertex.  The vertex with the
lowest degree is marked in yellow.  In the case where multiple vertices have the same degree value \MyFigureReference{fig:delete:degree:low:1},
random choice is used to select one vertex as the next one to be removed.
Removal of a vertex causes a reduction in the degree values of all of the removed vertex's neighbors.  This change in the
degreeness of potentially many vertices requires that the relative order of the vertices be evaluated after each removal.
 Some of the vertices are unlabeled because the attacker has not ``discovered'' them.}
{fig:delete:degree:low}
\end{figure}
\begin{figure}
\centering
\MyLocalSubfigure{First highest has been identified}{fig:delete:degree:high:0}{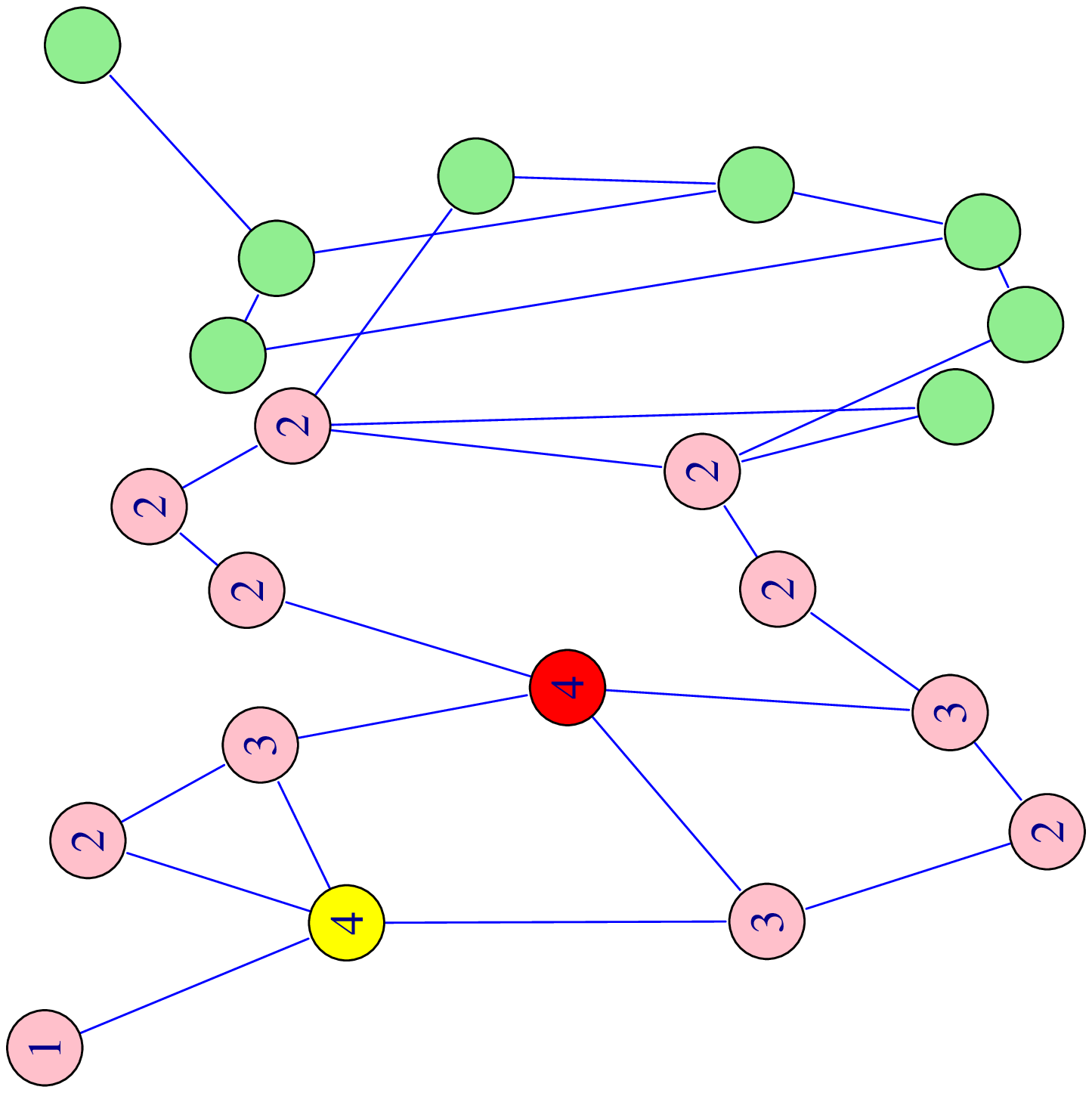}
\MyLocalSubfigure{Previous highest has been removed, new highest identified}{fig:delete:degree:high:1}{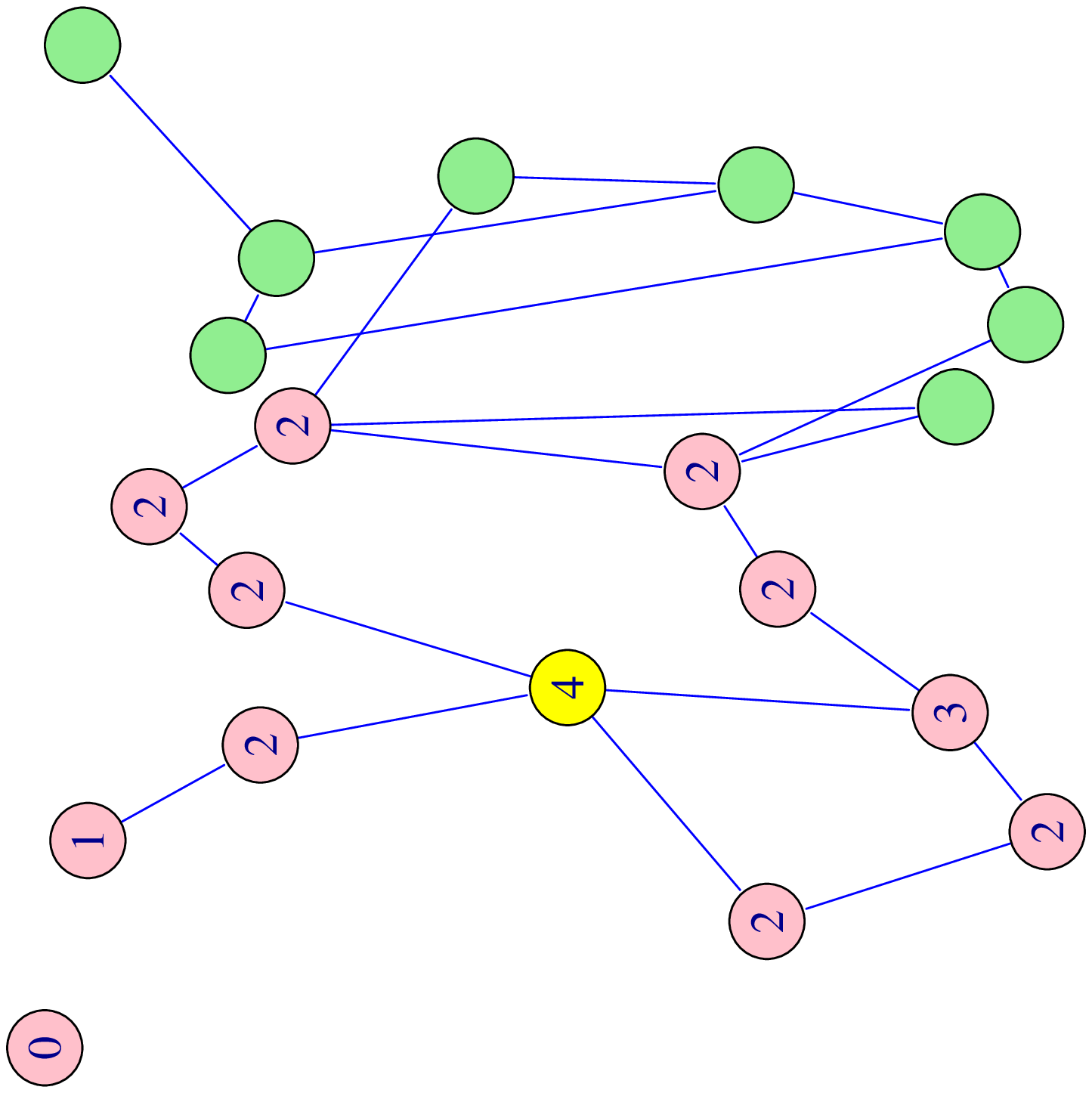}
\MyLocalSubfigure{Previous highest has been removed, new highest identified}{fig:delete:degree:high:2}{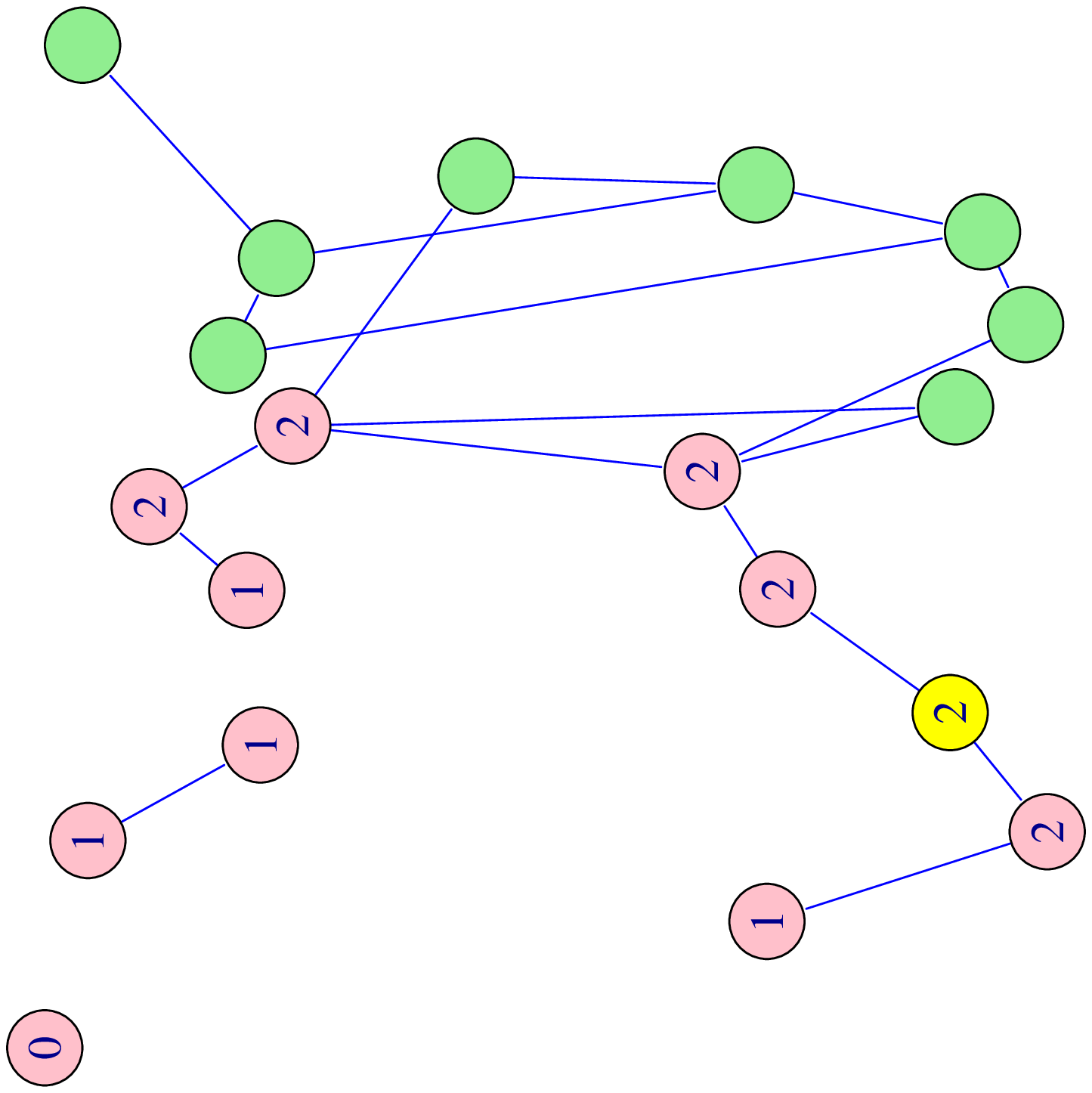}
\MyLocalSubfigure{Previous highest has been removed, new highest identified}{fig:delete:degree:high:3}{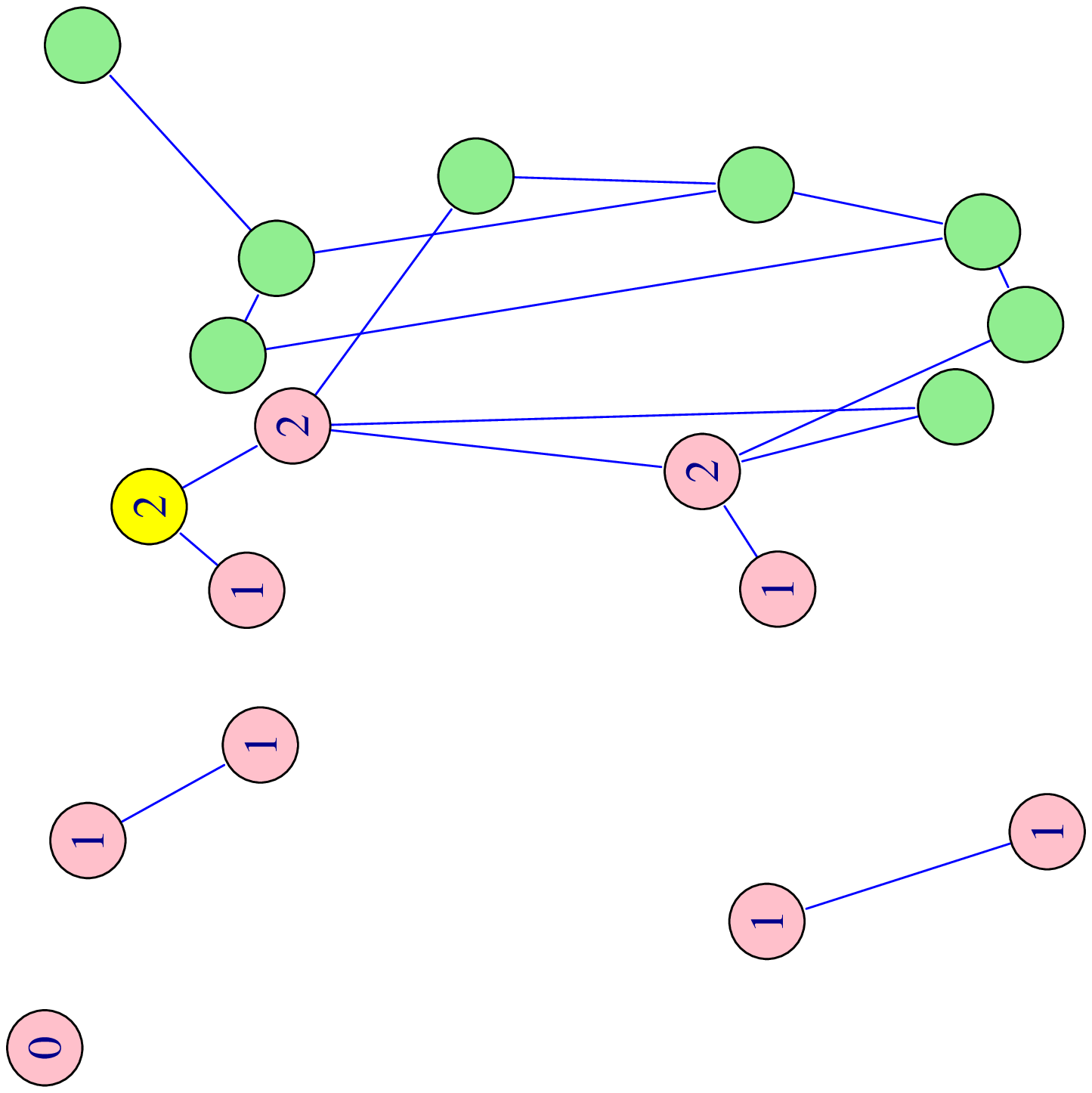}
\MyCaption{The effects of on \MyAttackNotation{D}{H} attack profile on the sample graph.}
{Vertex 5 (marked in red) is the center vertex.  The discovered graph is at a distance of 3 from the center vertex.  The vertex with the
highest degree is marked in yellow.  In the case where multiple vertices have the same degree value \MyFigureReference{fig:delete:degree:high:2},
random choice is used to select one vertex as the next one to be removed.
Removal of a vertex causes a reduction in the degree values of all of the removed vertex's neighbors.  This change in the
degreeness of potentially many vertices requires that the relative order of the vertices be evaluated after each removal.
 Some of the vertices are unlabeled because the attacker has not ``discovered'' them.}
{fig:delete:degree:high}
\end{figure}

\begin{table}
 \centering
\setlength{\extrarowheight}{3pt}
\begin{tabular} {|*{5}{c|}}
\MyHline
\multicolumn{1}{|p{0.75in}}{\textbf{Deletion}} &
\multicolumn{1}{|p{0.75in}}{\textbf{Local damage due to \MyAttackNotation{D}{H}}} &
\multicolumn{1}{|p{0.75in}}{\textbf{Global damage due to local damage by \MyAttackNotation{D}{H}}} &
\multicolumn{1}{|p{0.75in}}{\textbf{Local damage due to \MyAttackNotation{D}{L}}} &
\multicolumn{1}{|p{0.75in}|}{\textbf{Global damage due to local damage by \MyAttackNotation{D}{L}}} \\
\hline
0
&
0.00
&
0.00
&
0.00
&
0.00
\\
1
&
0.27
&
0.16
&
0.12
&
0.07
\\
2
&
0.61
&
0.37
&
0.24
&
0.14
\\
3
&
0.78
&
0.51
&
0.36
&
0.22
\\
4
&
0.88
&
0.62
&
0.47
&
0.28
\\
5
&
0.95
&
0.74
&
0.58
&
0.35
\\
6
&
0.97
&
0.75
&
0.66
&
0.41
\\
7
&
1.00
&
0.76
&
0.77
&
0.48
\\
8
&
1.00
&
0.76
&
0.83
&
0.54
\\
9
&
1.00
&
0.76
&
0.89
&
0.59
\\
10
&
1.00
&
0.76
&
0.93
&
0.64
\\
11
&
1.00
&
0.76
&
0.97
&
0.69
\\
12
&
1.00
&
0.76
&
1.00
&
0.77
\\
\MyHline
\end{tabular}

\MyCaption{Damage to the discovered subgraph of path length 3 based on  \MyAttackNotation{D}{*} attack profiles.}
{The degree of each vertex is computed after each deletion.  A vertex's degree value will change if one of it's
immediate neighbor vertices has been removed.  The removal of a neighbor will reduce the degreeness of all its
neighbors by one.  This change in the degreeness of all neighboring vertices may affect the relative order of all vertices
based on their respective degreeness.
The process is repeated again and again until all edges
are removed.}
{tbl:damage:degree}
\end{table}
\begin{figure}
 \centering
\includegraphics[width = 4.5in, angle = -90]{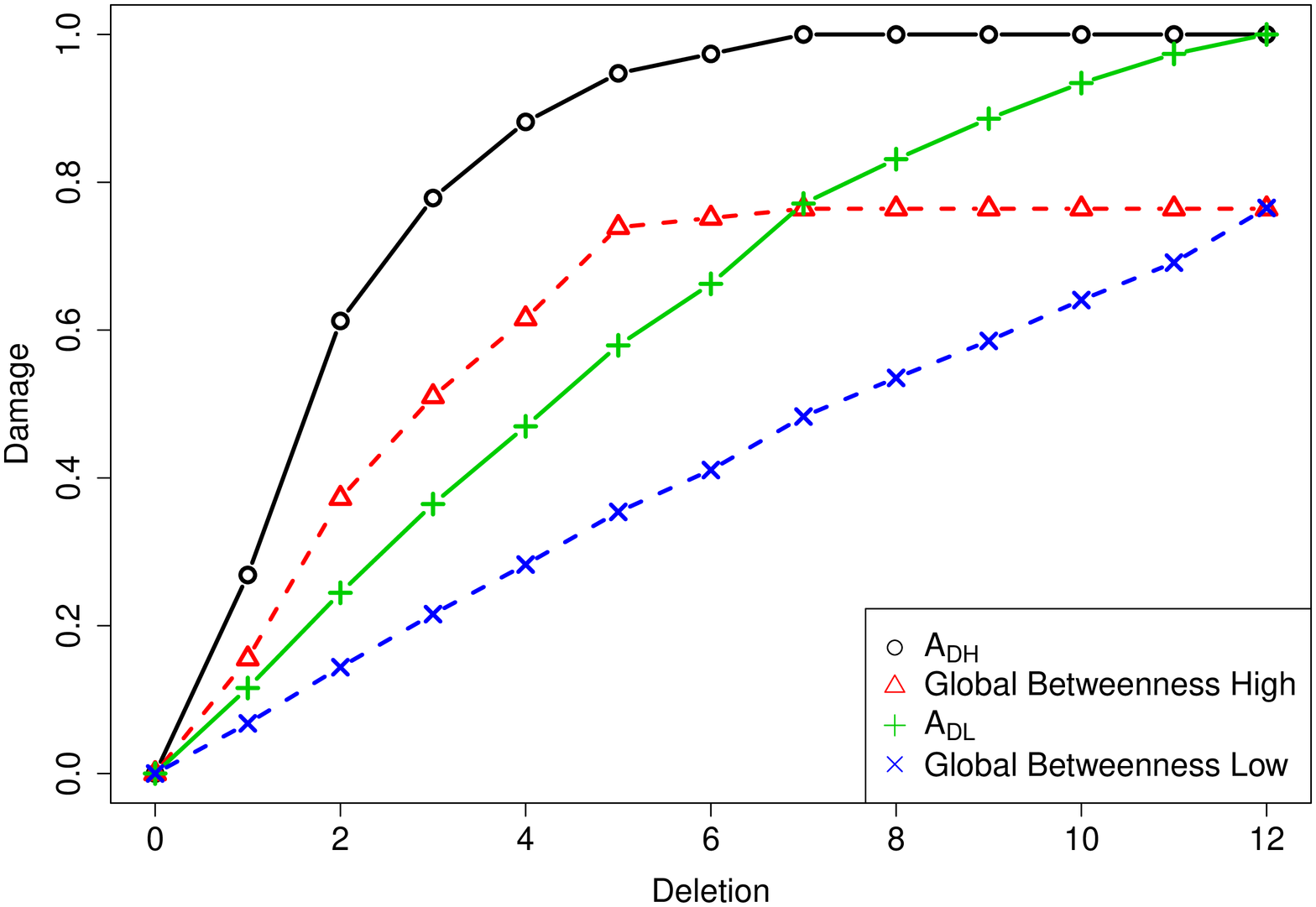}
\MyCaption{Damage to the discovered subgraph of path length 3 by based on \MyAttackNotation{D}{*} attack profiles.}
{The degree of each vertex is computed after each deletion.  A vertex's degree value will change if one of it's
immediate neighbor vertices have been removed.  The removal of a neighbor will reduce the degreeness of all its
neighbors by one.  This change in the degreeness of all neighboring vertices may affect the relative order of all vertices
based on their respective degreeness.
The process is repeated again and again until all vertices
are removed. The flat area on the Global High line 
is related to the discovered and global graphs becoming
disconnected \MyFigureReference{fig:damage:degree:explain}.}
{fig:damage:degree}
\end{figure}

\begin{figure}
 \centering
\MyLocalSubfigure{High degree}{fig:delete:degree:high:5}{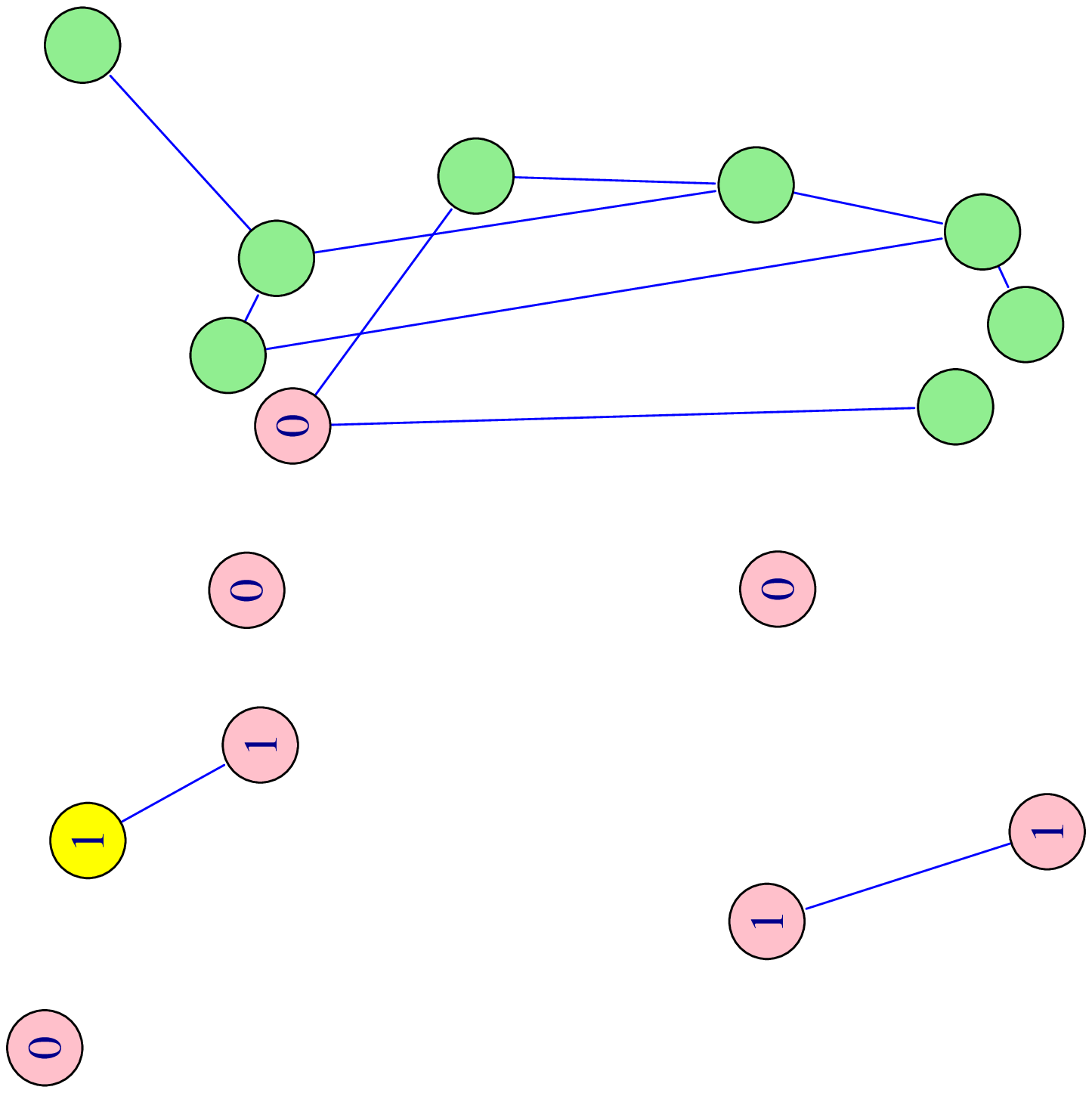}
\MyLocalSubfigure{Low degree}{fig:delete:degree:low:5}{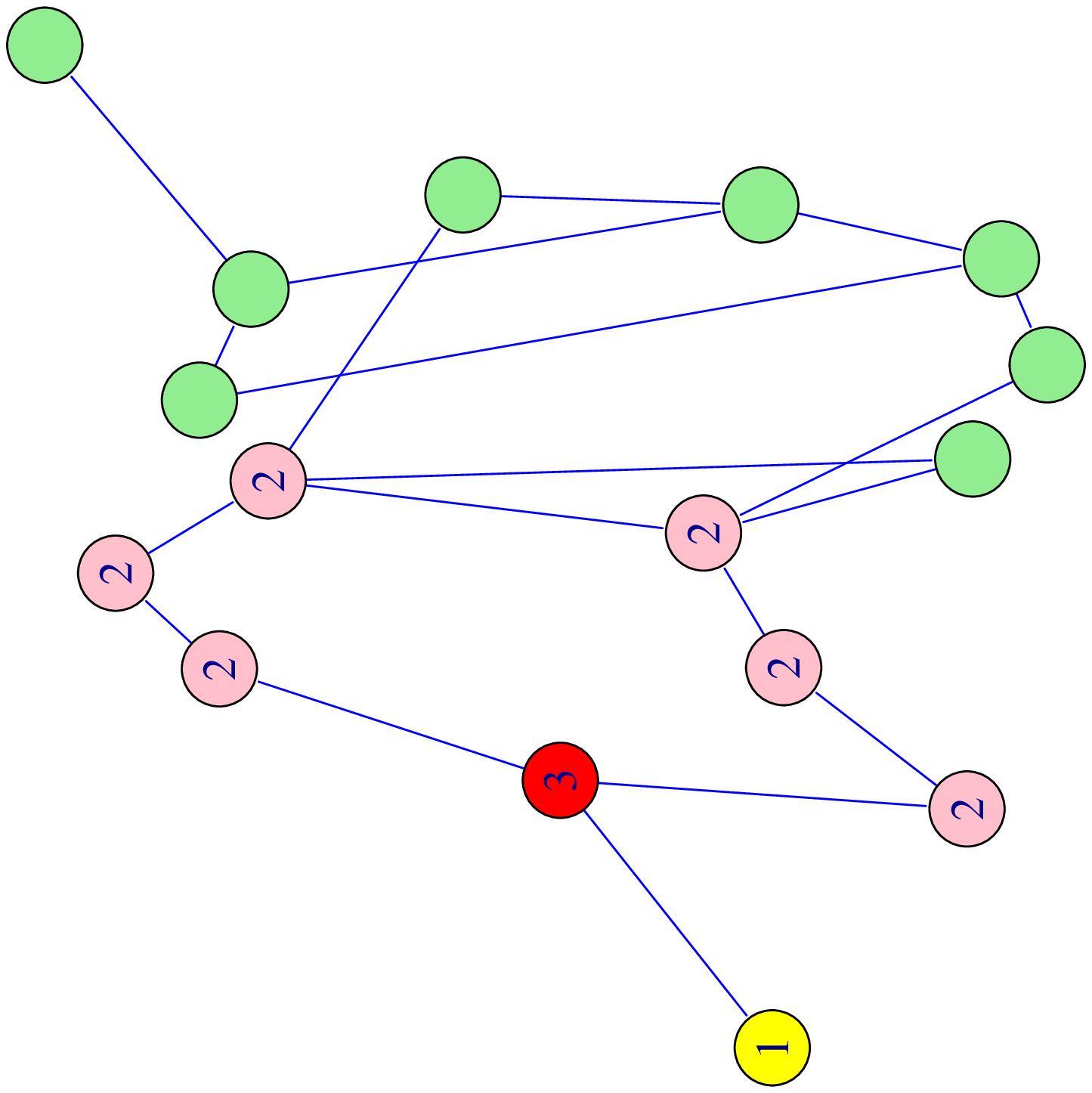}
\MyCaption{The sample graph after removing the fifth discovered node using 
\MyAttackNotation{D}{*} attack profiles.}
{The undiscovered graph is drawn in green. The central vertex, where it remains
is drawn in red  \MyFigureReference{fig:delete:degree:low:5}.  The vertex that will be deleted next
is drawn in yellow.  While each graph shows the effects of five deletions, selecting the
highest degreed node to delete results in a graph that is disconnected \MyFigureReference{fig:delete:degree:high:5}.
Focusing on the lowest degreed node results in damage to the periphery and a graph that is 
still connected \MyFigureReference{fig:delete:degree:low:5}.  Some of the vertices are unlabeled because the attacker has not ``discovered'' them.}
{fig:damage:degree:explain}
\end{figure}
\subsection{Attack profile conclusions}
All node based attacks (\MyAttackNotation{V}{*}, \MyAttackNotation{D}{*}) will 
totally destroy the discovered graph.  All edge based attacks \MyAttackNotation{E}{*}
will cause the discovered graph to be totally disconnected.  The two attack
philosophies differ in their efficacy and are summarized in Table \ref{tbl:attack:efficacy}.

If the attacker's goal is to disconnect the sample graph by repeated use of the same
attack profile, then the most effective profiles in order are: \MyAttackNotation{E}{H},
\MyAttackNotation{V}{H} and \MyAttackNotation{D}{H}.
\begin{table}
\centering
\setlength{\extrarowheight}{3pt}
\begin{tabular}{|c|p{2.75in}|}
\MyHline
\multicolumn{1}{|p{0.75in}}{\textbf{Attack Profile}} &
\multicolumn{1}{|c|}{\textbf{Efficacy}} \\
\hline
\MyAttackNotation{D}{H} & Tends to attack the core of the graph\\
\MyAttackNotation{D}{L} & Tends to attack the periphery of the graph\\
\MyAttackNotation{E}{H} & Tends to attack the core of the graph\\
\MyAttackNotation{E}{L} & Tends to attack the periphery of the graph\\
\MyAttackNotation{V}{H} & Tends to attack the core of the graph\\
\MyAttackNotation{V}{L} & Tends to attack the periphery of the graph\\
\MyHline
\end{tabular}
\MyCaption{Efficacy of various attack profiles.}
{In general, regardless of the attack profile utilized, 
attacking the highest valued component is the most destructive.}
{tbl:attack:efficacy}
\end{table}

\newpage
\end{document}